\documentclass[preprint,11pt]{elsarticle}
\usepackage{paper_draft_style}
\usepackage{MyNotations}
\usepackage{ulem}
\begin{document}

\begin{frontmatter}
    \title{On the limits of the energetic coupling between field dislocation mechanics and phase field crystal}

\author[add1]{Aymane Graini}
\author[add2]{Jorge Vi\~{n}als}
\author[add1]{Manas V. Upadhyay\corref{cor1}}

\affiliation[add1]{organization={Laboratoire de Mécanique des Solides (LMS), École Polytechnique, CNRS UMR 7649, Institut Polytechnique de Paris},
            addressline={Route de Saclay},
            city={Palaiseau},
            postcode={91128},
            country={France}}

\affiliation[add2]{organization={School of Physics and Astronomy, University of Minnesota},
            addressline={116 Church St. SE},
            city={Minneapolis},
            postcode={55455},
            country={MN, USA}}

\ead{manas.upadhyay@polytechnique.edu}
\cortext[cor1]{Corresponding author}

\begin{abstract}
This paper investigates the energetic coupling between Field Dislocation Mechanics (FDM) and the Phase Field Crystal (PFC) model proposed in Phys. Rev. B 102, 064109, 2020. 
While FDM correctly solves the initial boundary value problem of a continuum body with dislocation fields, PFC captures the underlying crystallographic structure. 
The coupling, which penalizes the $L^2$ distance between elastic distortion from FDM and configurational distortion from PFC in the $L^2$ sense, had been proposed to reconcile dislocation mechanics with crystallography in a single continuum framework. 
Variational analysis reveals that the coupling term acts as a divergence-driven forcing in the phase-field evolution that matches only the compatible (curl-free) parts of the distortion fields. 
Consequently, its contributions are insensitive to the incompatible (divergence-free) elastic distortion carrying all the information on dislocation topology. 
Furthermore, the nature of the configurational distortion causes mechanical boundary conditions to be transmitted diffusively from FDM to PFC rather than elastically. 
Numerical simulations demonstrate that this coupling cannot prevent the unnatural core spreading in FDM. 
Finally, it is shown that even in the most general case, an energetic coupling suffers from the same drawbacks, which limits its ability to integrate dislocation mechanics with crystallography.
\end{abstract}


\end{frontmatter}

\normalsize
\section{Introduction}
\sloppy
Dislocation-mediated plastic deformation in crystalline materials results from nucleation, transport, and interaction of dislocations with other defects, including other dislocations. 
These mechanisms are governed by multi-scale phenomena such as the local crystalline order and the associated energy landscape, resistance to evolution, imposed boundary conditions and elastic stress equilibrium or dynamics.
In turn, dislocations play a significant role in determining the local and macroscopic responses ranging from Type III residual stresses to non-linear geometric changes. 
Ideally, a plasticity model that resolves individual dislocations should capture all of these phenomena; however, their reconciliation has remained theoretically and computationally an open challenge due to the highly complex nature of the crystallographic dislocation mechanics problem.

Molecular dynamics \citep{zhouLargescalemolecular1998} simulations resolve individual crystallographic dislocations and appropriately capture their mechanics under applied loading. However, they also resolve the dynamics of individual atoms, which becomes computationally very demanding at length (micrometers) and time (more than a few nanoseconds) scales where dislocation structures form and evolve. 
Meanwhile, at the meso-scale, single crystal plasticity models have been developed to simulate elasto-plastic deformation of crystalline domains under quasi-static loading while capturing non-linear geometrical changes at the micrometer level due to dislocation slip \citep{wijnenDiscreteSlipPlane2021}; such approaches typically do not resolve individual dislocations but model their average response using physics-based or empirical laws.
As a trade-off between atomic-level and homogenized single crystal  models, discrete dislocation dynamics models were developed. 
They focus only on the dynamics of individually resolved dislocations, thus making significant gains in length and time scales over molecular dynamics simulations. 
They treat dislocations as singular lines whose transport is driven by the Peach-Koehler force \citep{devincreThreeDimensionalSimulations1992}. 
However, extending discrete dislocation dynamics models to finite deformations while appropriately capturing non-linear effects close to the dislocation core under quasi-static loading conditions remains elusive. 

Thus, there has been a long-standing need for a modeling framework that can thermodynamically consistently account for the evolution and interaction of individually resolved dislocations, and nonlinear mechanical responses under quasi-static loading conditions and dynamic loading used in mechanical testing.
To that end, there are two potential candidate models that have been developed independently since the start of the 21st century: Field Dislocation Mechanics (FDM) and the Phase Field Crystal (PFC). 

FDM involves representing dislocations continuously via a polar density field instead of a singular representation of dislocations as done in discrete dislocation dynamics models. 
The resulting elastic and kinematic fields remain finite everywhere in the domain, including the dislocation cores, and the initial boundary value problem (IBVP) can be well-defined and regularized in the entire domain.
Importantly, in FDM, the polar dislocation density field is evolved via the conservation of Burgers vector equation \citep{acharyamodelcrystal2001, acharyaMicrocanonicalEntropy2011}, extending the dynamical theory of moving dislocations \citep{muraContinuousdistribution1963, kosevichDYNAMICALTHEORY} to a fully tensorial formulation.
The remaining governing equations are derived from conservation of mass, momenta and energy and the constitutive equations are obtained from energetic and dissipation considerations akin to rational thermodynamics.
The framework has been shown to determine the evolution of elastic stresses and dislocation densities by solving the initial boundary value problem for a solid under quasi-static and dynamic loadings in finite deformations \citep{aroraFiniteelement2020}. 
FDM was shown to predict fundamental features of dislocation dynamics: dislocation annihilation, loop expansion and emission from Frank-Read sources \citep{varadhanDislocationtransport2006}, formation of shear bands and steps \citep{royFiniteelement2005},  size effects in confined plasticity \citep{aroraMechanicsMicropillarConfined2022}, among others.

However, crystallography, which is essential in studying dislocation dynamics, has thus far entered in FDM as a constitutive prescription in the free energy density; it is typically introduced via an additive non-convex energy term, which is a function of plastic distortion (e.g. see \citep{zhangsingletheory2015}). 
This non-convex term is motivated from the Peierls-Nabarro construction and it introduces a (conservative) resisting force to mimic the energetic barrier imposed by the local lattice structure to dislocation motion occurring under the Peach-Koehler force. 
Under equilibrium conditions, the trade-off between these competing forces gives a finite size to the dislocation cores. 
This non-convex energy term allows direction-specific plastic deformation and can capture localization phenomena such as shear banding \citep{zhangsingletheory2015, aroraComputationalApproximationMesoscale2019}. 
However, this approach relies on a scalar plastic-distortion variable associated with a predefined slip system, rather than on a general tensorial formulation in which crystallographically preferred directions arise naturally from the lattice energy.
Furthermore, introducing plastic distortion--a path-dependent field--as an independent variable into the free energy density--a state variable--introduces fundamental inconsistencies into the  thermodynamic framework.

PFC is a phase field theory that captures the non-equilibrium dynamics of defected crystals with an atomic-level resolution \citep{elderModelingElasticityCrystal2002, elderModelingElasticPlastic2004}. 
It introduces a scalar field $\psi$, representing the atomic density, and a free energy functional parametrized in $\psi$ that is minimized by equilibrium lattice symmetries such as face-centered or body centered cubic, among others depending on the chosen form of the functional.
Individual dislocations are stable topological defects in PFC \citep{berryDiffusiveatomistic2006, elderPhasefieldCrystalModeling2007,chanPlasticityDislocation2010} and conservation of Burgers vector is an outcome of the model, not a prescription \citep{skaugenDislocationdynamics2018}. 
However, PFC's main drawback that elastic behavior is incorrectly captured because its governing equation is solely obtained from dissipative considerations and evolves at a diffusive timescale \citep{skogvollphasefield2022}. 
Furthermore, because PFC uses the same scalar field to describe both atomic density and lattice distortion, it inherently links elastic relaxation to slow diffusive dynamics, which leads to unphysical mechanical behavior \citep{acharyaElasticityphase2022}.
Efforts have been made to address this shortcoming and they can be categorized into three main approaches.
One is a modification of the PFC governing equation by introducing a second-order time derivative to capture elasto-dynamics \citep{stefanovicPhaseFieldCrystalsElastic2006}.
This improves short-time stress relaxation, but remains phenomenological, does not correctly recover the full long-wavelength phonon behavior and is unsuitable for quasi-static loadings. 
Another approach \citep{skaugenSeparationElastic2018} supplements the phase field with a smooth correction at every timestep to ensure that mechanical equilibrium is established instantaneously (elastically) within the diffusive evolution of $\psi$. 
This approach requires the resolution of an auxiliary elasticity problem and cannot be formulated or justified thermodynamically.
Finally, the hydrodynamic PFC approach \citep{heinonenConsistentHydrodynamicsPhase2016, skogvollHydrodynamicphase2022} was proposed to couple PFC with macroscopic momentum conservation law through a velocity field. Even though this only improved modeling elastic relaxation in PFC, it did not address the ambiguity in describing both mass density and lattice distortion, which are physically independent quantities, with the same scalar order parameter $\psi$.

PFC and FDM are complementary approaches.
PFC contains sufficient crystallographic information to natively keep dislocation cores compact (prevent unphysical spreading during energy minimization) under equilibrium and non-equilibrium conditions, but it incorrectly captures elasticity. 
FDM, on the other hand, can correctly formulate and solve the dislocation mechanics IBVP but it requires prescribing crystallography to encode lattice energy and preserve localized cores. 
In view of this complementarity, Acharya and Vi\~{n}als \citep{acharyaFielddislocation2020} proposed a coupled PFC and FDM (PFCFDM) framework to address the aforementioned requirements of a dislocation mechanics model, while overcoming the limitations of each theory. 
In this unified PFCFDM model, the scalar field $\psi$ is treated as an indicator function of the crystalline order and defect structure. 
The configurational distortion arising from it does not directly provide the elastic energy; instead, the order parameter enters the free-energy density through a non-convex Swift-Hohenberg-type functional \citep{acharyaFielddislocation2020}. 
The coupling is introduced by penalizing the Frobenius norm of the mismatch between the configurational distortion inferred from the PFC field and the elastic distortion described by FDM. 
The purpose of this construction was twofold: endowing PFC with the correct elastic fields by solving FDM, and using PFC's ability to maintain compact dislocation cores to solve FDM without path-dependent quantities in the free-energy density.

This coupling was a first major attempt in introducing crystallography into FDM, and in general in dislocation mechanics models, without introducing path-dependent terms into the free energy density. 
The expectation from the coupling term was to treat the relaxation of the Swift-Hohenberg energy as a constrained minimization problem.
With a sufficiently large coupling coefficient, the phase field order parameter should relax faster to a local minimum corresponding to elastic equilibrium. 
This initial relaxation is expected to be followed by the slow evolution of defects driven by Peach-Koehler forces, while dislocation cores remain compact and localized due to the Swift-Hohenberg energy.

In this work, we perform a theoretical and numerical analysis of this coupling term and demonstrate that it does not yield the main expected outcomes and provide explanations to the reasons behind their limitations.
In doing so, we define the domains of applicability of this model.
The paper is organized as follows: Section 2 recalls the governing equations of PFC and FDM. Section 3 presents the coupled framework and the equations of the model. Section 4 discusses the mathematical structure of the problem and it's implications. Section 5 performs a numerical study of the implementation in different scenarios.

\section{Modeling framework}

\subsection{Small deformation field dislocation mechanics (FDM)}

The essential equations of FDM in the small-deformation limit are recalled below, starting with the kinematics of the problem. 
Consider a simply connected domain $\Omega$ with surface $\partial \Omega$ containing a distribution of individually resolved and continuously represented dislocations. 
The total displacement field is assumed to be smooth everywhere in $\Omega$ allowing defining the total distortion, $\Tens{U} = \nabla \vec{u}$, everywhere in $\Omega$. 
In the small deformation limit, $\Tens{U}$ can be additively decomposed into an elastic part $\Tens{U}^e$ and a plastic part $\Tens{U}^p$:
\begin{equation}\label{eqn:distortion}
    \Tens{U} = \nabla \vec{u} = \Tens{U}^e + \Tens{U}^p
\end{equation}

The total distortion is compatible i.e., $\nabla \times \Tens{U} = 0$, everywhere in the domain, whereas $\Tens{U}^e$ and $\Tens{U}^p$ will also contain an incompatibility (non-zero curl) at the location of dislocations. 
The incompatibility in the elastic distortion allows defining the Burgers vector
\begin{equation}\label{eqn:Burgers}
    \boldsymbol{b}:= \oint_C \boldsymbol{U^e} \cdot \text{d}\Tens{l} = \int_S \nabla \times \Tens{U}^e \cdot \Tens{n} \ \text{d}S
\end{equation}
where $C$ is an arbitrarily-shaped closed contour encircling a dislocation of Burgers vector $\Tens{b}$, and the integral over the surface $S$ bounded by $C$ is obtained from Stokes' theorem with $\Tens{n}$ being the normal to the surface.

The polar dislocation density tensor $\boldsymbol{\alpha}$ is defined as
\begin{equation}
    \boldsymbol{\alpha}:= \nabla \times \Tens{U}^e = - \nabla \times \Tens{U}^p
\end{equation}
where the second equality follows from \cref{eqn:distortion}.

The conservation of Burgers' vector (time derivative of \cref{eqn:Burgers}) yields the following transport equation for $\boldsymbol{\alpha}$ \citep{acharyaMicrocanonicalEntropy2011} in the absence of dislocation source/sink terms:
\begin{equation}\label{eq:Transport}
     \dot{\boldsymbol{\alpha}} + \nabla \times (\boldsymbol{\alpha} \times \vec{v}^d) = 0
\end{equation}
where $\vec{v}^d$ is the dislocation velocity field, which is constitutively prescribed.

An important and unique aspect of FDM is the use of the Stokes-Helmholtz-type decomposition of second-order tensors. 
The plastic distortion is uniquely decomposed into a compatible part $ \Tens{U}^{p \parallel} = \nabla \Tens{z^p}$ and an incompatible part $\Tens{U}^{p \perp} = \boldsymbol{\chi}^p$:
\begin{equation}
    \Tens{U}^p = \nabla \vec{z}^p + \boldsymbol{\chi}^p
\end{equation}

$\boldsymbol{\chi}^p$ respects the following conditions \citep{acharyaSizeeffects2006}:
\begin{equation}\label{eq:divcurl}
    \begin{cases}
    \nabla \times \boldsymbol{\chi}^p = -\boldsymbol{\alpha} & \text{in } \Omega \\
    \nabla \cdot \boldsymbol{\chi}^p = 0 & \text{in } \Omega \\
    \boldsymbol{\chi}^p \vec{n} = 0 & \text{on } \partial \Omega
    \end{cases}
\end{equation}
This set of equations is well-posed, yields a unique solution for $\boldsymbol{\chi}^p$, and reduces to $\boldsymbol{\chi}^p = 0$ when $\boldsymbol{\alpha} = 0$. 

$\Tens{U}^e$ can also be similarly decomposed into a compatible $\Tens{U}^{e\parallel}$ and an incompatible $\Tens{U}^{e\perp}$ part, with the latter being equal to $-\Tens{U}^{p \perp}$.

The compatible part $\nabla \vec{z}^p$, which represents the plastic history due to dislocation transport, evolves according to:
\begin{equation}\label{eq:parA_n}
    \nabla^2 \dot{\vec{z}}^p = \nabla \cdot (\boldsymbol{\alpha} \times \vec{v}^d)
\end{equation}
where $\nabla^2$ is the Laplacian operator.

Finally, the system is closed by the mechanical equilibrium equation:
\begin{equation}
    \nabla \cdot \, \boldsymbol{\sigma} = 0
\end{equation}
where $\boldsymbol{\sigma}$ is the Cauchy stress tensor that typically respects a linear constitutive relationship with elastic strain as
\begin{equation}
    \Tens{\sigma} = \mathbb{C}: \Tens{U}^e = \mathbb{C}: (\nabla \vec{u} - \Tens{U^p})
\end{equation}

\subsection{Small deformation phase field crystal (PFC)}
In the PFC framework, a scalar order parameter field $\psi(\vec{x}, t)$ and the phenomenological Swift-Hohenberg energy $\mathcal{F}_{sh}$ are used to describe the crystalline system. In two spatial dimensions, and for a hexagonal lattice, the free energy is given by \citep{elderModelingElasticityCrystal2002}:
\begin{equation}
    \mathcal{F}_{sh}[\psi] = \int_{\Omega} \left[ \frac{1}{2}\psi \left( -r + (\nabla^2 + 1)^2 \right) \psi + \frac{1}{4} \psi^4 \right] \text{d}\Omega
\end{equation}
where $r$ is a dimensionless parameter representing the quench depth, that is, the distance from the liquid state toward the ordered crystalline state. 
Physically, quenching refers to driving the system away from a homogeneous, liquid configuration into a regime where periodic density variations become energetically favorable. 
In this sense, increasing $r$ promotes crystalline ordering and is associated with a more pronounced lattice structure, which also affects the  stiffness of the system.
For $r < 0$, the ground state of $\mathcal{F}_{sh}$ is a spatially uniform field, representing a disordered liquid phase. 
When $r > 0$, the system relaxes to an ordered state. 
The critical value $r=0$ corresponds to the bifurcation point.

The resulting equilibrium structure depends on the values of $r$ and the average order parameter $\overline{\psi} = \frac{1}{|\Omega|} \int \psi \text{d}\Omega$. 
By constructing a phase diagram in the $(r, \overline{\psi})$ space, parameters can be selected such that either the hexagonal lattice or the striped phase is the global minimizer in two dimensions.

In the classical Phase Field Crystal model $\psi$ is interpreted as a normalized local atomic density \citep{elderModelingElasticityCrystal2002}, requiring the conservation of total mass. Consequently, the time evolution of $\psi$ follows conserved dynamics ($H^{-1}$ gradient flow):
\begin{equation}
    \dot{\psi} = \nabla^2 \frac{\delta \mathcal{F}_{sh}}{\delta \psi}
\end{equation}
However, we consider here mass density and lattice distortion (the latter represented by the phase field) as independent variables \citep{acharyaFielddislocation2020}. We model the phase field evolution as non-conserved dynamics under the constraint that the average order parameter $\overline{\psi}$ is constant,
\begin{equation}
    \dot{\psi} = -\frac{\delta \mathcal{F}_{sh}}{\delta \psi}
\end{equation}

Close to the bifurcation point $r \ll 1$, a multiple scale expansion leads to a representation of $\psi$ in terms of complex amplitudes $A_{n}$ that change slowly in space and time \citep{re:cross93}:
\begin{equation}\label{eq:psi_decomp}
    \psi = \overline{\psi} + \sum_n \left( A_ne^{i \vec{q}_n \cdot \vec{x}} + \text{c.c.} \right) + \ldots
\end{equation}
where the summation is over the $n$ reciprocal lattice vectors $\Tens{q}_n$ that are resonant at the bifurcation point \citep{skogvollphasefield2022}.

At every time step, the complex amplitudes are obtained via the following demodulation $\langle \cdot \rangle \;$:
\begin{equation} \label{eq:CmpA}
    A_n \approx \langle \psi e^{-i \vecc{q_n}\cdot \vecc{x}}\rangle
 \end{equation}
which is performed through a convolution with a Gaussian $\mathcal{G}_{a_0}$:
\begin{equation}\label{eq:demodulation}
    \langle X\rangle(\vecc{x}) = (X * \mathcal{G}_{a_0})(\vecc{x}) = \frac{1}{(2\pi a_0^2)^{d/2}}\int_\Omega dr' X(\vecc{x})  \: e^{-\cfrac{\|\vecc{x}-\vecc{x}'\|^2}{2a_0^2}}
 \end{equation}

Determining the complex amplitudes is fundamental to PFC's application in plasticity: tracking individual dislocations is reduced to identifying the topological singularities (zeros) of the complex fields $A_n$ \citep{skaugenDislocationdynamics2018, skogvollphasefield2022, mazenkoVortexVelocities1997}.

In the small deformation limit, distortion of surfaces of constant $\psi$ are described by the configurational distortion tensor $\Tens{Q}$, function of the amplitudes $A_n$ as,
\begin{equation}\label{eq:Q_def}
    \Tens{Q} = -\frac{d}{N} \sum_n \vec{q^n} \otimes \Im \frac{\nabla A_n}{A_n}.
\end{equation}

The polar dislocation density tensor can be constructed in this framework by interpreting each complex amplitude of the order parameter as a mapping from physical space to the complex plane $\mathbb{C}$ \citep{skogvollphasefield2022}.
The associated Jacobian is a conserved topological quantity, and this conservation is equivalent to the conservation of line defects, which in this context are dislocations \citep{Halperin1981, mazenkoVortexVelocities1997}. 
In this description, the polar dislocation density tensor $\widetilde{\Tens{\alpha}}$ is written as:
\begin{equation}
\label{eq:alphaT_def}
    \widetilde{\Tens{\alpha}}
    =
    -\frac{2\pi d}{N}
    \sum_n
    \delta(A_n)\,
    \vec{q}^{\,n}\otimes\vec{D}_n,
\end{equation}
where $d$ is the spatial dimension, $N$ is the number of resonant modes at the bifurcation, $\delta$ denotes Dirac's distribution, and 
\begin{equation}
    \vec{D}_n
    =
    \nabla \Re A_n \times \nabla \Im A_n
\end{equation}
is the Jacobian vector associated with the mapping from real space to complex amplitude space $A_n(\vec{x})$ \citep{skogvollUnifiedFieldTheory2023}.

The same topological conservation law yields a dislocation flux tensor $\Tens{\mathcal{J}}^\psi$, so that $\widetilde{\Tens{\alpha}}$ satisfies the conservation equation
\begin{equation}
\label{eq:current}
    \dot{\widetilde{\Tens{\alpha}}}
    +
    \nabla\times \Tens{\mathcal{J}}^\psi
    =
    0,
    \qquad
    \Tens{\mathcal{J}}^\psi
    =
    -\frac{2\pi d}{N}
    \sum_n
    \vec{q}^{\,n}\otimes
    \Im\!\left(\dot{A}_n \,\nabla A_n\right)
\end{equation}
Furthermore, it can be shown that the curl of the configurational distortion tensor $\Tens{Q}$ introduced in \cref{eq:Q_def} coincides with $\widetilde{\Tens{\alpha}}$,
\begin{equation}
\label{eq:alpha_T_def}
    \widetilde{\Tens{\alpha}}
    =
    \nabla \times \Tens{Q}.
\end{equation}  
 
\section{Coupling PFC and FDM via elastic distortions}

In the PFC and FDM coupling proposed by Acharya and Vi\~{n}als \citep{acharyaFielddislocation2020}, an energetic penalty on the squared difference between $\boldsymbol{Q}$ and $\boldsymbol{U}^e$ is introduced.
For the same dislocation configuration, under non-equilibrium conditions, $\Tens{Q}$ differs from $\Tens{U}^e$; $\Tens{U}^e = \Tens{Q}$ holds only at equilibrium.
This mismatch, demonstrated in \citep{upadhyayCouplingPhaseField2024}, is the foundation of the PFCFDM model \citep{acharyaFielddislocation2020}.

The Helmholtz free energy functional $\mathcal{F}[\psi, \Tens{U}^e]$ proposed in \citep{acharyaFielddislocation2020} involves the sum of three contributions: the elastic energy of the solid $\mathcal{F}_{el}[\Tens{U}^e]$, the Swift-Hohenberg energy $c_{sh} \mathcal{F}_{sh}[\psi]$, and a penalty term $c_{pen} \mathcal{F}_{pen}[\Tens{Q},\Tens{U}^e]$ representing the mismatch between $\Tens{U}^e$ and $\Tens{Q}$, which are treated as independent variables:
\begin{equation}\label{eq:mainFreeEnergy}
    \mathcal{F}[\psi, \Tens{U}^e] = \int_\Omega \varphi[\psi, \Tens{U}^e] \ \text{d}\Omega= \mathcal{F}_{el}[\Tens{U}^e] + c_{sh}\mathcal{F}_{sh}[\psi] + c_{pen}\mathcal{F}_{pen}[\Tens{Q}[\psi], \Tens{U}^e]
\end{equation}
with
\begin{eqnarray}
    \mathcal{F}_{el} &=& \int_\Omega \varphi_{el} \ \text{d}\Omega \\ 
    \mathcal{F}_{pen} &=& \int_\Omega f_{pen} \ \text{d}\Omega = \int_\Omega \frac{1}{2} \| \Tens{U}^e - \Tens{Q} \|^2 \ \text{d}\Omega
\end{eqnarray}
where $\varphi$ is the Helmholtz free energy density, $\varphi_{el}$ is the elastic energy density, and $c_{sh}$ and $c_{pen}$ are energetic coefficients. An additional penalty term was also included in \citep{acharyaFielddislocation2020} as $c_\rho/2 \int (\rho-\psi)^2$ where $\rho$ is the mass density. However, this term allows for the presence of vacancies, which are not considered in the present work.

For the given system of equations, non-negativity of dissipation implies:
\begin{equation}\label{eq:D_def}
    \mathcal{D} = \int_\Omega \Tens{\sigma}: \dot{\Tens{U}} \dd{\Omega} - \int_\Omega \dot{\varphi} \dd{\Omega} \quad \geq 0
\end{equation}

\cref{eq:mainFreeEnergy} gives:
\begin{equation}
    \int_\Omega \dot{\varphi} \dd{\Omega} = \int_\Omega \left[ \frac{\partial \varphi_{el}}{\partial \Tens{U}^e}: \dot{\Tens{U}^e}+ c_{sh}\frac{\delta \mathcal{F}_{sh}}{\delta \psi} \dot \psi + c_{pen} \frac{\delta \mathcal{F}_{pen}}{\delta \Tens{U}^e}: \dot{\Tens{U}^e}  + c_{pen} \frac{\delta \mathcal{F}_{pen}}{\delta \psi} \dot \psi \right] \dd{\Omega}
\end{equation}

Substituting \cref{eqn:distortion} in the total dissipation gives:
\begin{equation}\label{eq:const_ana}
    \begin{aligned}
        &\text{Modified Cauchy stress:} \quad \Tens{\sigma} = \frac{\partial \varphi_{el}}{\partial \Tens{U}^e} + c_{pen} \left( \frac{\delta \mathcal{F}_{pen}}{\delta \Tens{U}^e}\right)_{\mathrm{sym}}\\
        &\text{Phase field evolution:}  \quad \dot{\psi}= - \left[c_{sh} \frac{\delta \mathcal{F}_{sh}}{\delta \psi}  + c_{pen} \frac{\delta \mathcal{F}_{pen}}{\delta \psi} \right] \\
        &\text{Total dissipation:} \quad \mathcal{D} = \int_\Omega \Tens{\sigma}: \dot{\Tens{U^p}} \dd{\Omega} + \int_\Omega \dot{\psi}^2\dd{\Omega} \geq 0
    \end{aligned}
\end{equation}

The term $c_{pen} \delta \mathcal{F}_{pen}/{\delta \psi}$ enters in the elastic constitutive relationship as well as a forcing term in the evolution of the order parameter. Cross terms between forces and currents (e.g., between $\dot{\psi}$ and $\frac{\partial \varphi_{el}}{\partial \Tens{U}^e}$) are neglected. 

In order to close the model, an additional equation is needed for plastic distortion rate $\dot{\Tens{U^p}}$, that can either be FDM-driven or PFC-driven; the latter is a constitutive assumption made in the present work that was not considered in \cite{acharyaFielddislocation2020}.

\subsubsection*{FDM-driven plasticity}

In the first case considered of FDM-driven dislocation motion, \cref{eq:Transport} is solved and requires a prescription of the dislocation velocity field $\vec{v^d}$. In this case, the dissipation becomes,
\begin{equation}
    \mathcal{D} = \int_\Omega \Tens{\sigma}:(\Tens{\alpha} \times \vec{v^d})  \dd{\Omega} + \int_\Omega \dot{\psi}^2\dd{\Omega} \geq 0
\end{equation}
which can be re-arranged to yield an explicit formula for the Peach-Koehler force acting on the dislocations:
\begin{equation}
    \int_\Omega \Tens{\sigma}: \dot{\Tens{U^p}} \dd{\Omega} = \int \Tens{\sigma}:(\Tens{\alpha} \times \vec{v^d})  \dd{\Omega}  = \int (\underbrace{\Tens{\sigma} \Tens{\alpha}: \Tens{X}}_{\text{Peach-Koehler force}}) \cdot \vec{v^d} \dd{\Omega}
\end{equation}
Where $\Tens{X}$ is the third order Levi-Civita permutation tensor. The following expression of the dislocation velocity field $\vec{v^d}$ can be assumed in order to guarantee non-negative dissipation:
\begin{equation} \label{eq:vel_Field}
    \vec{v^d} = \frac{1}{B} \vec{f^d}=\frac{1}{B} \Tens{\sigma^T} \Tens{\alpha}: \Tens{X} = \frac{1}{B}  \left[\Bigl(\frac{\partial \varphi_{el}}{\partial \Tens{U}^e_{,\mathrm{sym}}} + c_{pen} \frac{\partial f_{pen}}{\partial \Tens{U}^e}\Bigr)^T\Tens{\alpha} \right]: \Tens{X} 
\end{equation}
where the second equality is obtained by substituting the expression for stress from \cref{eq:const_ana} and $1/B$ being a scalar mobility.
Through the elastic constitutive law, the coupling term enters as an additional forcing term $c_{pen} \partial f_{pen}/\partial \Tens{U}^e$ in the dislocation velocity field.

\subsubsection*{PFC-driven plasticity}

In the second case, the PFC is chosen as the driver of plasticity. 
Motivated by the conservation law \cref{eq:current}, $\dot{\Tens{U^p}} = \Tens{\mathcal{J}^\psi}$ is chosen, which is equivalent to the incompatibility condition $\nabla \times \Tens{U^p} = - \widetilde{\Tens{\alpha}}$. 
The latter is explicitly enforced instead of solving the dislocation transport \cref{eq:Transport}. 
In this paradigm, the total dissipation reads:
\begin{equation}\label{eq:PFCDissipation}
    \mathcal{D} = \int_\Omega \Tens{\sigma}: \Tens{\mathcal{J}^\psi} \dd{\Omega} + \int_\Omega \dot{\psi}^2\dd{\Omega} \geq 0 \qquad \text{(PFC-driven)}
\end{equation}

In contrast to the FDM-driven approach, defect evolution emerges naturally in the PFC-driven approach through the time evolution of the order parameter $\psi$. 
The conservative flux $\Tens{\mathcal{J}}^\psi$ is computed as data in order to update the plastic distortion, ensuring that it matches the current dislocation state. This is performed by separately updating the compatible part of $\Tens{U}^p$ from $\Tens{\mathcal{J}}^\psi$ and the incompatible part from the density  $\widetilde{\Tens{\alpha}}$.

Note here that if the governing equations in this case remain \cref{eq:const_ana}, there is no guarantee the PFC-driven plastic dissipation $\Tens{\sigma}: \Tens{\mathcal{J}^\psi}$ will always be positive.
Since $\Tens{\mathcal{J}^\psi}$ only depends on the slowly varying amplitudes $A_{n}$ whereas the remaining contribution to dissipation depends directly on $\psi$, a consistent evolution equation for $\dot{\psi}$ that guarantees the positivity of dissipation cannot be obtained.

\section{Theoretical analysis of the penalty term}

The penalty term is expressed in terms of the elastic and configurational distortions, the latter function of the amplitudes $A_{n}$. Both of them vary on the same slow scale relative to the lattice parameter. However, the order parameter $\psi$ changes in a fast scale, on the order of the lattice spacing, \cref{eq:const_ana}. Therefore, further manipulation of the dissipation inequality can only be done approximately. We analyze in this section the consequences on the evolution of $\psi$ that derive from the energy penalty term introduced.

\subsection{Impact on evolution of $\psi$}

Whereas the coarse graining from $\psi$ to the amplitudes $A_{n}$ is well established, we argue here in terms of a formal inverse mapping from the amplitudes to $\psi$, given the explicit convolution used numerically, \cref{eq:demodulation}. The conclusion is independent of the inverse mapping used.

The functional derivative of the penalty term with respect to the order parameter $\psi$ is:
\begin{equation}\label{eq:var^psi_def}
    \frac{\delta \mathcal{F}_{pen}}{\delta \psi (\vec{x})} = \sum_n \int_{\Omega} \left( \frac{\delta\mathcal{F}_{pen}}{\delta A_n(\vec{x}')} \frac{\delta A_n(\vec{x}')}{\delta \psi(\vec{x})} + \frac{\delta\mathcal{F}_{pen}}{\delta A_n^*(\vec{x}')} \frac{\delta A_n^*(\vec{x}')}{\delta \psi(\vec{x})} \right) \text{d}\vec{x}' = 2 \Re \sum_n \int_{\Omega} \frac{\delta\mathcal{F}_{pen}}{\delta A_n(\vec{x}')} \frac{\delta A_n(\vec{x}')}{\delta \psi(\vec{x})} \text{d}\vec{x}'
\end{equation}

From \cref{eq:CmpA} and \cref{eq:demodulation} we have:
\begin{equation}\label{eq:An^psi}
    \frac{\delta A_n(\vec{x}')}{\delta \psi(\vec{x})} = e^{-i \vec{q}_n \cdot \vec{x}} \mathcal{G}_{a_0}(\vec{x}'- \vec{x})
\end{equation}

Substituting this into \cref{eq:var^psi_def}, the functional derivative becomes:
\begin{equation} \label{eq:var^psi_opt}
    \frac{\delta \mathcal{F}_{pen}}{\delta \psi (\vec{x})} = 2 \Re \sum_n \Bigl[ \underbrace{\int_{\Omega} \frac{\delta\mathcal{F}_{pen}}{\delta A_n(\vec{x}')} \mathcal{G}_{a_0}(\vec{x}'- \vec{x}) \text{d}\vec{x}'}_{\text{Complex Envelope}} \, e^{-i \vec{q}_n \cdot \vec{x}} \Bigr]
\end{equation}

\noindent For the $L^2$ form of the penalty term, $\delta\mathcal{F}_{pen}/\delta A_n$ derived in the Appendix \ref{app:Var_der_derivation} yields
\begin{equation}\label{eq:Func_derv_Fu}
    \begin{cases}
        \cfrac{\delta \mathcal{F}_{pen}}{\delta A_n}
        = \cfrac{d}{2N \mathrm{i}|A_n|^2} \
        \vec{q}^{n} \cdot \left[
        \nabla \cdot \left(
            \Tens{Q}-\Tens{U}^e \right) \right] A_n^* \qquad &\text{in }  \, \Omega\\
            \\
        \vec{n}\cdot \left(\Tens{Q}-\Tens{U}^e\right)^T  \cdot \vec{q^n} = 0
        \qquad &\text{on } \partial\Omega
        \end{cases}
        \end{equation}
        
Combining \cref{eq:Func_derv_Fu}, \cref{eq:var^psi_def} and \cref{eq:An^psi}, we get:
\begin{equation}\label{eq:Pen_bulk}
\begin{aligned}
\frac{\delta \mathcal{F}_{pen}}{\delta \psi(\vec{x})} &= - \frac{d}{N}\,\Im \left\{\sum_n e^{-i \vec{q}_n \cdot \vec{x}}\int_{\Omega}\frac{\vec{q}^{\,n} \cdot\left[\nabla \cdot\left(\Tens{Q}(\vec{x}')-\Tens{U}^e(\vec{x}')\right)\right]}{A_n(\vec{x}')}\mathcal{G}_{a_0}(\vec{x}'-\vec{x})\,\mathrm{d}\vec{x}'\right\} \\
&= - \frac{d}{N}\,\Im \left\{\sum_n e^{-i \vec{q}_n \cdot \vec{x}}\left\langle\frac{\vec{q}^{\,n} \cdot\left[\nabla \cdot\left(\Tens{Q}-\Tens{U}^e\right)\right]}{A_n}\right\rangle (\vec{x})\right\}
\end{aligned}
\end{equation}

\begin{figure}[h!]
  \centering

  \begin{subfigure}[t]{0.5\textwidth}
    \centering
    \caption{}\label{fig:dfqpsi_psi}
    \includegraphics[width=0.99\linewidth]{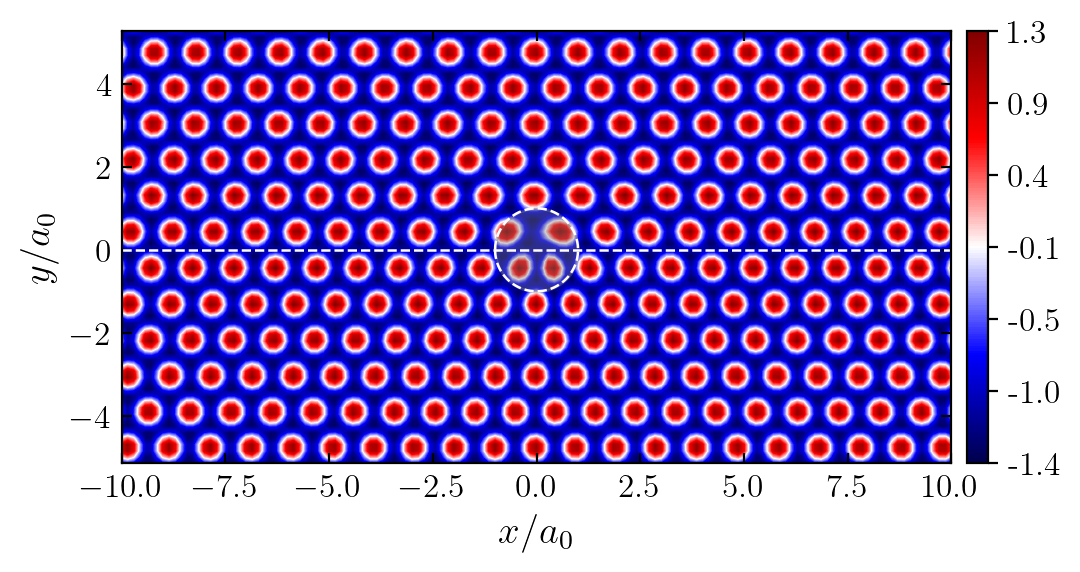}
  \end{subfigure}

  \vspace{0.3em}

  \begin{subfigure}[t]{0.48\textwidth}
    \centering
    \caption{}\label{fig:dfqpsi_alpha}
    \includegraphics[width=\linewidth]{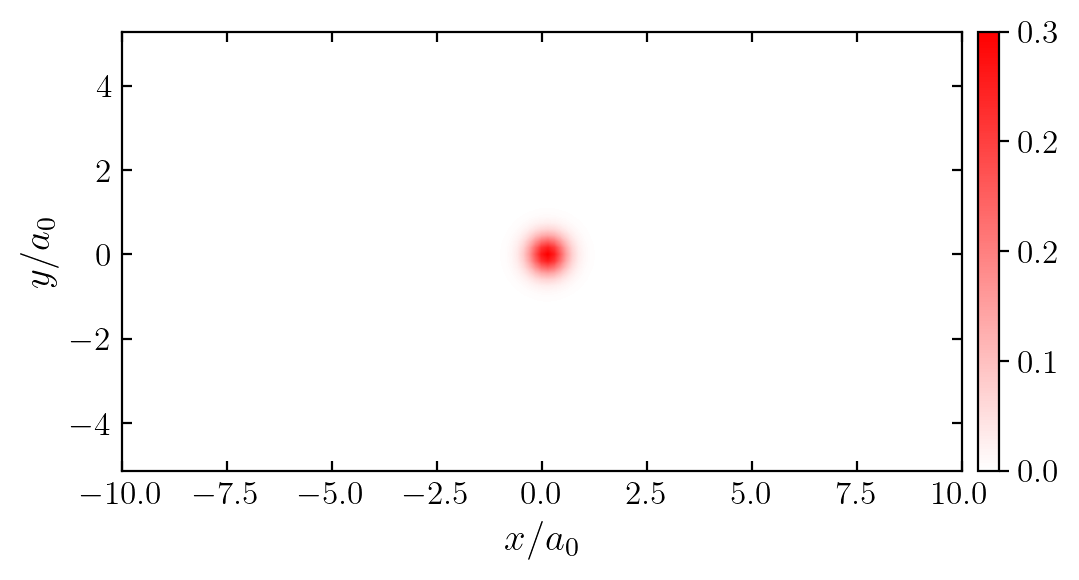}
  \end{subfigure}\hfill
  \begin{subfigure}[t]{0.48\textwidth}
    \centering
    \caption{}\label{fig:dfqpsi_slice}
    \includegraphics[width=\linewidth]{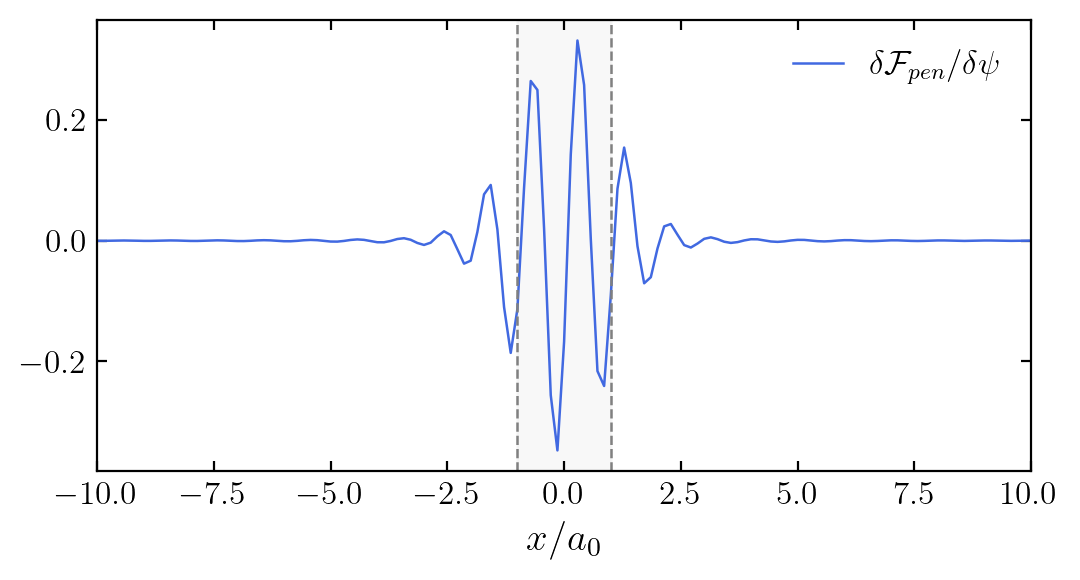}
  \end{subfigure}

  \vspace{0.3em}

  \begin{subfigure}[t]{0.48\textwidth}
    \centering
    \caption{}\label{fig:dfqpsi_field}
    \includegraphics[width=\linewidth]{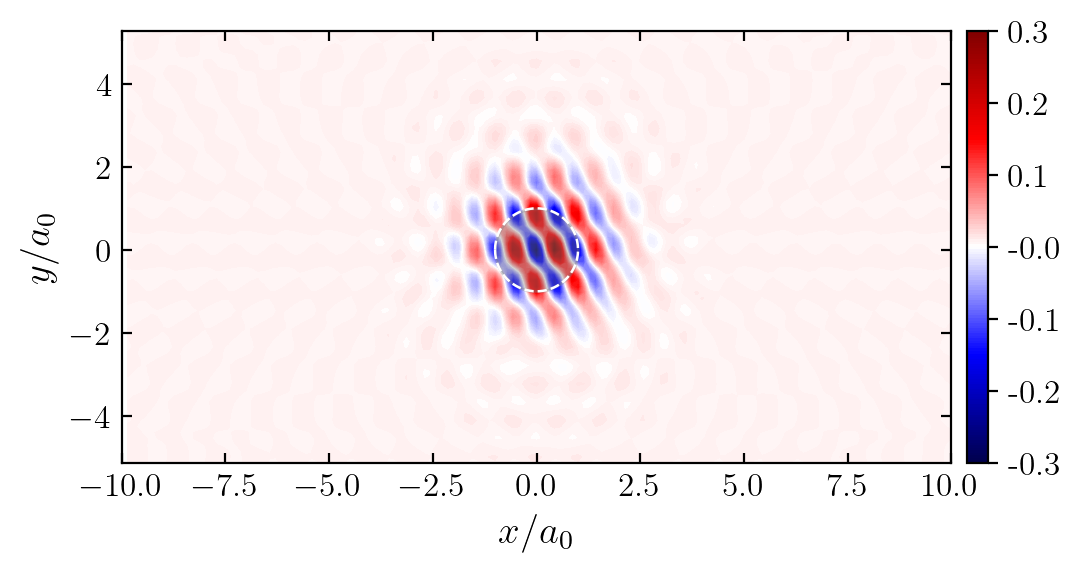}
  \end{subfigure}\hfill
  \begin{subfigure}[t]{0.48\textwidth}
    \centering
    \caption{}\label{fig:dfqpsi_QvsU}
    \includegraphics[width=\linewidth]{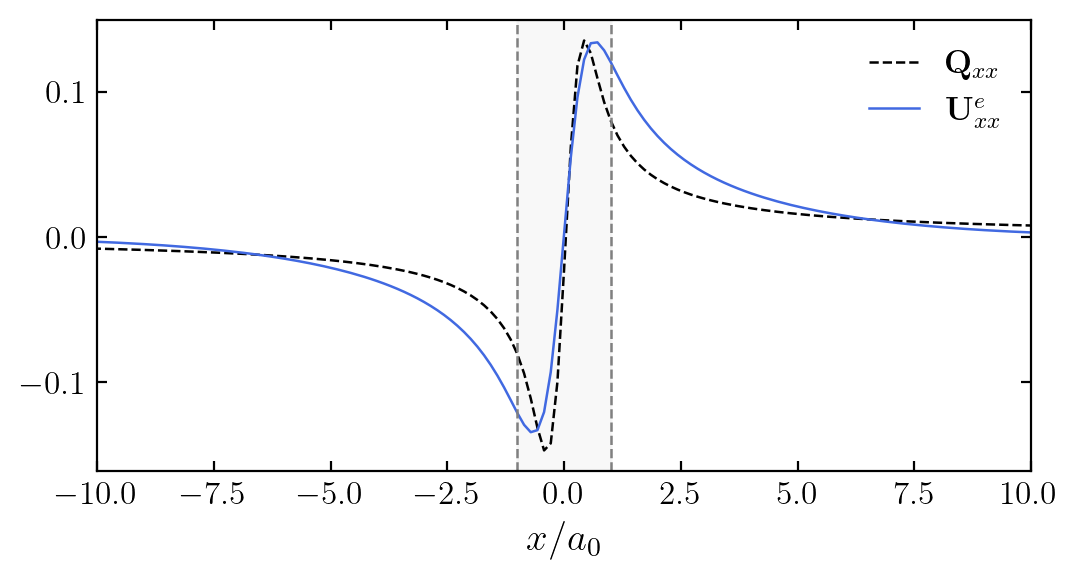}
  \end{subfigure}

  \caption{\subref{fig:dfqpsi_psi} Order parameter field in the presence of a single edge dislocation at the center of a 2D hexagonal lattice. \subref{fig:dfqpsi_alpha}  Corresponding dislocation density $\alpha_{xz}$. \subref{fig:dfqpsi_field}  Contour plot of the penalty-induced forcing for this configuration. 
  In both, the gray circle shows the extent of the core. 
  \subref{fig:dfqpsi_slice}  A horizontal slice along $y=0$ of the penalty forcing whereas \subref{fig:dfqpsi_QvsU}  show the behavior of the $xx$ of $\Tens{Q}$ and $\Tens{U}^e$. The vertical strip corresponds to the extent of the dislocation core.}
  \label{fig:Penalty_plot}
\end{figure}

In order to illustrate the spatial evolution of the variational derivative $\delta \mathcal{F}_{pen}/{\delta \psi}$, we consider a hexagonal lattice with an edge dislocation at its center shown in \cref{fig:Penalty_plot}. 
This state is generated by applying a displacement field $\vec{u}_0$ from Volterra's construction \citep{voltera1907} to the ground-state density, $\psi(\vec{x}) \to \psi(\vec{x} - \vec{u}_0(\vec{x}))$, followed by PFC relaxation with no penalty coupling. 
The dislocation core extends to $\sim 2a_0$ and the mismatch between $Q_{xx}$ and $U^e_{xx}$ occurs in the vicinity of the core.

The coupling term $\delta \mathcal{F}_{{pen}} / \delta \psi$ has a wave-packet structure, as already suggested by the complex envelope in \cref{eq:var^psi_opt}. 
When $a_0$ is small relative to the length scale on which $\delta \mathcal{F}_{{pen}} / \delta A_n$ varies, the convolution kernel $\mathcal{G}_{a_0}$ approaches a Dirac delta distribution giving:
\begin{equation}
\label{eq:var^psi_simpl}
    \frac{\delta \mathcal{F}_{{pen}}}{\delta \psi(\vec{x})}
    \approx
    2 \,\Re\!\left(
        \sum_n
        \frac{\delta \mathcal{F}_{{pen}}}{\delta A_n(\vec{x})}
        e^{-i \vec{q}_n \cdot \vec{x}}
    \right).
\end{equation}

Thus, \cref{eq:var^psi_opt,eq:var^psi_simpl} show that the penalty term can be interpreted as a modulated Fourier expansion on the microscopic length scale of $\psi$. 
The lattice Fourier modes are weighted by complex envelopes determined by the Gaussian-smoothed functional derivatives $\delta \mathcal{F}_{{pen}} / \delta A_n$. 
In this way, $\delta \mathcal{F}_{{pen}} / \delta \psi$ provides the bridge across scales in this coupled framework, transferring mismatch in distortion from the macroscopic scale of the amplitude scale to the finer scale of $\psi$.

\subsubsection{Loss of incompatible distortions in $\delta\mathcal{F}/\delta\psi$}

The divergence operator in \cref{eq:Pen_bulk} eliminates the incompatible (divergence-free) parts of $\Tens{Q}$ and $\Tens{U}^e$ as well as any  homogeneous distortions from the penalty driving force, making the evolution of $\psi$ independent of them. 
If the incompatible components evolve through distinct mechanisms, then the distance between the incompatible parts of $\Tens{Q}$ and $\Tens{U}^e$ will not diminish over time, implying that $\nabla \times \Tens{Q} \neq \nabla \times \Tens{U}^e$ i.e., the polar dislocation densities defined by these tensors are unequal, $\Tens{\widetilde{\alpha}} \neq \Tens{\alpha}$. 
Due to \cref{eq:Pen_bulk} and \cref{eq:Func_derv_Fu}, equilibrium is reached when:
\begin{equation} \label{eq:sameDiv}
    \nabla \cdot \Tens{Q} = \nabla \cdot \Tens{U}^e
\end{equation}
The penalty term enforces a matching between the divergences of the two fields, which affects only their compatible components. 
Therefore, it cannot induce any topological changes. 
When equilibrium (\cref{eq:sameDiv}) is reached, then according to the Stokes-Helmholtz decomposition:
\begin{equation}
    2\mathcal{F}_{pen} =\int_\Omega \|\Tens{Q} - \Tens{U}^e\|^2 \, \mathrm{d}x
    = \int_\Omega \|\Tens{Q}^{\perp} - \Tens{U}^{e\perp}\|^2 \, \mathrm{d}x + \int_\Omega ||\Tens{C}(\vec{x})||^2\mathrm{d}x > 0
\end{equation}
where $\Tens{C}$ is a divergence-free and a curl-free tensor. 
The last equality holds because the compatible and incompatible parts are orthogonal in the $L^2$-sense. 
The current penalty formulation thus does not allow reconciling the incompatible parts of $\Tens{Q}$ and $\Tens{U}^e$ and it requires an additional prescription, which can be achieved through the PFC-driven approach but not through the FDM-driven approach.

\subsubsection{Unphysical diffusive transmission of elasticity}

The structure of \cref{eq:Func_derv_Fu} also shows that an externally imposed loading on the boundaries does not statically or dynamically allow transmitting the information from the boundaries throughout the bulk. 
For the $L^2$ penalty, the boundary condition associated with \cref{eq:Func_derv_Fu} can be rewritten as:
\begin{equation}\label{eq:dF_Bcs}
     \left(\Tens{U}^e - \Tens{Q}\right)\cdot \vec{n}   = 0 \quad \text{on } \partial\Omega
\end{equation}

Far from the defect cores, \cref{eq:Q_def} simplifies to $\Tens{Q} = -\frac{d}{N} \sum_n \vec{q}_n \otimes \nabla \theta_n$, where $\theta_n$ denotes the phase of each complex amplitudes $A_n$. Hence \cref{eq:dF_Bcs} is a Neumann-type boundary condition for the phases and \cref{eq:Func_derv_Fu} takes the form (for the $L^2$ norm penalty):
\begin{equation} \label{eq:Diffusion_phase}
\frac{1}{A_n^*}\cfrac{\delta \mathcal{F}_{pen}}{\delta A_n} =
\ii \frac{d^2}{2N^2 |A_n|^2} \left[ \sum_m (\vec{q}_m \cdot \vec{q}_n) \Delta \theta_m \right] +
\frac{d}{2N |A_n|^2} \nabla \cdot \left( \Tens{U}^{e,T} \vec{q}_n \right)
\end{equation} 

Away from the cores, the moduli $|A_n|$ are approximately constant and evolution is governed by the phase shifts $\dot{\theta}_n$. The Laplacian $\Delta \theta_m$ in \cref{eq:Diffusion_phase} shows that the penalty term acts as diffusion on the phases. Assuming each amplitude evolves as $\dot{A}_n = -c_{{pen}}\, \delta \mathcal{F}_{{pen}}/\delta A_n^*$, one obtains a coupled set of diffusion equations $\dot{\theta}_n = \sum_m D_{mn} \Delta \theta_m + f_n$, with diffusion coefficients
\begin{equation} \label{eq:Diffusion_Coeff_PEN}
 D_{mn} \propto  \frac{1}{3|A_n|^2} c_{{pen}} 
\end{equation}

Mechanical loading is transmitted from the boundary into the bulk by phase diffusion rather than by elastic relaxation, which is physically unacceptable. $D_{mn}$ are approximately constant only far from the cores. Near a core, two of the amplitudes $|A_n|$ vanish, so the corresponding diffusion coefficient diverges. Consequently, the diffusion from the boundary becomes increasingly faster as one approaches the defect core, and the phase diffusion breaks down near the core.

\subsection{Alteration of elastic stress}

The penalty term also contributes to the elastic stress in \cref{eq:const_ana} and dislocation velocity in \cref{eq:vel_Field}. 
For a linearly elastic material with elastic tensor $\mathbb{C}$, the stress expression in \cref{eq:const_ana} simplifies to: 
\begin{equation} \label{eq:elastic_stress_simple}
       \frac{\partial f_{pen}}{\partial \Tens{U}^e} = \Tens{U}^e -\Tens{Q} \quad \Rightarrow \quad  \boldsymbol{\sigma} = \mathbb{C}: \Tens{U}^e + c_{pen} (\Tens{U}^e - \Tens{Q})_{\text{sym}}
\end{equation}
The penalty parameter $c_{pen} > 0$ introduces two distinct mechanical regimes. 
In regions where $\Tens{Q} \neq \Tens{U}^e$, the material exhibits an increased effective stiffness, characterized by an effective shear modulus $\mu_{\text{eff}} = \mu + c_{pen}/2$. 
Conversely, regions where $\Tens{U}^e \approx \Tens{Q}$, linear elasticity is recovered. 
While a large $c_{pen}$ accelerates the minimization of $\mathcal{F}_{pen}$, this spatially heterogeneous stiffening alters the dislocation response to both self-stress and external loads.

\subsection{Regularity of the penalty term}

To ensure that the penalty term and its resulting functional derivatives remain finite and well-defined, $f_{pen}$ must exhibit sufficient regularity. 
For instance, in the case of a single dislocation in a two dimensional body $\Tens{Q}$ and $\Tens{U}^e$ both scale as $1/r$ \citep{andersonTheoryDislocations2017}, with $r$ measured from the center of the dislocation core.
If the scaling factors between $\Tens{U}^e$ and $\Tens{Q}$ are not matching, e.g. when the incompatible parts of $\Tens{U}^e$ and $\Tens{Q}$ are not the same, then $\Tens{S}:=\Tens{Q}-\Tens{U}^e$ exhibits a $1/r$ singularity which renders $\mathcal{F}_{pen}= 1/2 \int \Tens{S}:\Tens{S}\, r\dd{r} \dd{\theta}$ infinite. 
This can happen in the FDM-driven plasticity where $\nabla \times \Tens{U}^e$ and $\nabla \times \Tens{Q}$ evolve differently even if they were initially equal.

This is also the case at equilibrium, where the penalty enforces $\nabla \cdot \Tens{S}=0$ only and $\nabla \times \Tens{S} \neq 0$ is permissible. 
Since each row of $\Tens{S}$ is divergence-free, there exists a vector field $\vec{h}$ such that $ S_{ij}=\varepsilon_{jk}\partial_k h_i$.
Taking the tensor curl yields $\nabla \times \Tens{S}=\Delta \vec{h}$. 
If we write $\nabla \times \Tens{S}= \vec{\widetilde{b}}\,\delta(\vec{x})$ \footnote{$\Tens{S}$ is a $2\times2$ tensor in this case and its curl can be identified as a vector} with $\vec{\widetilde{b}} \in\mathbb{R}^2$ the residual burgers vector and Dirac delta's distribution at the origin. 
Then, $\vec{h}$ satisfies $\Delta \vec{h}=\vec{b}\,\delta(\vec{x})$. 
Using the Green function of the Laplacian in two dimensions, one obtains $\vec{h}(\vec{x})=\frac{\widetilde{b}}{2\pi}\log |\vec{x}|+O(1)$. Therefore, close to the core $|\Tens{S}(\vec{x})|^2 \sim \frac{|\vec{b}|^2}{4\pi^2}\frac{1}{r^2}$, with $r=|\vec{x}|$. 
Since $f_{pen}=\frac12 \Tens{S}:\Tens{S}=\frac12 |\Tens{S}|^2,$ it follows that the penalty energy has a logarithmic divergence:
\begin{equation}
f_{pen}(\vec{x})\sim \frac{|\vec{\widetilde{b}}|^2}{8\pi^2}\frac{1}{r^2} \quad \Rightarrow \mathcal{F}_{pen}= \int f_{pen} r \dd{r} \dd{\theta} \sim \frac{|\vec{\widetilde{b}}|^2}{4\pi} \int \frac{1}{r} dr
\end{equation}

\subsection{Generalizing the penalty term}

One might argue that previous conclusions are a result of the chosen penalty function and be tempted to employ a higher-order $p$-norm. 
For instance, a family of penalty densities:
 \begin{equation}\label{eq:p-norm_def}
 f_{pen} = \frac{1}{2p} \left[(\Tens{U}^e - \Tens{Q}):\mathbb{A}:(\Tens{U}^e - \Tens{Q})\right]^p
 \end{equation}
 where $\mathbb{A}$ is a symmetric positive definite $4^{\text{th}}$-order tensor. Near the core, for non-matching incompatible parts, $f_{pen}\sim 1/r^{2p}$. 
 The total energy is integrable as $r\rightarrow0$ only if $p<1$ in 2D and $p<3/2$ in 3D and it is only convex when $p \ge 1/2$, which highly restricts the choice of $p$. In all cases, for the $p$-norm introduced in \cref{eq:p-norm_def}, the functional derivative \cref{eq:Func_derv_Fu} simplifies to:
\begin{equation}
\frac{\delta \mathcal{F}_{pen}}{\delta A_n} =
\frac{d}{2N \mathrm{i} |A_n|^2} \vec{q}_n \cdot \nabla \cdot
\Big[ \left[(\Tens{U}^e - \Tens{Q}):\mathbb{A}:(\Tens{U}^e - \Tens{Q})\right]^{p-1} \mathbb{A}:(\Tens{Q} - \Tens{U}^e) \Big] A_n^*
\end{equation}
This expression fundamentally describes the same phase diffusion process, albeit with a space varying diffusivity. Thus, the choice of $p$-norm does not alter the underlying transport physics of the model. The constitutive law becomes highly nonlinear and introduces significant mathematical and numerical complexity, see \cref{eq:elastic_stress_NL},  without a clear theoretical advantage. 
Consequently, the resulting elastic stress would take the form:
\begin{equation} \label{eq:elastic_stress_NL}
\Tens{\sigma} = \frac{\partial \varphi_{el}}{\partial \Tens{U}^e} + c_{pen} \left(\frac{\delta \mathcal{F}_{pen}}{\delta \Tens{U}^e}\right)_{\!\mathrm{sym}} = \frac{\partial \varphi_{el}}{\partial \Tens{U}^e} + c_{pen} \left[(\Tens{U}^e - \Tens{Q}):\mathbb{A}:(\Tens{U}^e - \Tens{Q})\right]^{p-1} \mathbb{A}:(\Tens{Q} - \Tens{U}^e)
\end{equation}

Thus, to maintain model simplicity and minimize free parameters, we follow the original formulation in \citep{acharyaFielddislocation2020} and set $p=1$ and $\mathbb{A} = \mathbb{I}$.


\section{Numerical results}\label{sec:numerics}
We now present consequences of the coupling through a sequence of numerical tests. We first model a static edge dislocation dipole so as to isolate the effect of the penalty term and examine the equilibrium between $\Tens{Q}$ and $\Tens{U}^e$. We next allow the core to evolve in the absence of applied loading and investigate the model's ability to maintain a localized core once transport is solved in FDM. We finally consider the glide and annihilation of a dislocation dipole, comparing the behavior when plastic evolution is governed by FDM and then by PFC.

In order to retain a smooth and finite value of $f_{pen}$, a numerical smoothing of $\Tens{Q}$ is used by convolving it with a gaussian of width $a_0/4$. 
It results in regularizing the singularity in $\Tens{Q}$ near the core and a numerical mollification of $\widetilde{\Tens{\alpha}}$.

\subsection{Immobile dislocation dipole}\label{sec:immobile}

\begin{figure}[H]
\centering

\begin{subfigure}[t]{0.55\textwidth}
  \centering
  \vspace{0pt}
  \caption{Initial configuration of the immobile simulation $\psi(\vec{x},0)$.}
  \label{fig:psiClimbt0}
  \includegraphics[width=0.9\linewidth]{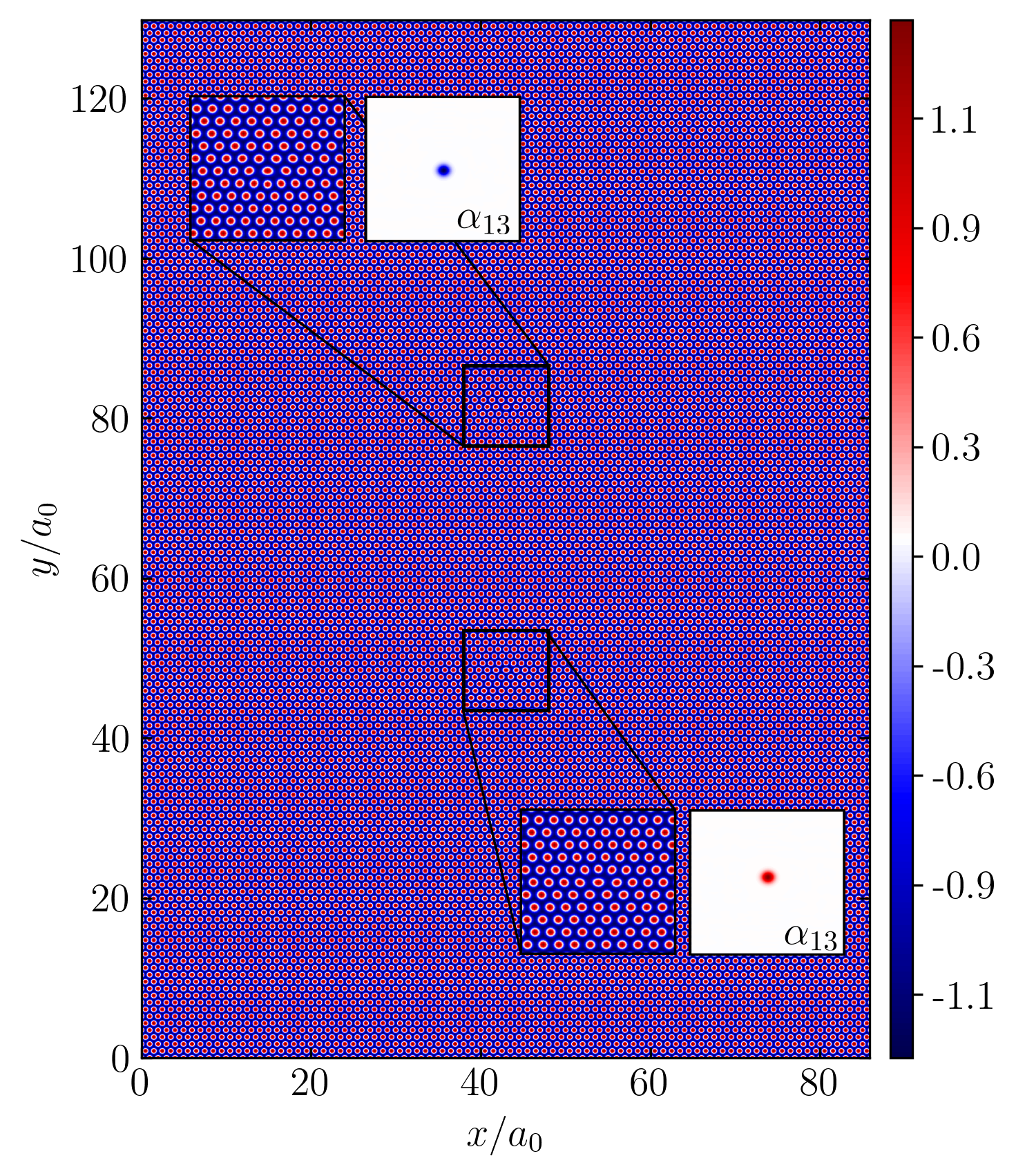}
\end{subfigure}
\hfill
\begin{subfigure}[t]{0.4\textwidth}
  \centering
  \vspace{0pt}
    
  \caption{Evolution of $\|\Tens{U}^e-\Tens{Q}\|$}
  \includegraphics[width=0.72\linewidth]{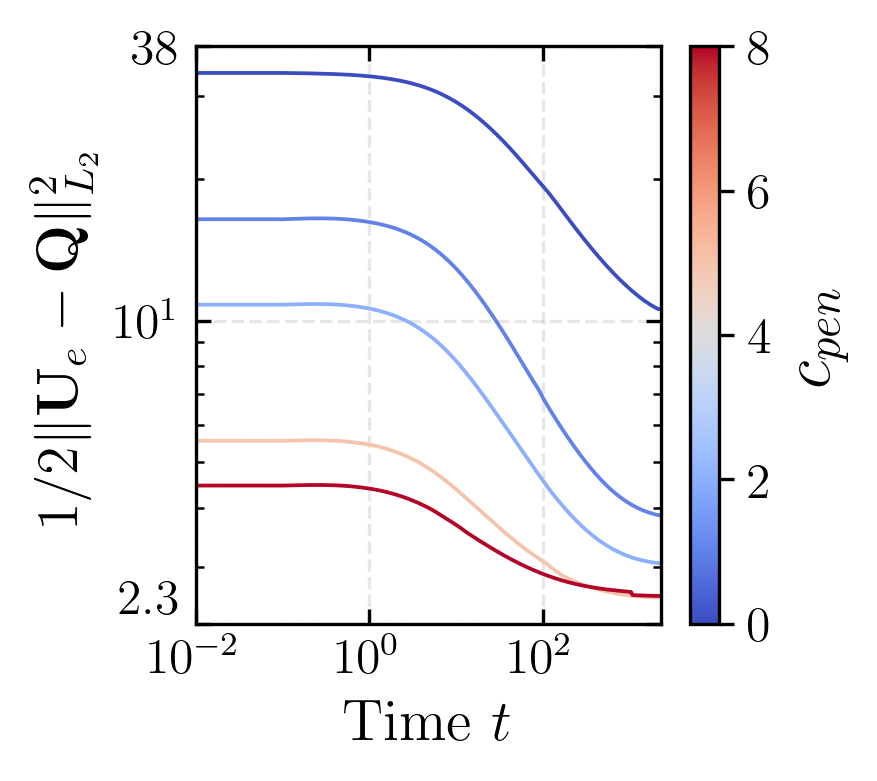}
  \label{fig:PenClimb}

  \vspace{0.2em}

  \caption{Evolution of $\|\nabla \cdot(\Tens{U}^e-\Tens{Q})\|$}
  \includegraphics[width=0.73\linewidth]{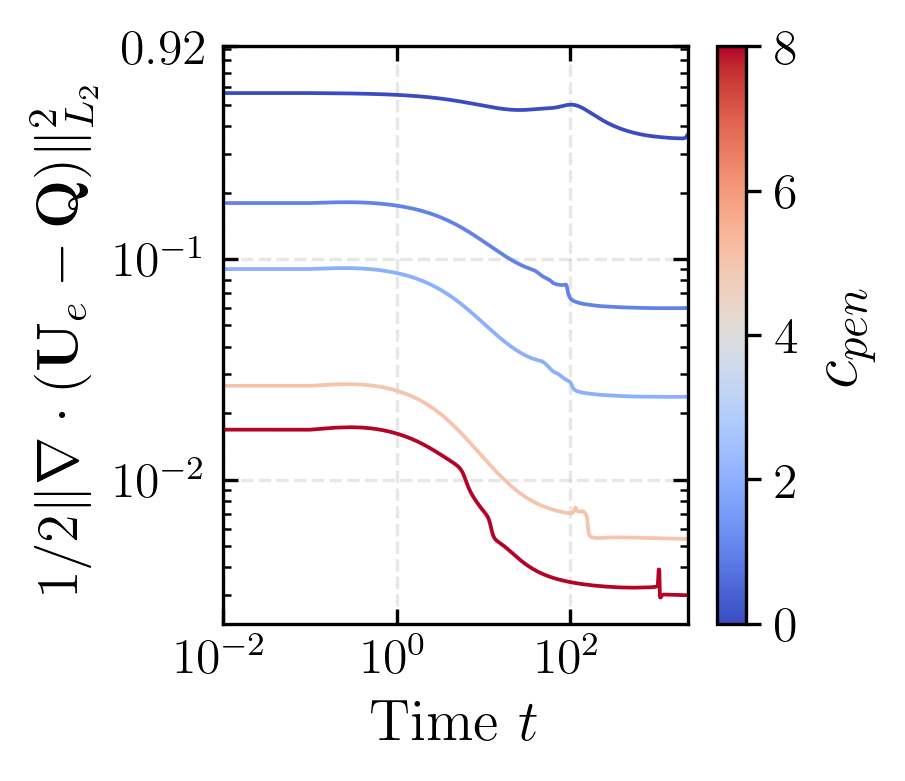}
  \label{fig:divClimb}
\end{subfigure}

\caption{Immobile edge dislocation dipole in Climb configuration. \subref{fig:psiClimbt0}: Order parameter $\psi$ at $t=0$ and the dislocation density tensor $\Tens{\alpha}$. \subref{fig:PenClimb} relaxation of $\| \Tens{U}^e - \Tens{Q} \|$ with increasing $c_{pen}$. \subref{fig:divClimb}, corresponding evolution of $\|\nabla \cdot(\Tens{U}^e-\Tens{Q})\|$.}\label{fig:climbInit}
\end{figure}

We examine an edge dislocation dipole in a climb configuration within a 2D hexagonal lattice, following the studied case in \citep{upadhyayCouplingPhaseField2024} where the cores remain immobile. 
The periodic simulation domain of size $602 \times 960$ uses grid spacings $dx = a_0/7$ and $dy = \sqrt{3}a_0/12$, with $a_0 = 4\pi/\sqrt{3}$. 
Simulations were performed using spectral methods (see Appendix \ref{app:Methods} for details) and were run until reaching equilibrium using a timestep $dt=0.1$. 
The initial configuration, shown in \cref{fig:psiClimbt0}, is generated by displacing a perfect hexagonal lattice:
\begin{equation}
    \psi(t=0) = \overline{\psi} + \sum_n A_n \exp[i \vec{q}_n \cdot (\vec{x} - \vec{u}_v)], \quad \vec{u}_v = \sum_p \vec{u}_\infty(\vec{x}-\vec{x}^p),
\end{equation}
where $\vec{u}_\infty$ is the displacement field of an edge dislocation in an infinite body and $\vec{x}^p$ is the position of the $p$-th dislocation. 
In this section, no transport equation is solved; instead, we enforce $\nabla \times \Tens{U}^p \leftarrow \widetilde{\Tens{\alpha}}$ at every timestep to isolate the impact of $\delta \mathcal{F}_{pen}/\delta\psi$. 
We set $c_{sh}=1$ and vary the coupling strength $c_{pen}$. The PFC parameters $r=-1.2$ and $\overline{\psi}=-0.5$ correspond to an isotropic crystal with Lamé coefficients $\lambda = \mu = 3A_0^2 \approx 1.08$.

\begin{figure}[H]
  \centering

  \begin{subfigure}[t]{0.45\textwidth}
    \centering
    \caption{$\Tens{c_{pen}=0}$, $\Tens{t=0}$}\label{fig:cw0t0}
    \includegraphics[width=\linewidth]{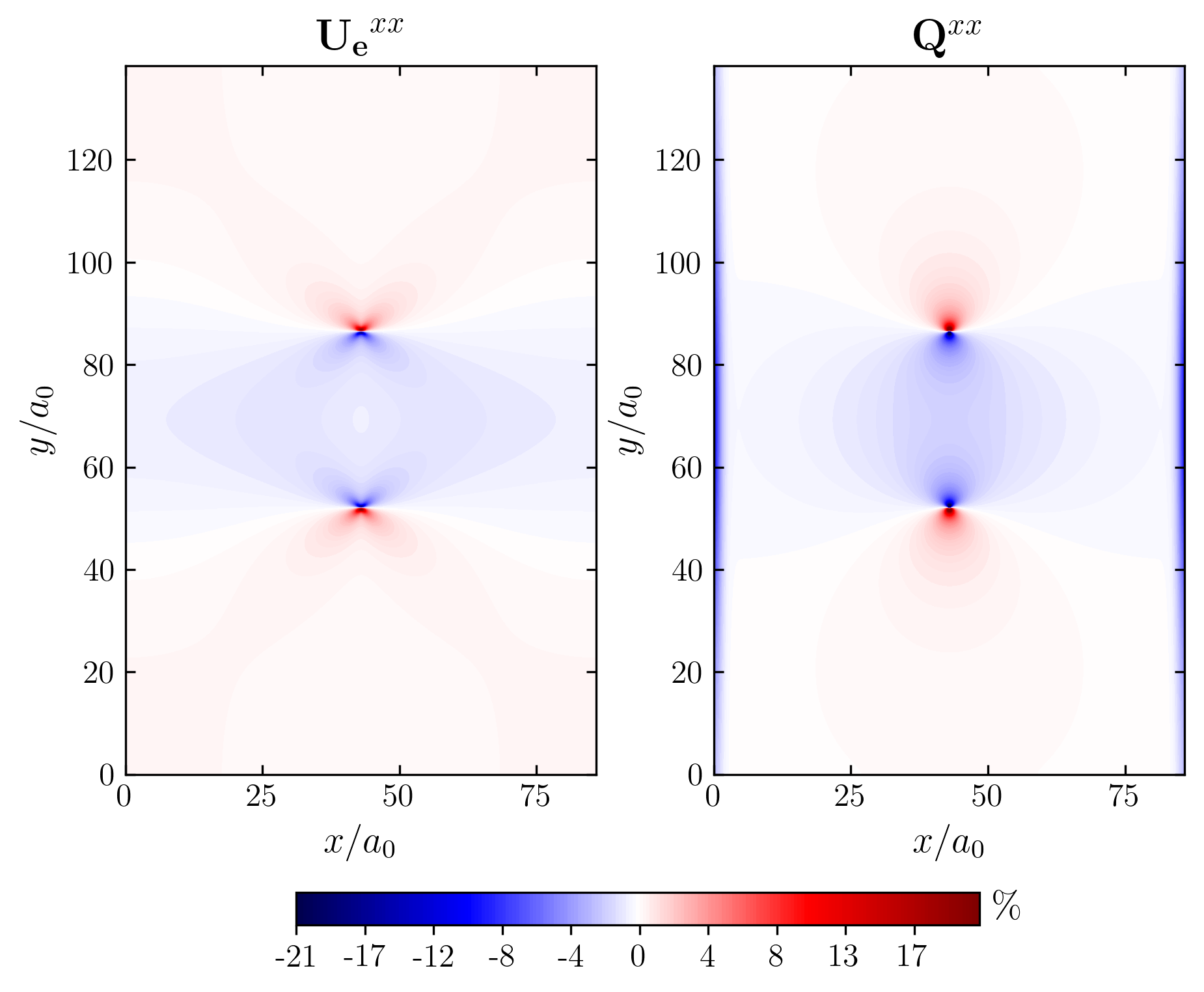}
  \end{subfigure}\hfill
  \begin{subfigure}[t]{0.45\textwidth}
    \centering
    \caption{$\Tens{c_{pen}=2}$, $\Tens{t=0}$}\label{fig:cw2t0}
    \includegraphics[width=\linewidth]{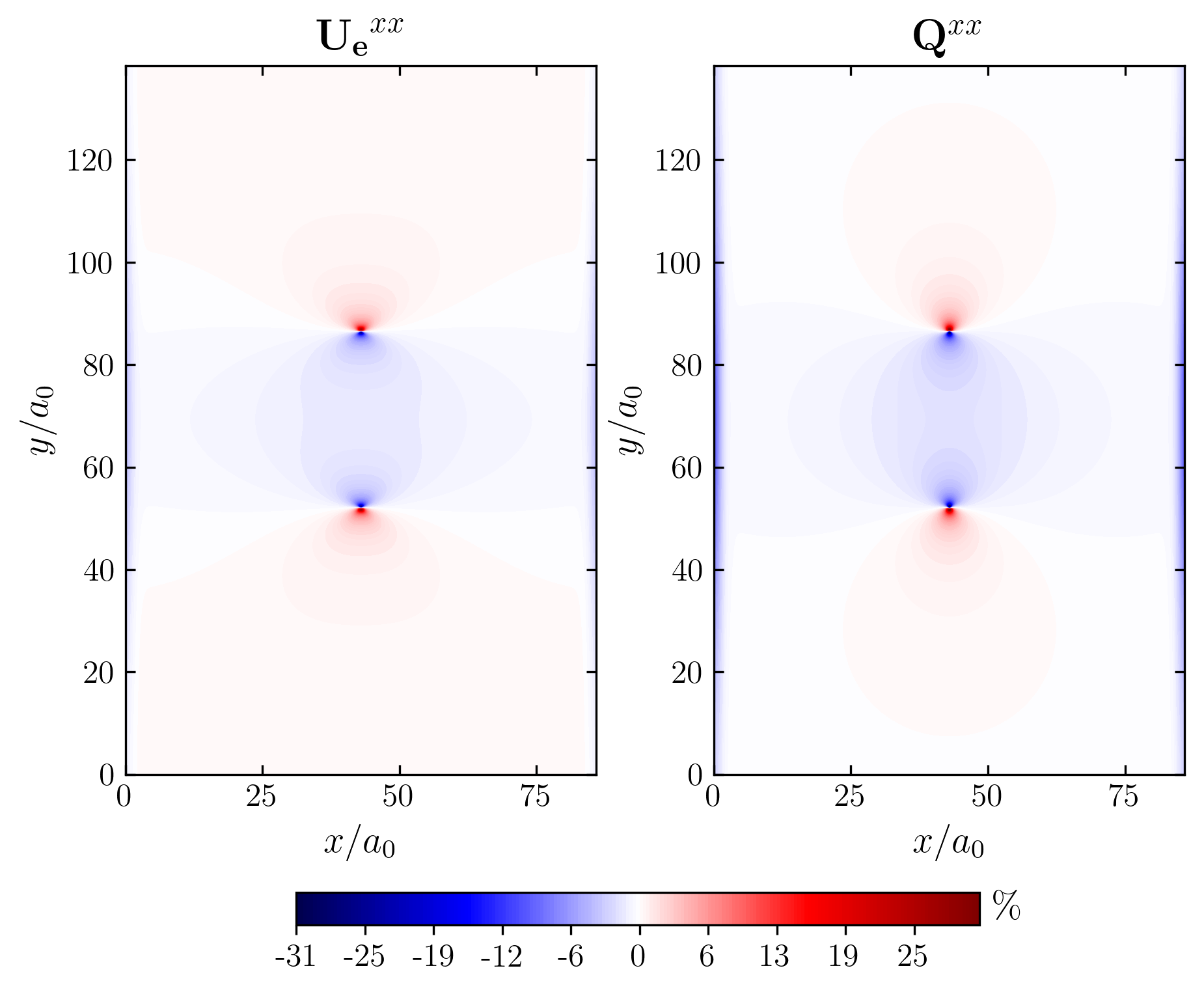}
  \end{subfigure}

  \vspace{0.6em}

  \begin{subfigure}[t]{0.45\textwidth}
    \centering
    \caption{$\Tens{c_{pen}=0}$, $\Tens{t=t_f}$}\label{fig:cw0tf}
    \includegraphics[width=\linewidth]{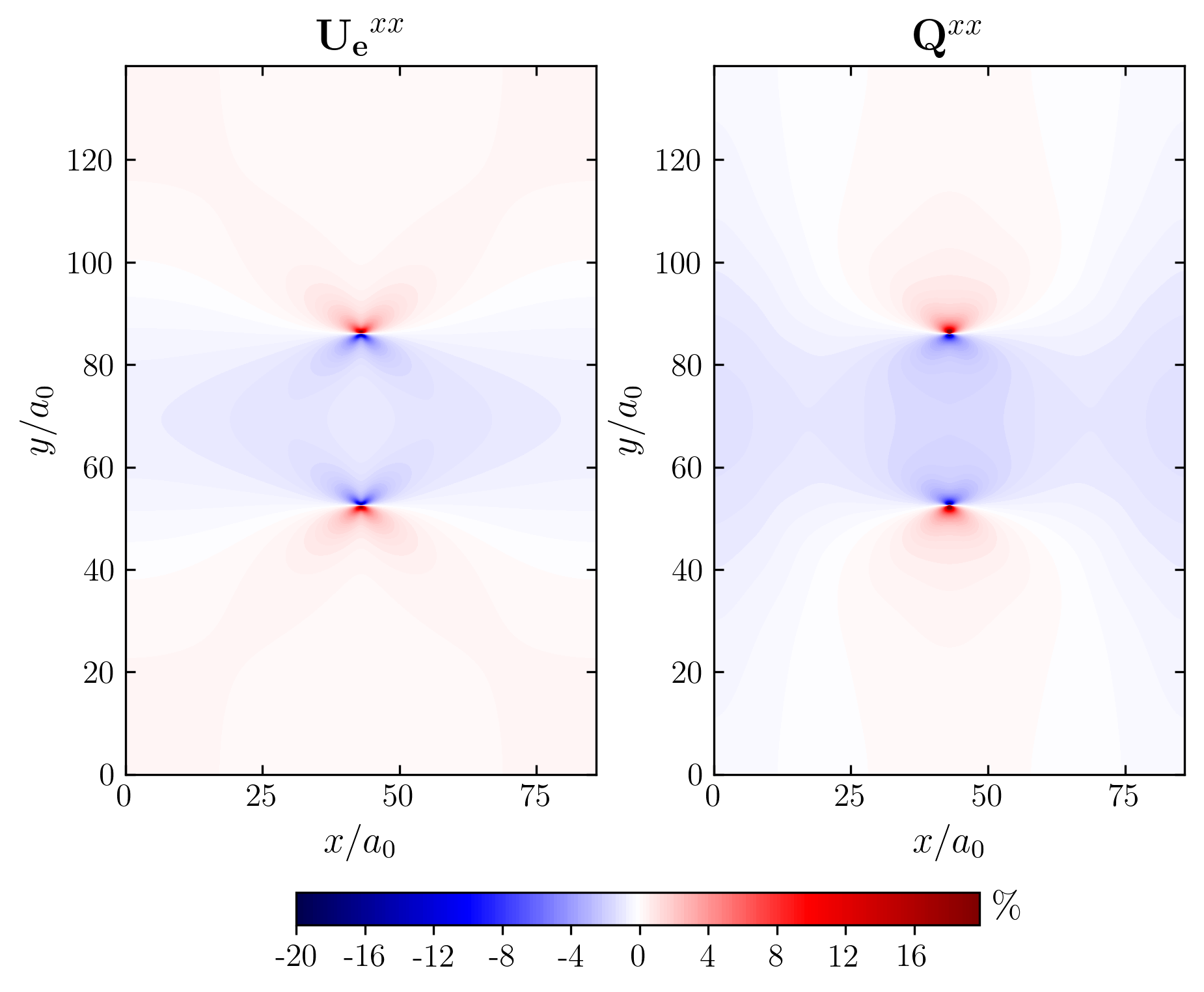}
  \end{subfigure}\hfill
  \begin{subfigure}[t]{0.45\textwidth}
    \centering
    \caption{$\Tens{c_{pen}=2}$, $\Tens{t=t_f}$}\label{fig:cw2tf}
    \includegraphics[width=\linewidth]{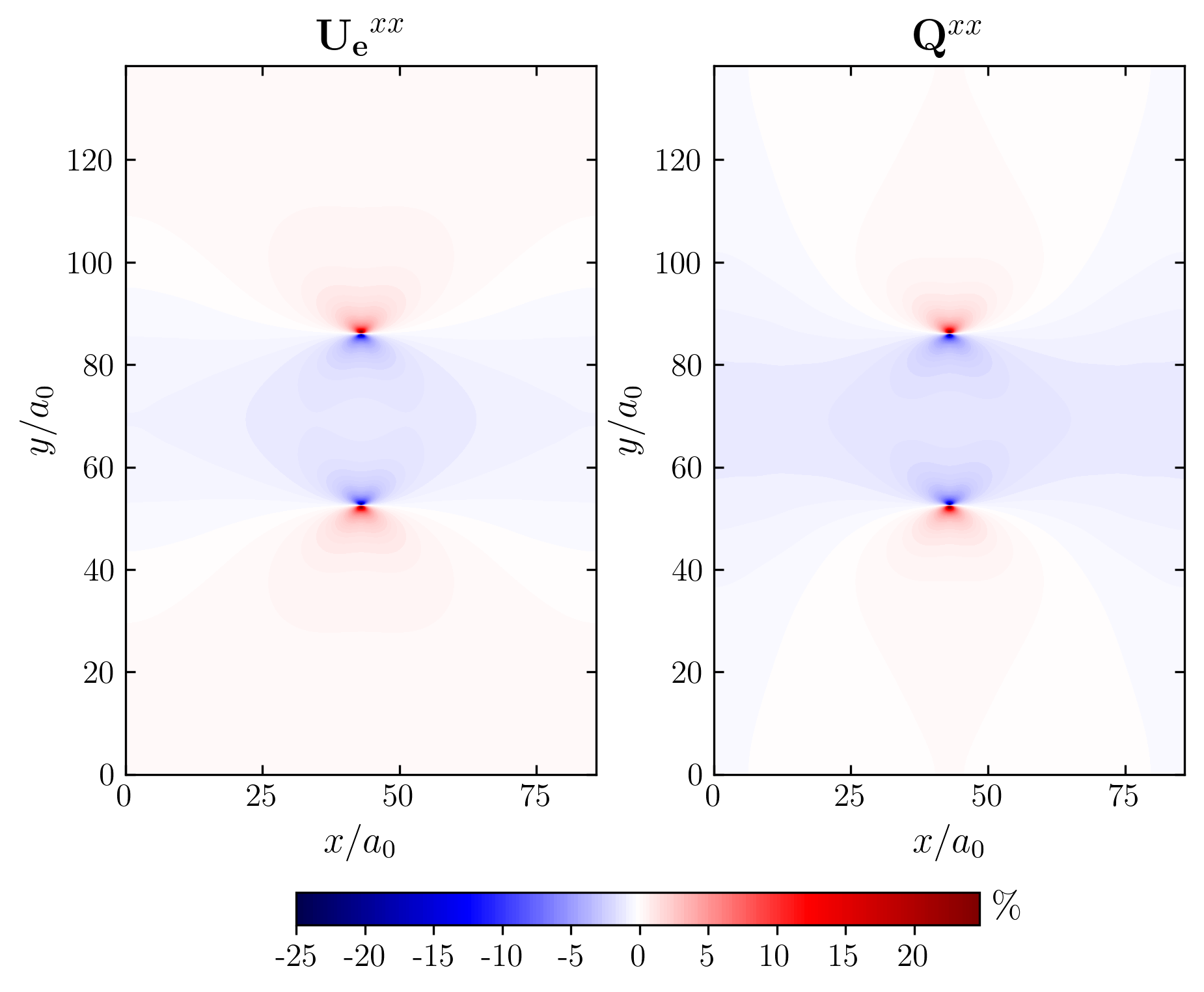}
  \end{subfigure}
\caption{Comparison of $\Tens{U}^e_{xx}$ and $\Tens{Q}_{xx}$ in uncoupled and coupled simulations: \subref{fig:cw0t0} uncoupled, $t=0$; \subref{fig:cw2t0} coupled, $t=0$; \subref{fig:cw0tf} uncoupled, $t=t_f$; \subref{fig:cw2tf} coupled, $t=t_f$.} \label{fig:COntourplots}
\end{figure}

\begin{figure}[H]
  \centering
    \includegraphics[width=0.25\linewidth]{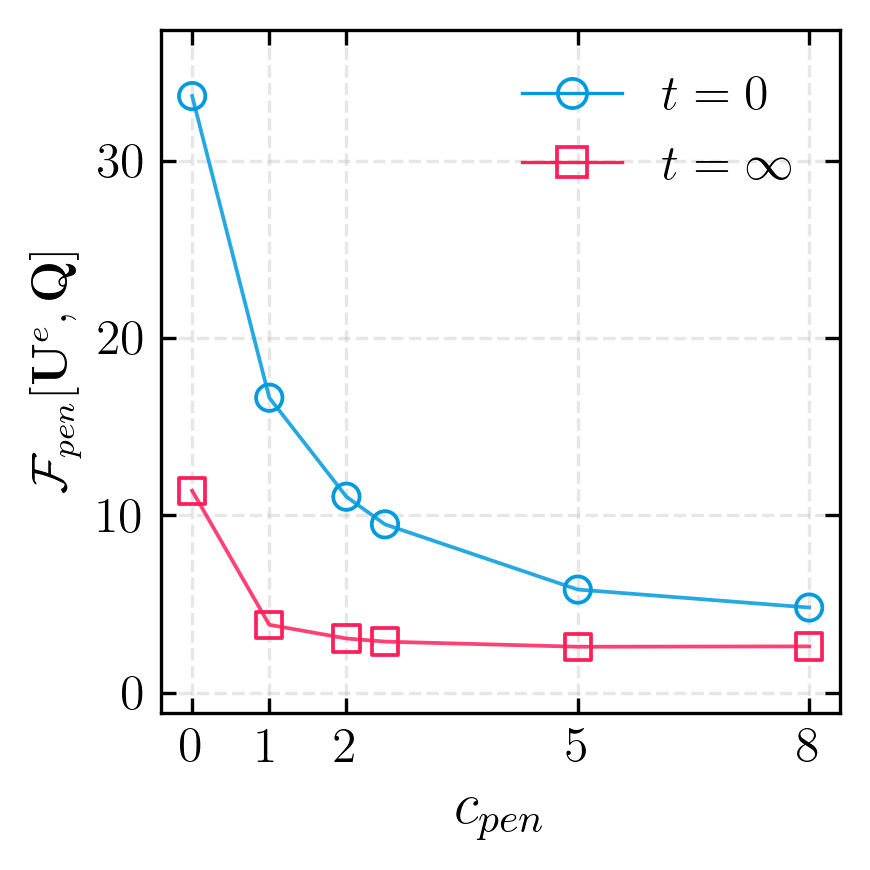}
    \caption{$\mathcal{F}_{pen}$ at $t=0$ and $t=\infty$ for different $c_{pen}$}\label{fig:climb_summary}
  \end{figure}

At $t = 0$, in the absence of coupling ($c_{pen} = 0$), the elastic distortion $\Tens{U}^e$ and the configurational lattice distortion $\Tens{Q}$ exhibit different spatial distributions(\cref{fig:cw0t0}), as previously demonstrated in \citep{upadhyayCouplingPhaseField2024}. The spatial distribution around the core differs and $\Tens{U}^e$ has quadruple shape. When the coupling term is introduced ($c_{pen} = 2$), the initial shape of $\Tens{U}^e$ changes (\cref{fig:cw2t0}). While $\Tens{Q}$ remains identical to the $c_{pen} = 0$ case, $\Tens{U}^e$ shifts significantly towards $\Tens{Q}$ due to the expression of the elastic stress in \cref{eq:elastic_stress_simple} resulting in a much smaller initial mismatch between the two fields. In the uncoupled case, the phase field then relaxes diffusively towards equilibrium (\cref{fig:cw0tf}). With coupling, $\Tens{Q}$ and $\Tens{U}^e$ rapidly converge towards similar profiles  (\cref{fig:cw2tf}). This behavior is summarized in \cref{fig:climb_summary}, where $\mathcal{F}_{pen}(t=0)$ decreases with increasing $c_{pen}$. Consequently, the distance traveled, ($\mathcal{F}_{pen}(t=0)-\mathcal{F}_{pen}(t=\infty)$), becomes smaller for larger values of $c_{pen}$.
 
\begin{figure}[H]
  \centering

  \begin{subfigure}[t]{0.45\textwidth}
    \centering
    \caption{$\bm{t=0}$}\label{fig:sliceXcw2t0}
    \includegraphics[width=\linewidth]{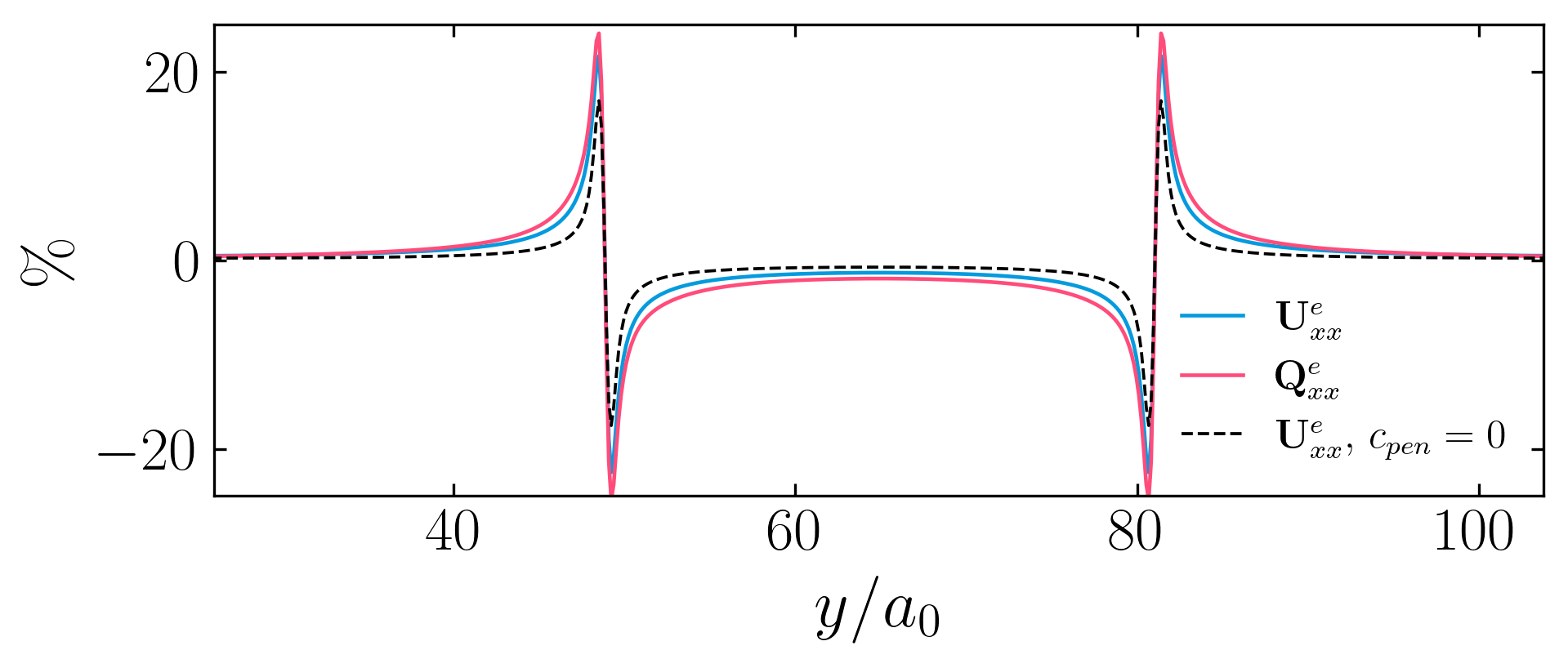}
  \end{subfigure}
    \begin{subfigure}[t]{0.45\textwidth}
    \centering
    \caption{$\bm{t=t_f}$}\label{fig:sliceXcw2tf}
    \includegraphics[width=\linewidth]{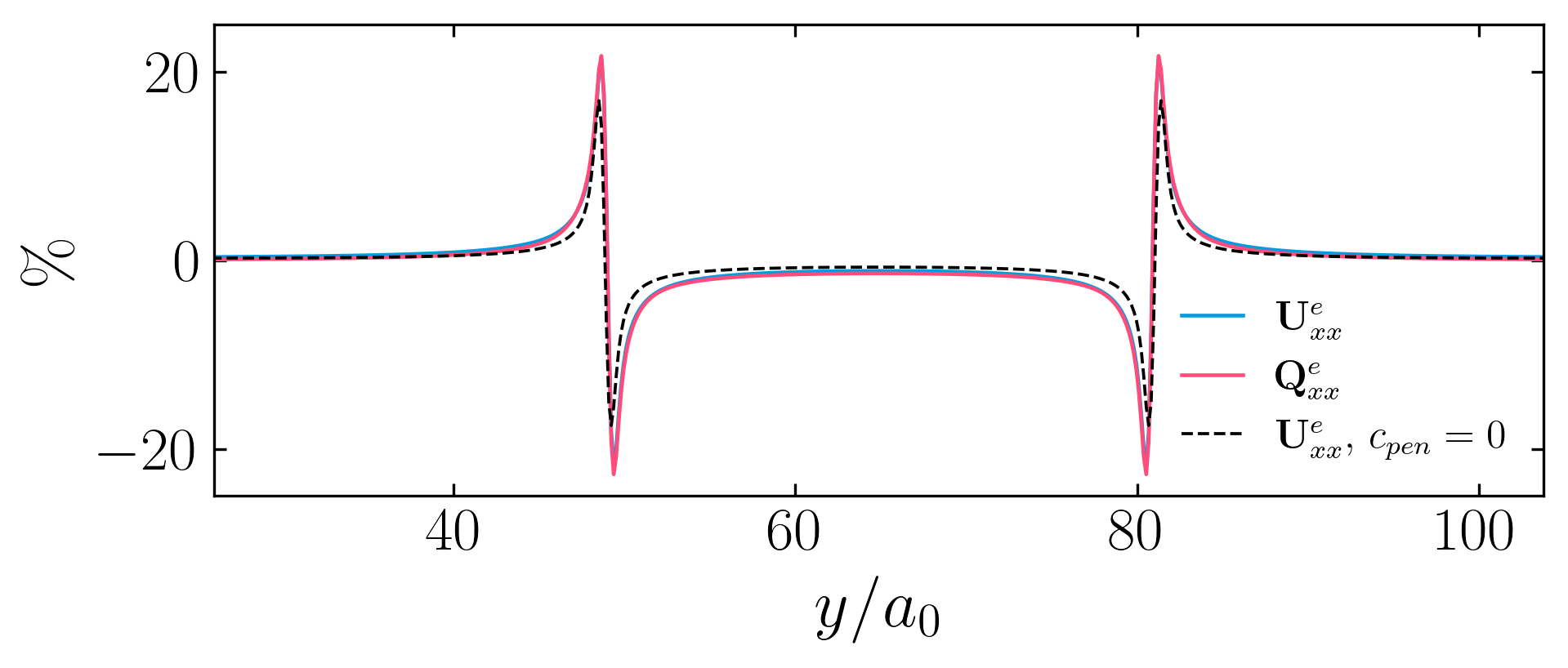}
  \end{subfigure}

  \vspace{0.6em}
    \begin{subfigure}[t]{0.45\textwidth}
    \centering
    \caption{$\bm{t=0}$}\label{fig:sliceycw2t0}
    \includegraphics[width=\linewidth]{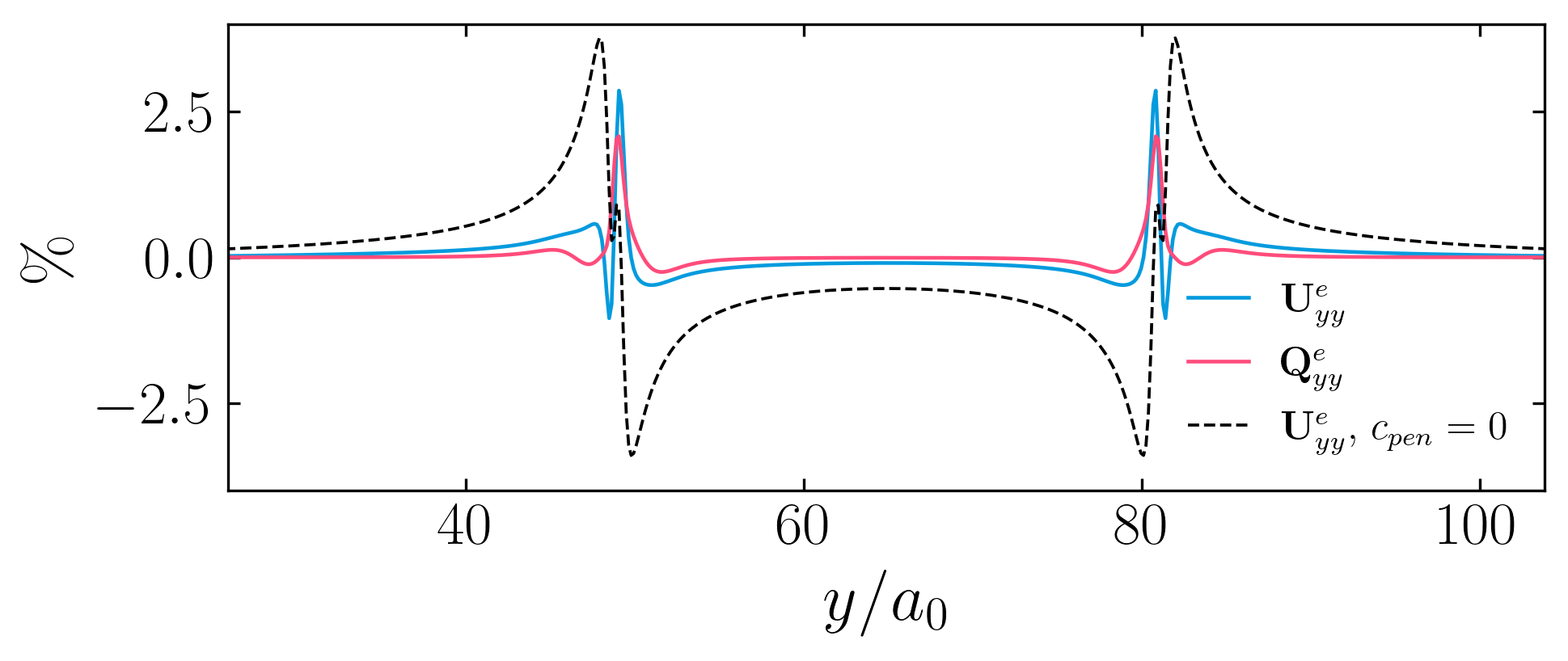}
  \end{subfigure}
  \begin{subfigure}[t]{0.45\textwidth}
    \centering
    \caption{$\bm{t=t_f}$}\label{fig:sliceycw2tf}
    \includegraphics[width=\linewidth]{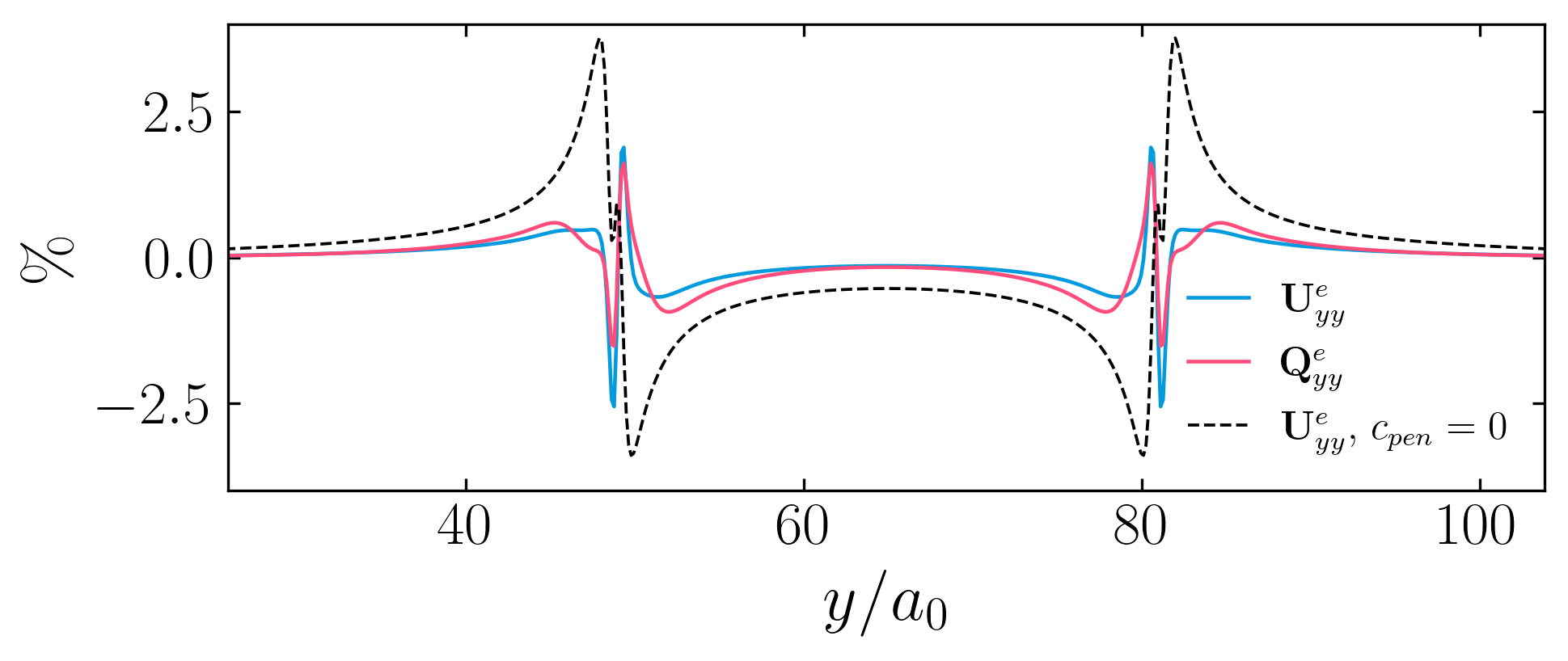}
  \end{subfigure}
\caption{Vertical profiles at $x=L/2$ of the distortion fields for $c_{pen}=2$, compared with the uncoupled elastic solution for the same defect configuration: \subref{fig:sliceXcw2t0} $xx$ component at $t=0$; and at $t=t_f$ in \subref{fig:sliceXcw2tf}. \subref{fig:sliceycw2t0} $yy$ component at $t=0$; and  at $t=t_f$ in \subref{fig:sliceycw2tf}.} \label{fig:slices_horizontal}
\end{figure}

\begin{figure}[h!]
  \centering

  \begin{subfigure}[t]{0.48\textwidth}
    \centering
    \caption{$\bm{t=0}$}\label{fig:sliceXYcw2t0}
    \includegraphics[width=\linewidth]{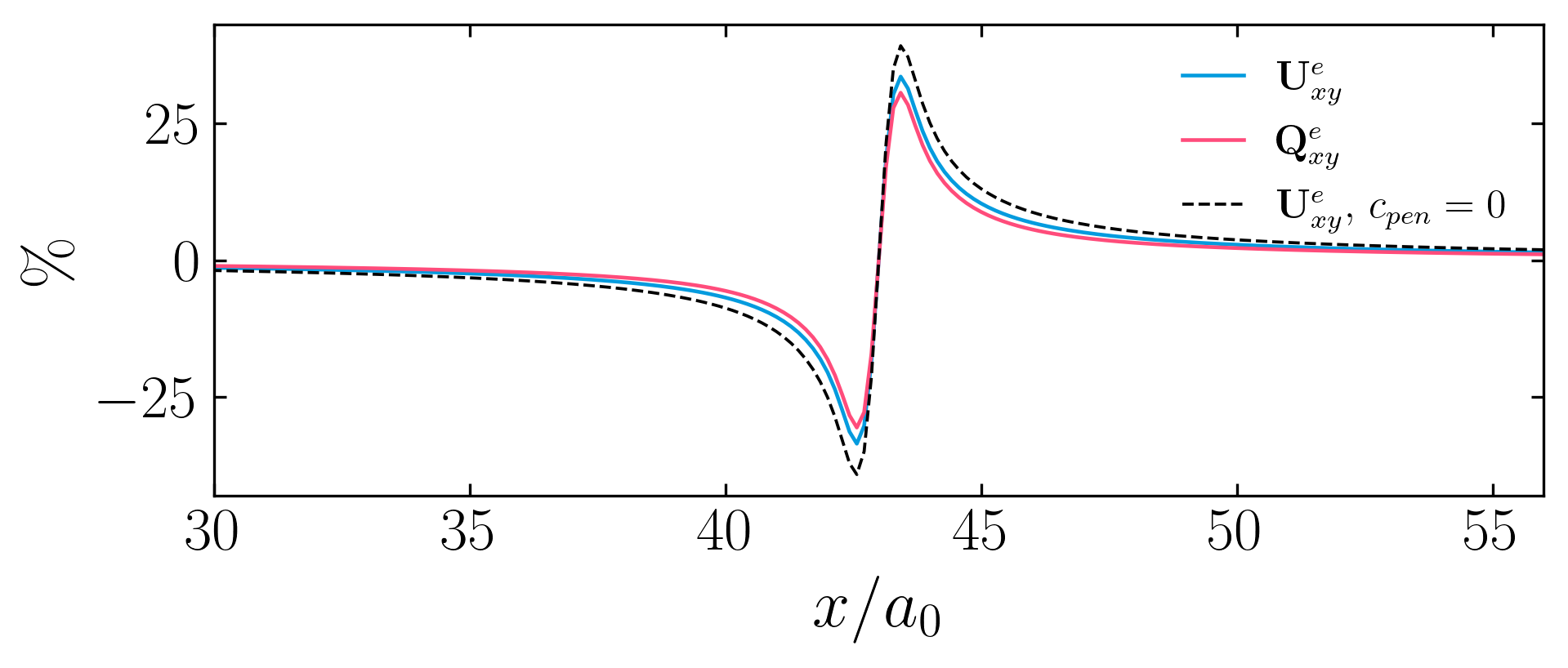}
  \end{subfigure}
    \begin{subfigure}[t]{0.48\textwidth}
    \centering
    \caption{$\bm{t=t_f}$}\label{fig:sliceXYcw2tf}
    \includegraphics[width=\linewidth]{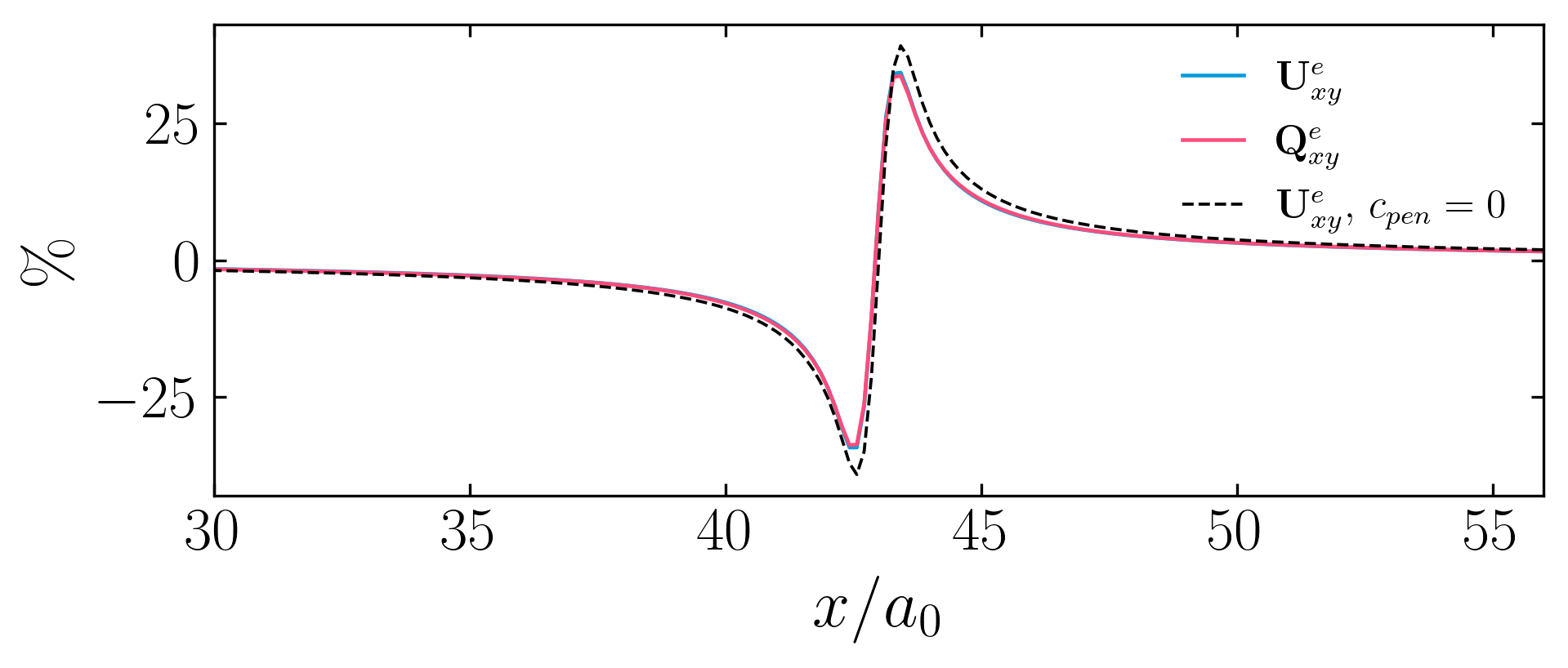}
  \end{subfigure}

  \vspace{0.6em}

    \begin{subfigure}[t]{0.48\textwidth}
    \centering
    \caption{$\bm{t=0}$}\label{fig:sliceyxcw2t0}
    \includegraphics[width=\linewidth]{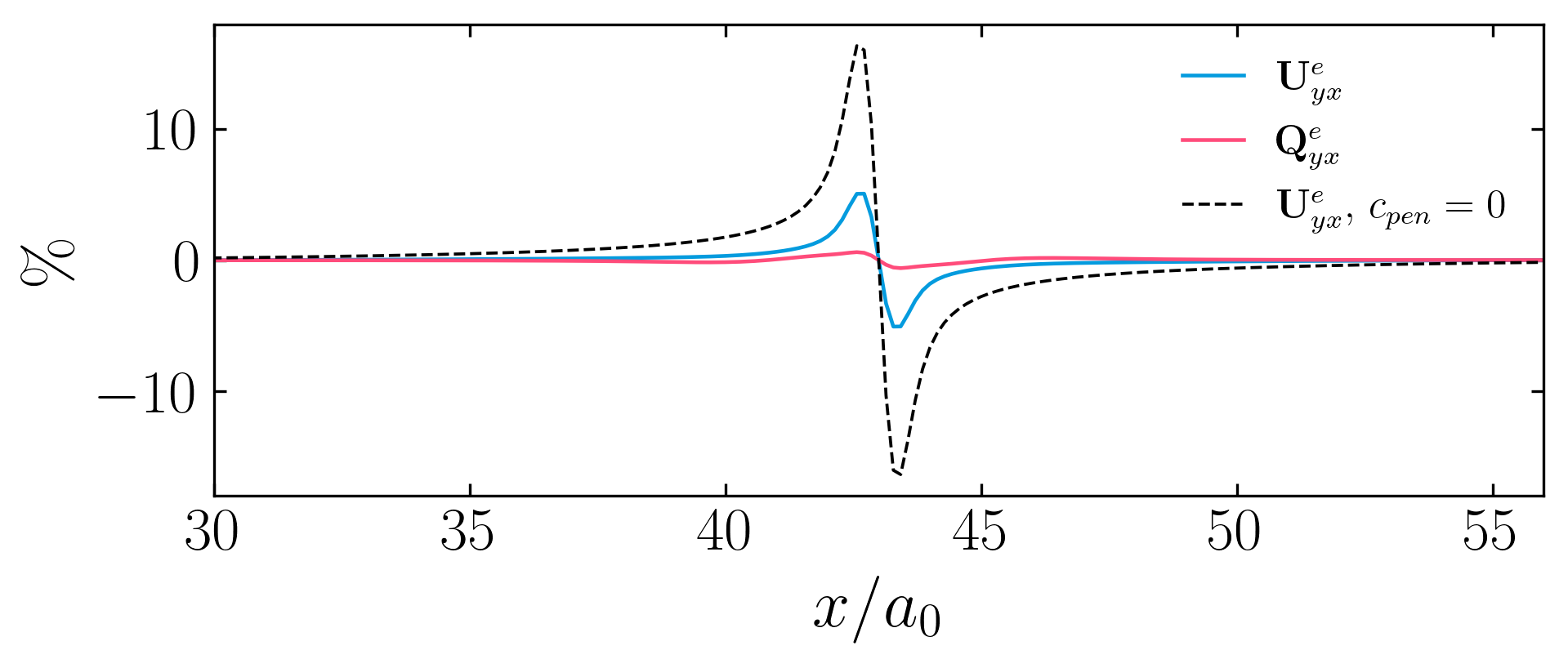}
  \end{subfigure}
  \begin{subfigure}[t]{0.48\textwidth}
    \centering
    \caption{$\bm{t=t_f}$}\label{fig:sliceyxcw2tf}
    \includegraphics[width=\linewidth]{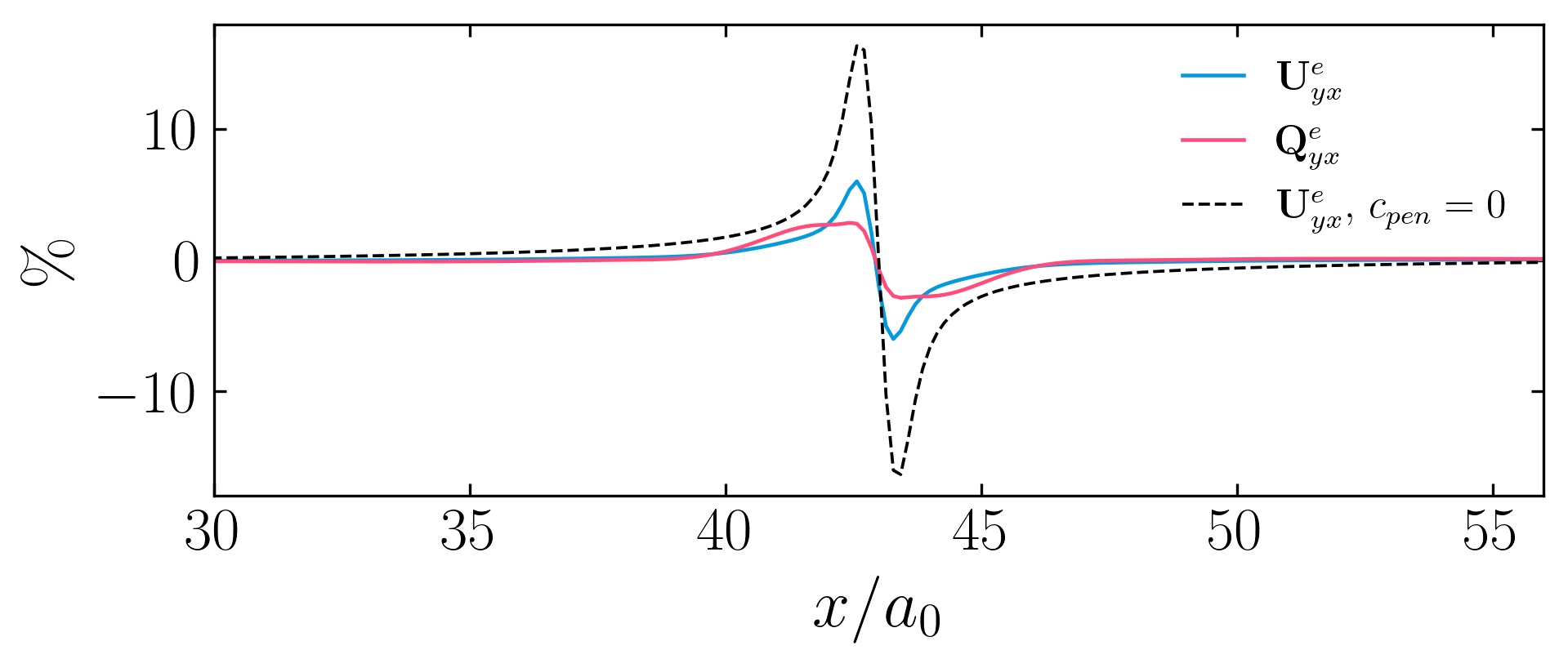}
  \end{subfigure}
\caption{Horizontal profiles at $y=3H/8$ of the distortion fields for $c_{pen}=2$, compared with the uncoupled elastic solution for the same defect configuration: \subref{fig:sliceXYcw2t0} $xy$ component at $t=0$; and at $t=t_f$ in \subref{fig:sliceXYcw2tf}. \subref{fig:sliceyxcw2t0} $yx$ component at $t=0$; and at $t=t_f$ in \subref{fig:sliceyxcw2tf}.} \label{fig:slices_vertical}
\end{figure}

At equilibrium, the two distortions have similar $xx$ components \cref{fig:cw2tf}. However, their shapes deviate from the classical solution for an edge dislocation, as shown in the slices in \cref{fig:slices_horizontal,fig:slices_vertical}. We emphasize that, once equilibrium is reached, the system remains at a local minimizer characterized by the condition $ \nabla \cdot \Tens{Q} = \nabla \cdot \Tens{U}^e$, rather than by the stronger condition $\Tens{U}^e = \Tens{Q}$. The former is sufficient to arrest the evolution, as shown in \cref{fig:divClimb}. As a result, a visible mismatch persists in the ($yy$) and ($yx$) components (\cref{fig:sliceycw2tf,fig:sliceyxcw2tf}), since the evolution only enforces agreement of the compatible parts of the distortion fields. Indeed, the penalty depends on the divergence of the differences $\partial_x (U^e_{xx}-Q_{xx}) + \partial_y (U^e_{xy}-Q_{xy})$ and $\partial_x (U^e_{yx}-Q_{yx}) + \partial_y (U^e_{yy}-Q_{yy})$, whereas the curls depend on different and independent set of combinations $\partial_x (U^e_{xy}-Q_{xy}) - \partial_y (U^e_{xx}-Q_{xx})$ and $
\partial_x (U^e_{yy}-Q_{yy}) - \partial_y (U^e_{yx}-Q_{yx})$. Nonetheless, a close match is achieved for the ($xx$) and ($xy$) components (\cref{fig:sliceXcw2tf,fig:sliceXYcw2tf}), which is sufficient for the study of two-dimensional transport of an edge dislocation considered in this work.

\subsection{Evolving dislocation core in a periodic domain without macroscopic loading}
Before investigating dislocation motion under an applied loading, we first examine the model's ability to maintain a localized core, a feature missing in classical FDM. 
For this analysis, the previous spatial discretizations with $N_x=490$ and $N_y=360$ is used in a periodic box containing a single edge dislocation centered in the domain. 
A timestep $dt=0.001$ is used to respect the CFL condition for the dislocation transport equation. 
The dislocation mobility is set at $B=1$ and the coefficient $c_{pen}$ is varied while the coefficient $c_{sh}$ is fixed at a value of $10$ in order to slow down PFC's dynamics and allowing the penalty term to act.

Unlike the previous setup, the dislocation densities are allowed to evolve separately by relaxing the assumption that $\nabla \times \Tens{U}^e = \nabla \times \Tens{Q}$. An FDM-driven approach is followed in which the transport equation (\cref{eq:Transport}) is solved to evolve the density $\Tens{\alpha} = \nabla \times \Tens{U}^e$, while $\widetilde{\Tens{\alpha}} = \nabla \times \Tens{Q}$ evolves independently via the relaxation of the order parameter. 
In the initial configuration, $\Tens{\alpha} = \widetilde{\Tens{\alpha}}$ (a Gaussian profile) is obtained from a relaxed core in PFC.

\begin{figure}[h!]
  \centering

  \begin{subfigure}[t]{0.32\textwidth}
    \centering
    \caption{$\bm{c_{pen}=0}$, $\bm{t=0}$}\label{fig:alpha_0_0}
    \includegraphics[width=\linewidth]{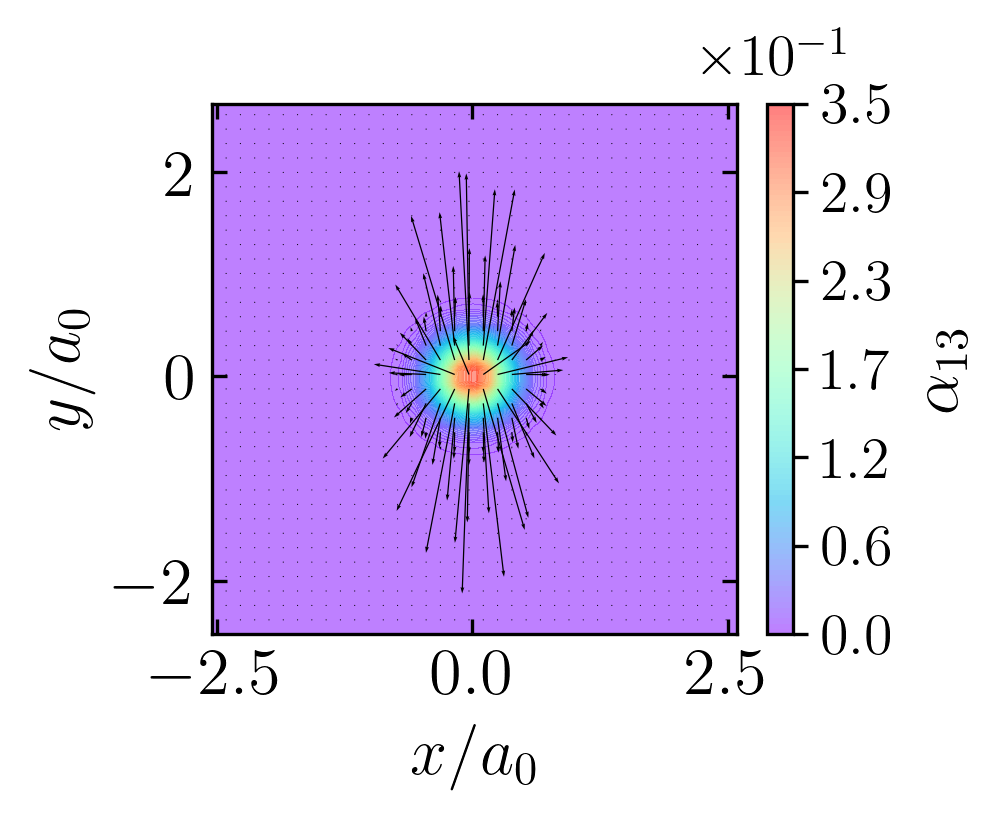}
  \end{subfigure}
  \hspace{0.05cm}
  \begin{subfigure}[t]{0.32\textwidth}
    \centering
    \caption{$\bm{c_{pen}=2}$, $\bm{t=t_f/5}$}\label{fig:alpha_2_0}
    \includegraphics[width=\linewidth]{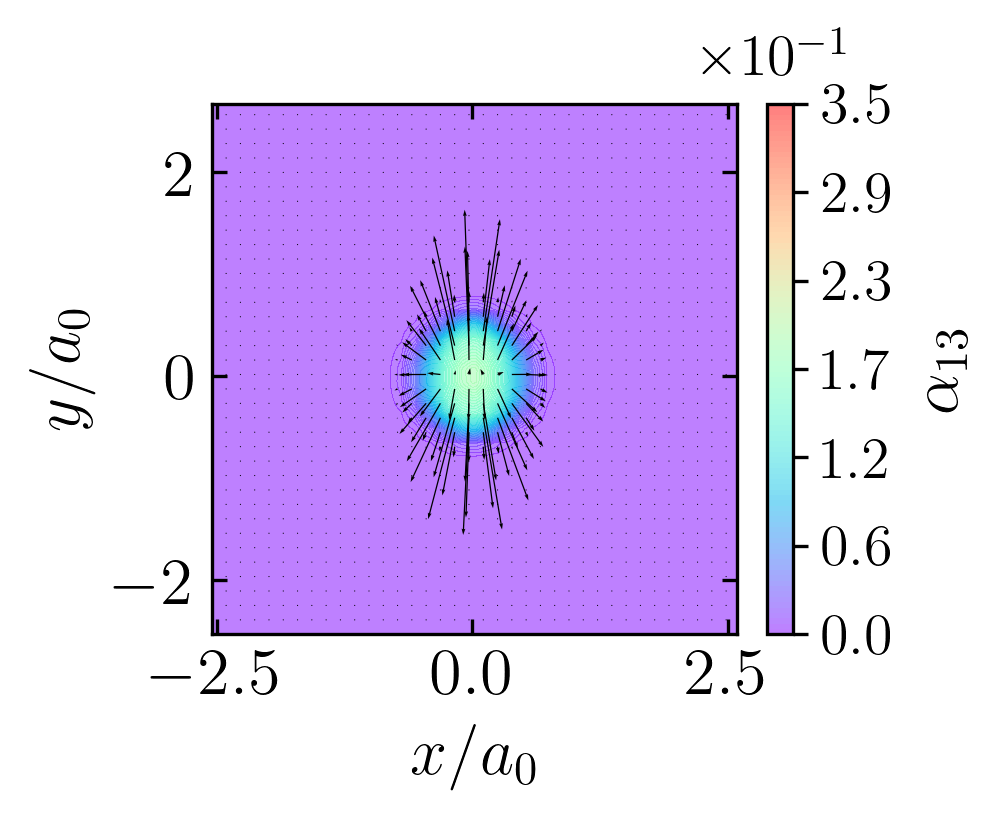}
\end{subfigure}
\hspace{0.05cm}
  \begin{subfigure}[t]{0.32\textwidth}
    \centering
    \caption{$\bm{c_{pen}=8}$, $\bm{t=t_f/5}$}\label{fig:alpha_8_0}
    \includegraphics[width=\linewidth]{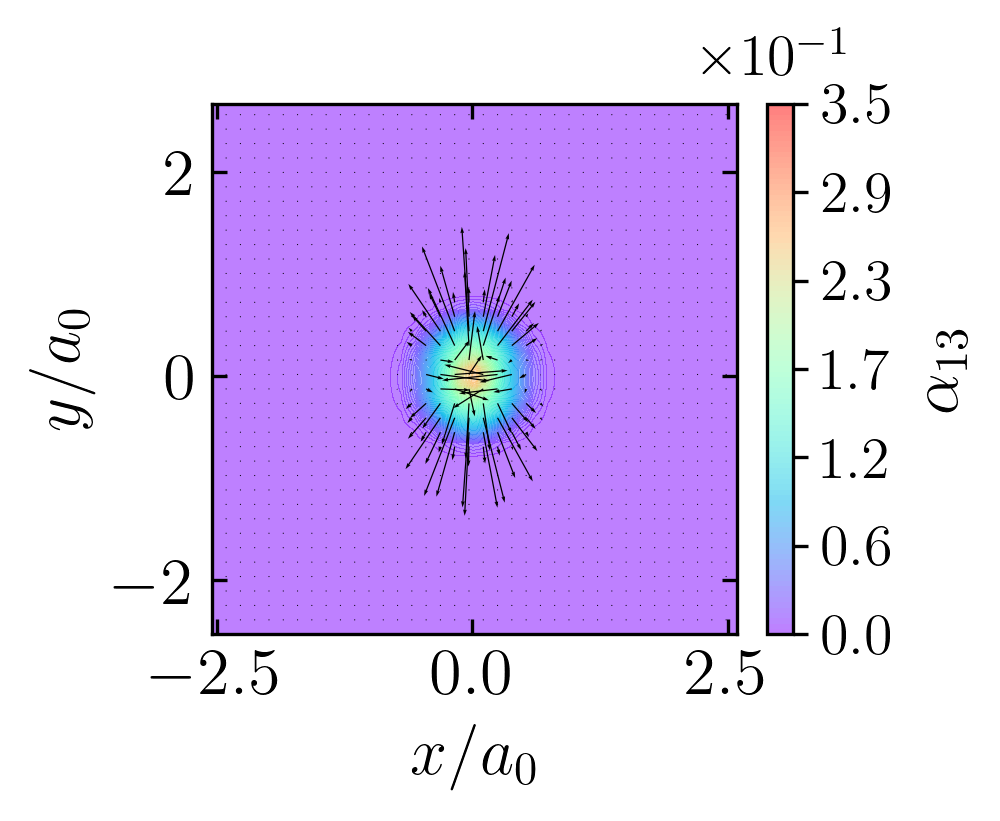}
  \end{subfigure}

  \vspace{0.3em}

  \begin{subfigure}[t]{0.32\textwidth}
    \centering
    \caption{$\bm{c_{pen}=0}$, $\bm{t=t_f}$}\label{fig:alpha_0_f}
    \includegraphics[width=\linewidth]{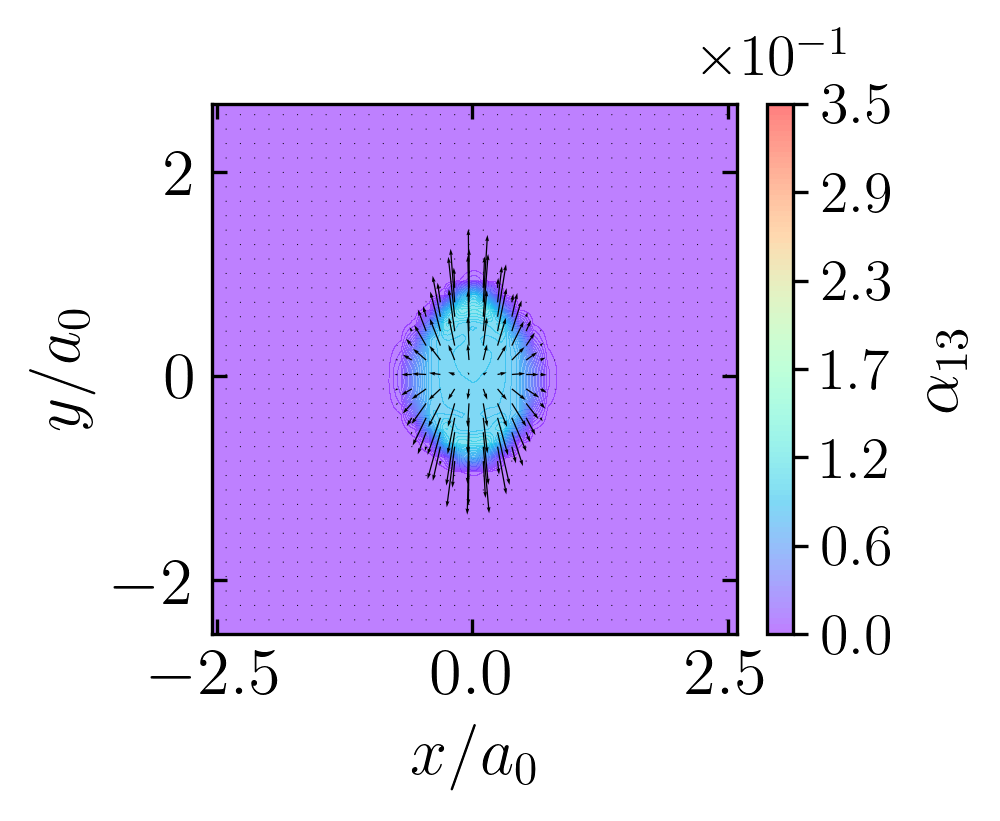}
  \end{subfigure}
  \hspace{0.05cm}
  \begin{subfigure}[t]{0.32\textwidth}
    \centering
    \caption{$\bm{c_{pen}=2}$, $\bm{t=t_f}$}\label{fig:alpha_2_f}
    \includegraphics[width=\linewidth]{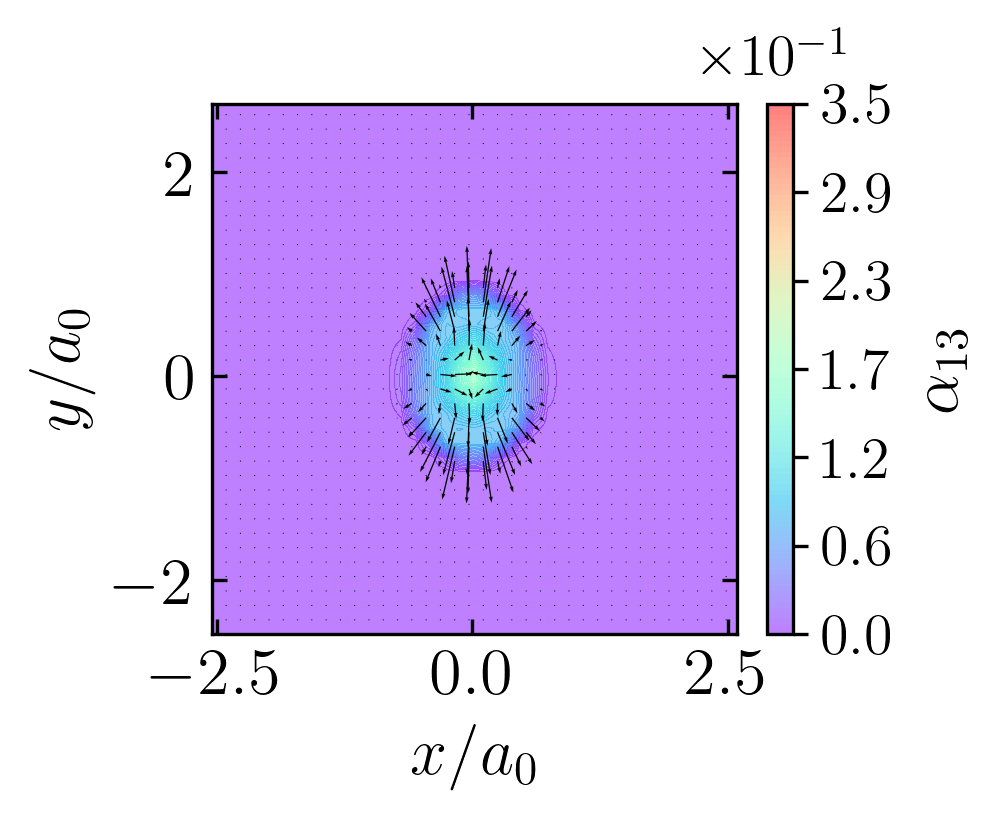}
\end{subfigure}
\hspace{0.05cm}
  \begin{subfigure}[t]{0.32\textwidth}
    \centering
    \caption{$\bm{c_{pen}=8}$, $\bm{t=t_f}$}\label{fig:alpha_8_f}
    \includegraphics[width=\linewidth]{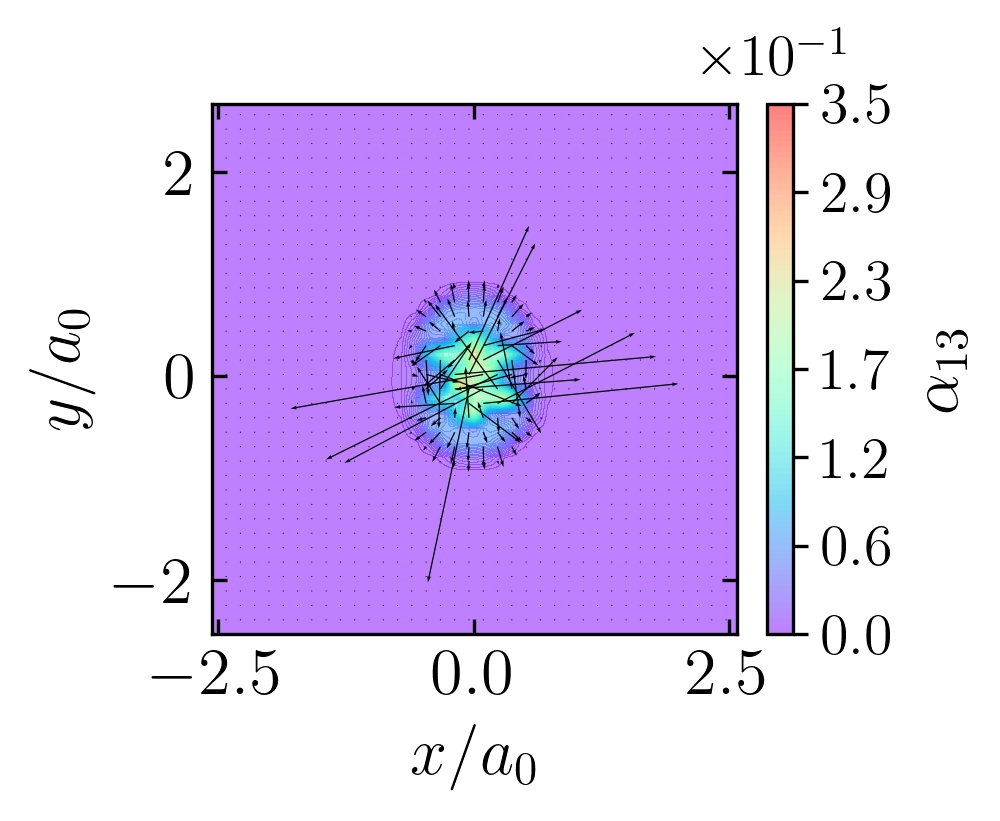}
  \end{subfigure}

    \vspace{0.3em}

    \begin{subfigure}[t]{0.32\textwidth}
    \centering
    \caption{$\bm{c_{pen}=0}$, $\bm{t=t_f}$}\label{fig:alpha_slice_0}
    \includegraphics[width=0.9\linewidth]{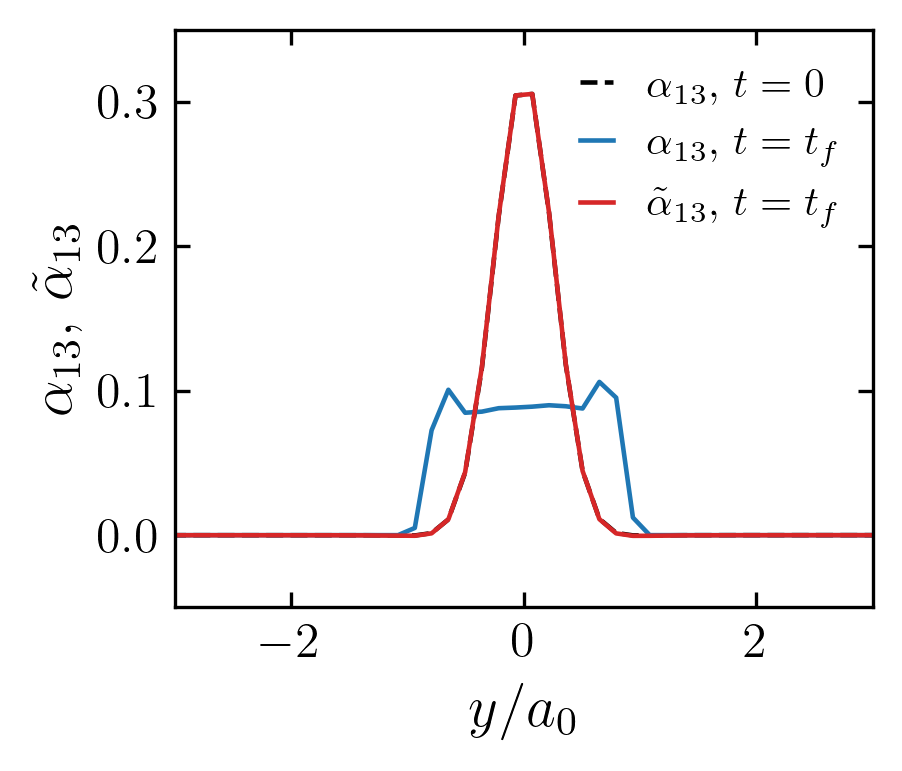}
  \end{subfigure}
  \hspace{0.05cm}
  \begin{subfigure}[t]{0.32\textwidth}
    \centering
    \caption{$\bm{c_{pen}=2}$, $\bm{t=t_f}$}\label{fig:alpha_slice_2}
    \includegraphics[width=0.9\linewidth]{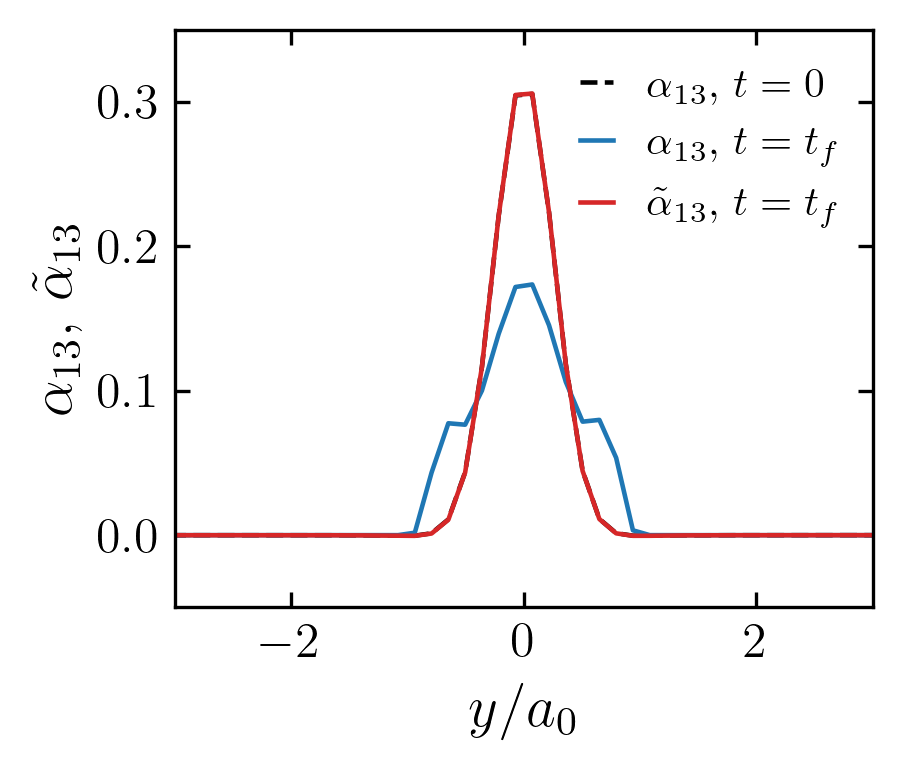}
  \end{subfigure}
  \hspace{0.05cm}
  \begin{subfigure}[t]{0.32\textwidth}
    \centering
    \caption{$\bm{c_{pen}=8}$, $\bm{t=t_f}$}\label{fig:alpha_slice_8}
    \includegraphics[width=0.9\linewidth]{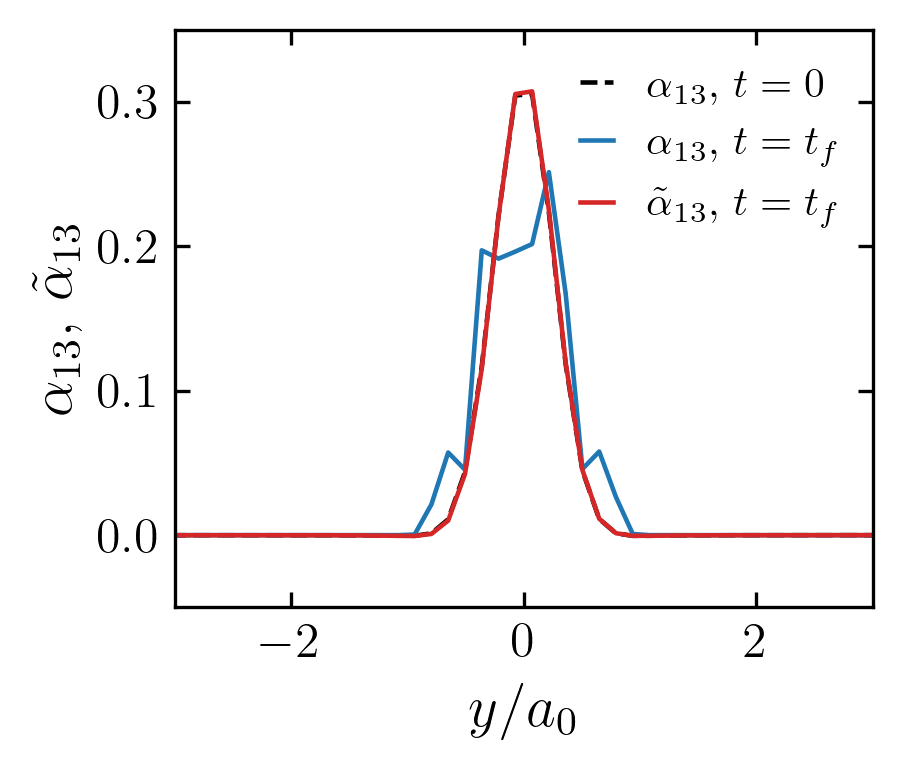}
  \end{subfigure}

    \caption{Evolution of the dislocation density fields $\Tens{\alpha}_{xz}$ and $\widetilde{\Tens{\alpha}}_{xz}$ and the dislocation velocity field $\vec{v}^{d}$ for different values of $c_{pen}$ while keeping $c_{sh}=10$. Panels \subref{fig:alpha_0_0} to \subref{fig:alpha_8_0} show contour plots of the dislocation density together with the dislocation velocity field $\vec{v}^{,d}$ at the initial state ($c_{pen}=0$, $t=0$) and at the intermediate time $t=t_f/5$ for $c_{pen}=2$ and $c_{pen}=8$, respectively. Panels \subref{fig:alpha_0_f} to \subref{fig:alpha_8_f} show the corresponding fields at the final time $t=t_f$ for $c_{pen}=0$, $2$, and $8$. Panels \subref{fig:alpha_slice_0} to \subref{fig:alpha_slice_8} present vertical profiles along $x=0$ of $\Tens{\alpha}_{xz}$ and $\widetilde{\Tens{\alpha}}_{xz}$ at $t=t_f$.}
  \label{fig:alpha_comp}
\end{figure}

In \cref{fig:alpha_comp}, we compare the evolution of the dislocation fields and plot the spatial distribution of the dislocation velocity field, as derived from the constitutive analysis in \cref{eq:vel_Field}, for different values of $c_{pen}$. In the absence of coupling ($c_{pen} = 0$), \cref{fig:alpha_0_f} shows the expected endless relaxation of the core, resulting in an evenly spread distribution along the $\vec{x}'$ direction at $t=t_f$. 
In \cref{fig:alpha_slice_0}, we observe the expected mismatch between $\Tens{\alpha}_{xz}$ and $\widetilde{\Tens{\alpha}}_{xz}$ that remained intact.

As we activate the coupling with $c_{pen}=2$, the behavior of the system remains qualitatively similar (\cref{fig:alpha_2_0,fig:alpha_2_f}). 
The spreading of the core is still noticeable, but it is less evenly distributed than in the $c_{pen}=0$ case; instead, a peak remains in the middle, as seen in the line plot in \cref{fig:alpha_slice_2}. 
The spreading of $\Tens{\alpha}$ causes $\Tens{U^p}$ and $\Tens{U}^e$ to drop, whereas $\Tens{Q}$ maintains its value. 
The presence of the term $c_{pen} (\Tens{U}^e- \Tens{Q})_{sym}$ in \cref{eq:elastic_stress_simple} ensures that even if $\Tens{\alpha} \rightarrow 0$, the elastic distortion obtained from mechanical equilibrium does not vanish completely, though a mismatch persists where $\Tens{Q}^\perp \neq \Tens{U}^{e,\perp}$. 
The plastic distortion, however, does not see this coupling and drops alongside with $\Tens{\alpha}$. 

This mismatch $\Tens{\alpha} \neq \widetilde{\Tens{\alpha}}$ can be interpreted as two different evolutions of the plastic information in the system with different plastic distortion rates, $\Tens{\alpha} \times \vec{v^d} \neq \Tens{\mathcal{J}^\psi}$. Unless we enforce $\nabla \times \Tens{U}^e = \nabla \times \Tens{Q}$ and $\dot{\Tens{U^p}} = \Tens{\mathcal{J}^\psi}$, this coupled model will always exhibit two different dislocation densities as it shall be demonstrated in what follows.

\subsection{Dislocation annihilation}\label{sec:anni}
\begin{figure}[h!]
\begin{subfigure}[t]{0.45\textwidth}
  \centering
    
  \caption{Zoomed view of $\psi(t=0)$}
  \label{fig:psiZGlide}
\includegraphics[width=0.9\linewidth]{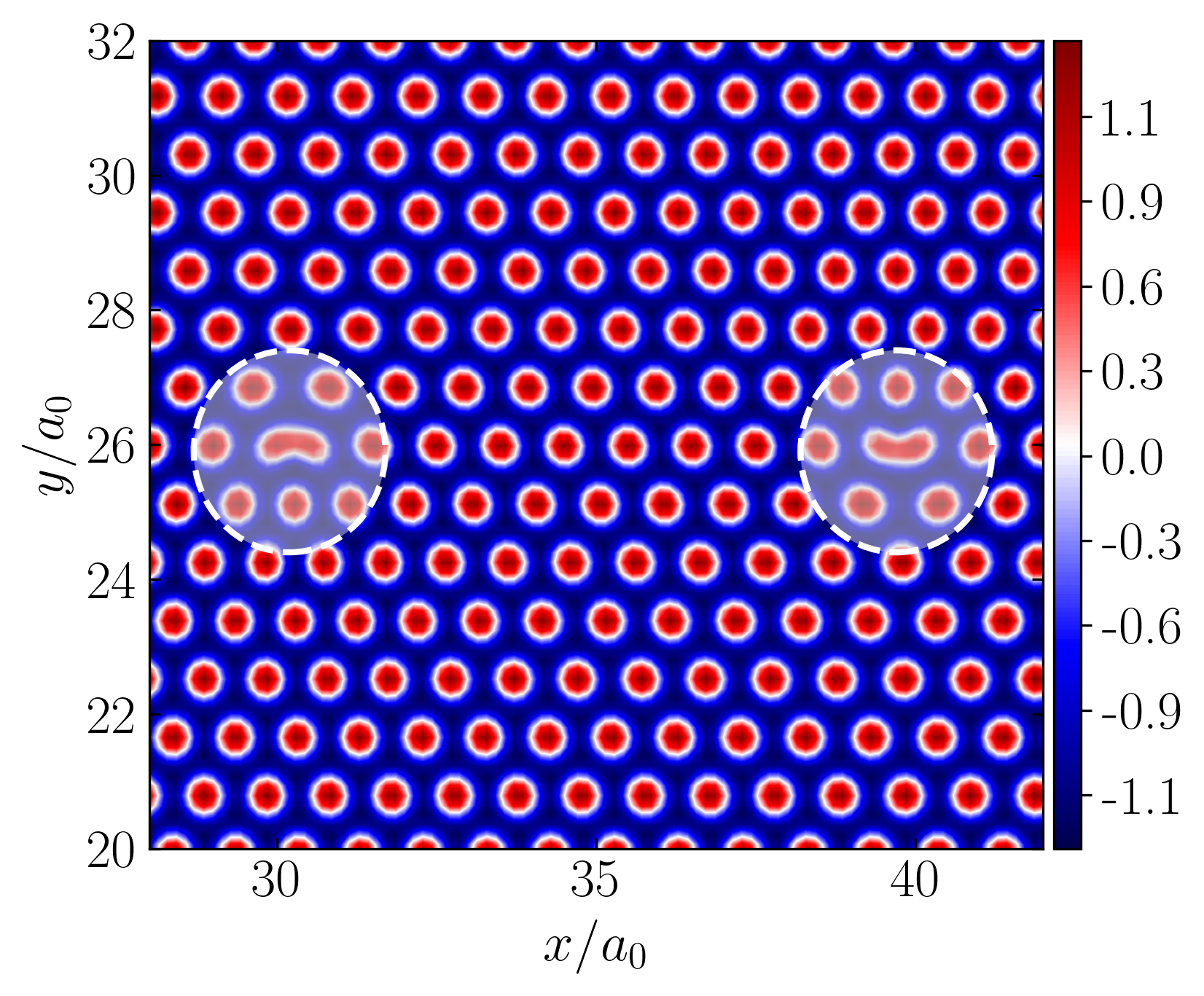}

\end{subfigure}
\hfill
\begin{subfigure}[t]{0.45\textwidth}
  \centering

  \caption{$\widetilde{\Tens{\alpha}}_{xz}$}\label{fig:alphaGlide}
  \includegraphics[width=0.9\linewidth]{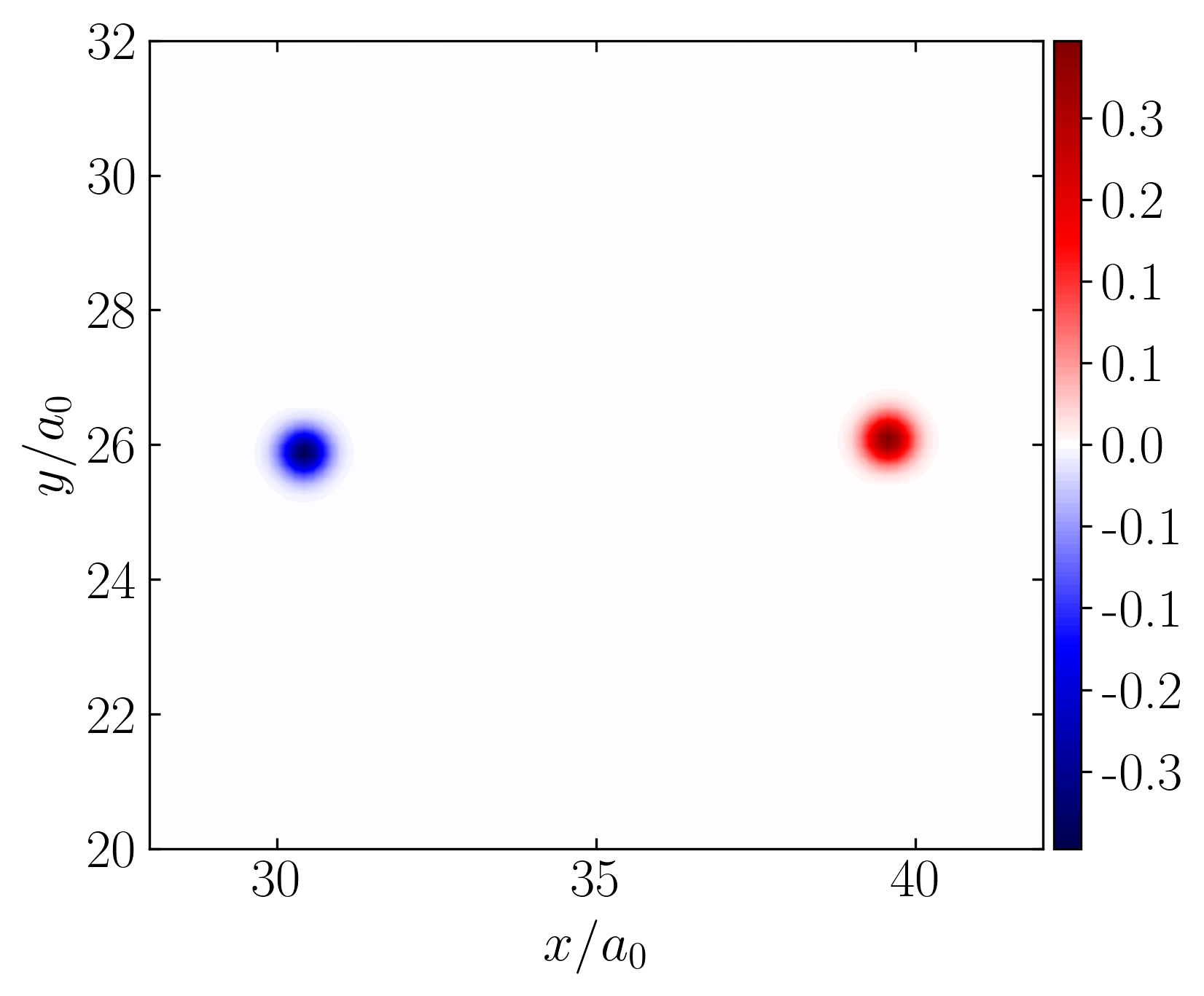}
  
\end{subfigure}
\caption{Initial configuration of the dislocation annihilation simulation. \subref{fig:psiZGlide} A zoom around the center of the domain : $\psi$ in the background and the circles highlight the dislocations position. \subref{fig:alphaGlide}  The corresponding initial dislocation density.  }\label{fig:init_glide}
\end{figure}

In this section, the evolution of a dislocation dipole in a hexagonal lattice is studied for different values of the model parameters $(B,c_{pen})$. 
A $490\times360$ domain with the same spatial discretization as before is considered and a constant timestep of $dt = 1.10^{-3}$ is selected. 
The initial configuration (\cref{fig:psiZGlide}) is given by inserting two oppositely signed edge dislocations at $13L/20$ and $17L/30$ such that the initial separation of the dipole is $d\approx9a_0$. 

\subsubsection{FDM-driven motion}

We now investigate the fully coupled FDM-driven scheme, in which the dislocation transport equation is solved explicitly, and assess its behavior for a range of $B$ and $c_{{pen}}$.

In the uncoupled limit ($c_{{pen}}=0$, $c_{\mathrm{sh}}=100$), the relaxation of $\mathcal{F}_{\mathrm{sh}}$ alone drives both cores to glide and annihilate, restoring the perfect-lattice configuration that minimizes the Swift-Hohenberg energy. The resulting dislocation separation $d_{{ref}}(t)$ provides a baseline against which the coupled simulations can be calibrated: assuming the classical Peach-Koehler interaction $f_{{ref}} = \mu b^2/[2\pi(1-\nu)d]$ and a linear drag relation 
$v_d = f_{{ref}}/B$ (with $2v_d = -\dot{d}$), integration yields
\begin{equation}\label{eq:PK_distance}
    d_{{ref}}(t) = \sqrt{d_0^{\,2} - \frac{2\mu b^2}{\pi(1-\nu)B}\,t}
\end{equation}
Fitting this expression to the PFC trajectory (\cref{app:PKfit2PFC}) gives $B_{{ref}} \approx 0.095$, which we adopt as the reference mobility for the FDM-driven simulations below.

\begin{figure}[h!]
  \centering
  \begin{minipage}[t]{0.33\textwidth}
    \centering
    \begin{subfigure}[t]{\textwidth}
      \centering
      \caption{$B=B_{{ref}}=0.095$}
      \label{fig:ann_fdm_B095}
      \includegraphics[width=0.78\linewidth]{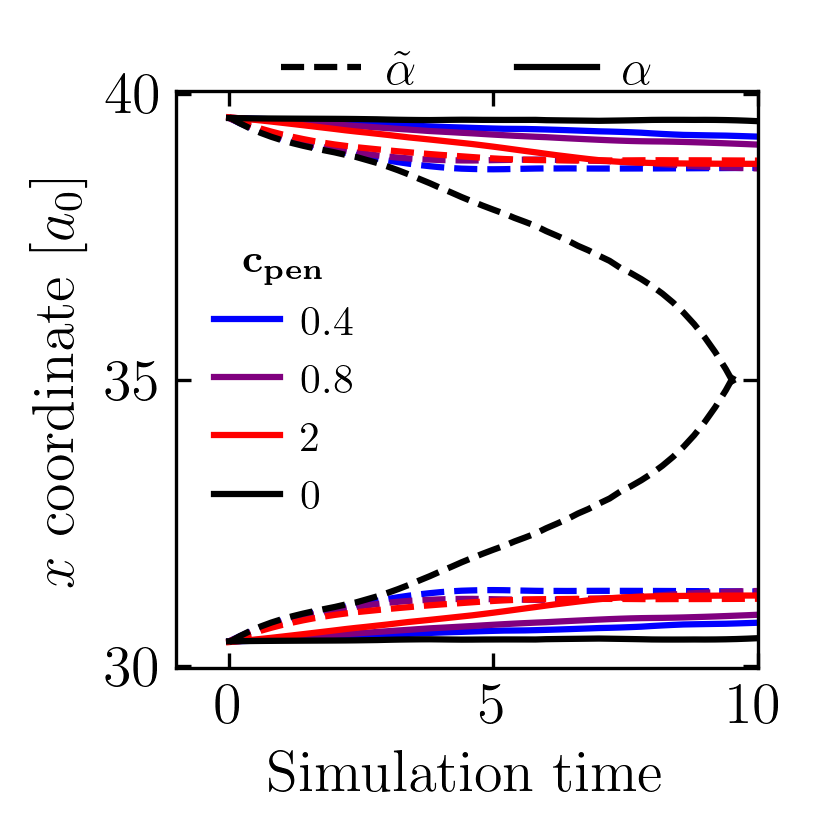}
    \end{subfigure}
    \vspace{0.6em}
    \begin{subfigure}[t]{\textwidth}
      \centering
      \caption{$B = B_{{ref}}/95=0.001$}
      \label{fig:ann_fdm_B001}
      \includegraphics[width=0.78\linewidth]{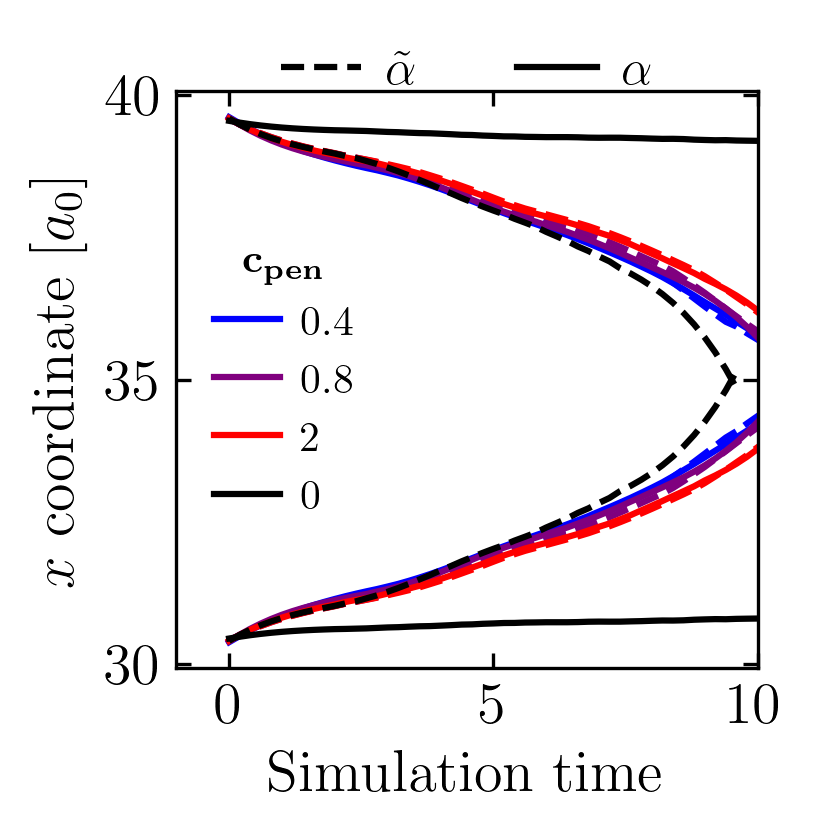}
    \end{subfigure}
  \end{minipage}
  \begin{minipage}[t]{0.32\textwidth}
    \centering
    \begin{subfigure}[t]{\textwidth}
      \centering
      \caption{Snapshot of $\Tens{\alpha}$ at $t=5$}
      \label{fig:ann_fdm_alpha_stacked}
      \includegraphics[width=\linewidth]{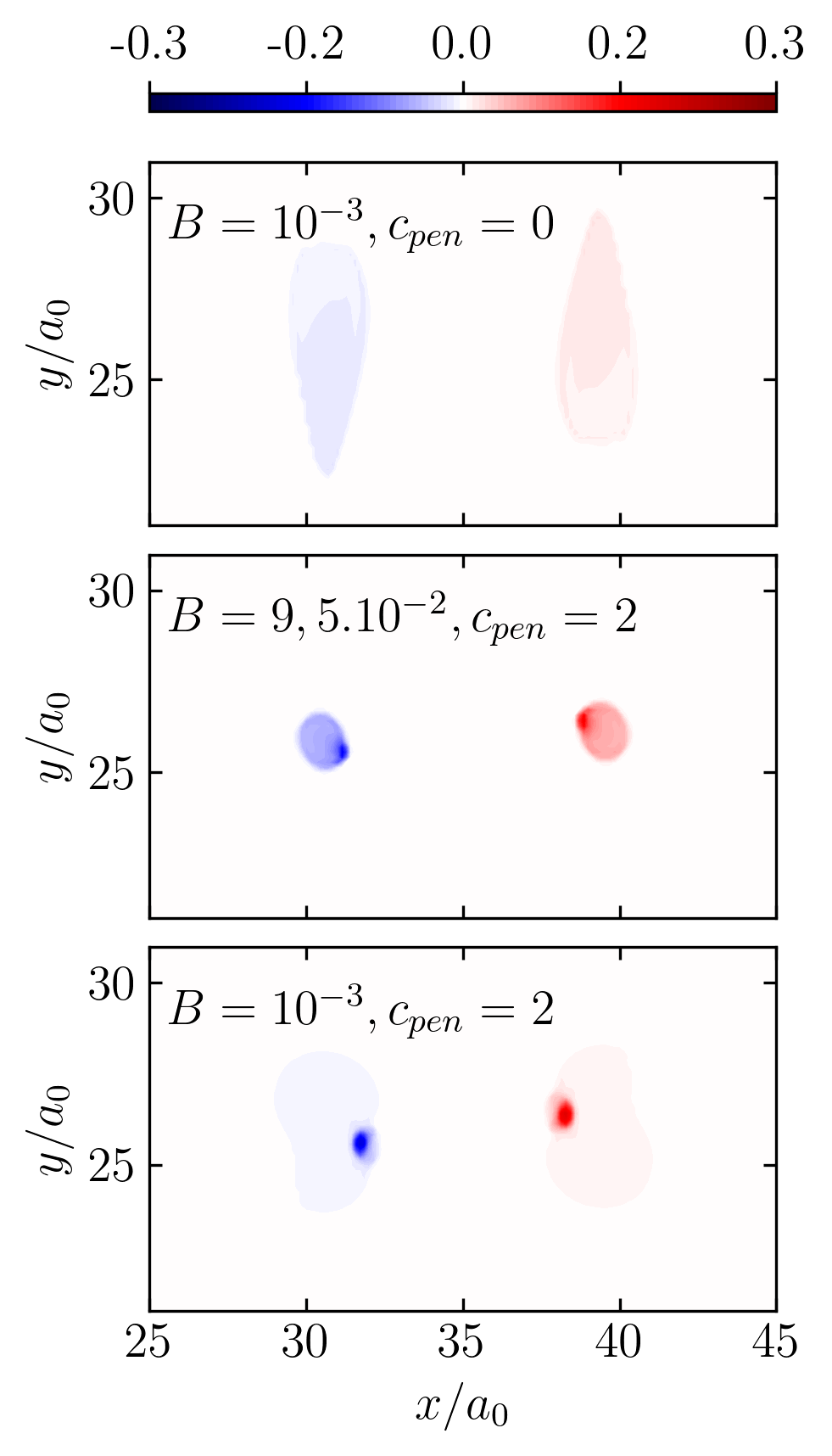}
    \end{subfigure}
  \end{minipage}
    \begin{minipage}[t]{0.32\textwidth}
    \centering
    \begin{subfigure}[t]{\textwidth}
      \centering
      \caption{Snapshot of $\widetilde{\Tens{\alpha}}$ at $t=5$}
      \label{fig:ann_fdm_alphaT_stacked}
      \includegraphics[width=\linewidth]{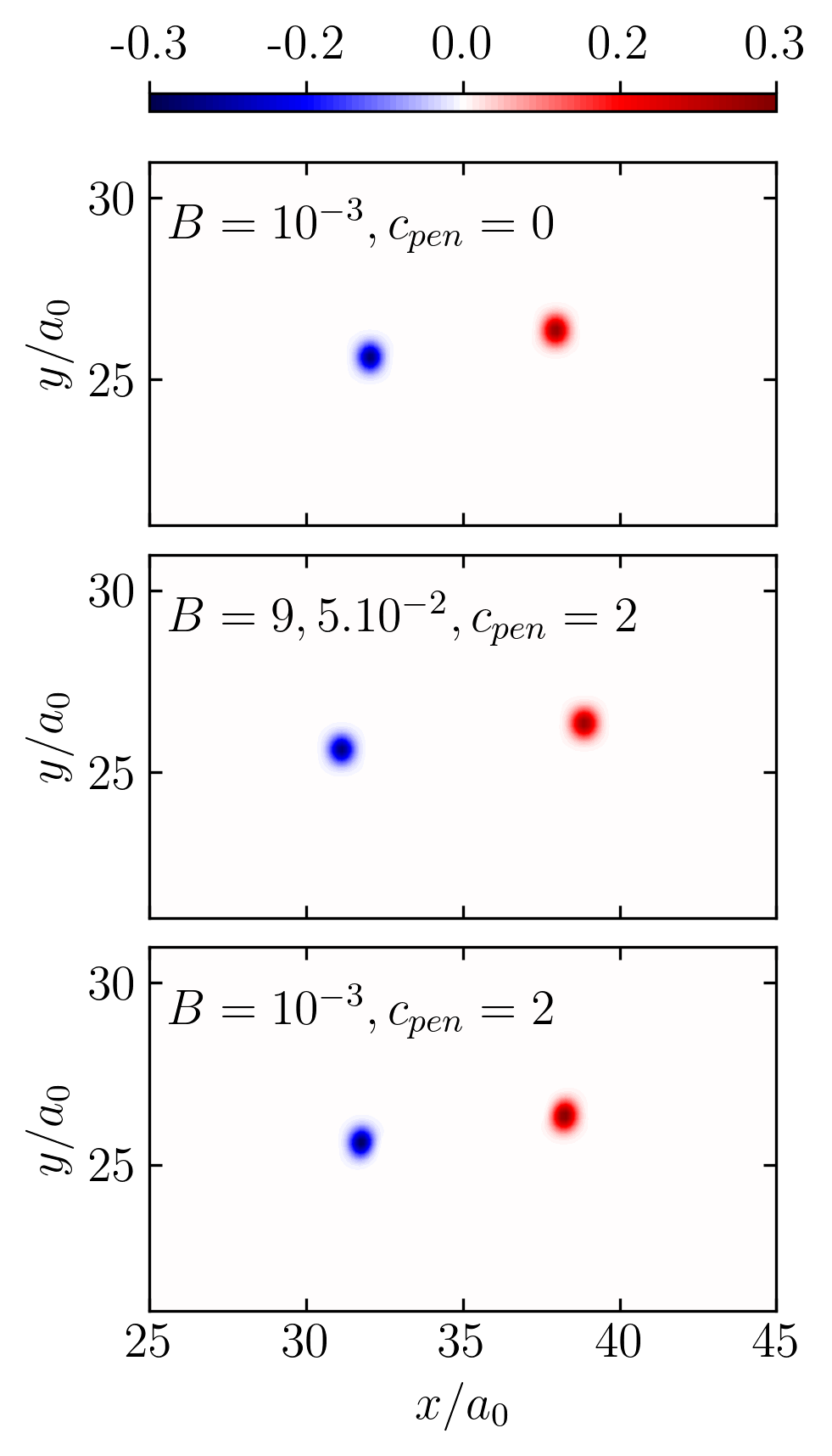}
    \end{subfigure}
    
  \end{minipage}
  \caption{In \subref{fig:ann_fdm_B095} and \subref{fig:ann_fdm_B001},Time evolution of the dislocation core positions for two values of $B$.Dashed lines track the PFC density $\widetilde{\Tens{\alpha}}$ and solid lines the FDM density $\Tens{\alpha}$. \subref{fig:ann_fdm_alpha_stacked} shows the snapshot of FDM cores for different values of $B$ and $c_{pen}$, \subref{fig:ann_fdm_alphaT_stacked} shows the PFC cores for the same parameters.}
  \label{fig:ann_fdm_B_comparison}
\end{figure}

Activating the coupling ($c_{{pen}}>0$) in the absence of external load strongly suppresses the dynamics. At $B = B_{{ref}}$, the cores advance by at most one atomic spacing before locking in place (\cref{fig:ann_fdm_B095}). Reducing the drag by a factor of 95 restores motion for small $c_{{pen}}$ (\cref{fig:ann_fdm_B001}), but the resulting glide is not faster than in the uncoupled PFC 
reference. 
Furthermore, $\Tens{\alpha}$ fails to remain compact during transport: as shown in \cref{fig:ann_fdm_alpha_stacked}, the sharp peak follows $\widetilde{\Tens{\alpha}}$, while $\Tens{\alpha}$ trails behind in a diffuse wake. 
The additional $c_{pen}(\Tens{U}^e - \Tens{Q})_{sym}$ in FDM therefore mitigate but do not eliminate the spreading of the FDM density.

Next, a target macroscopic stress $\overline{\Tens{\Sigma}}$ is imposed by controlling the macroscopic applied distortion. Starting from $\overline{\Tens{\Sigma}} = \overline{\mathbb{C}:\Tens{U}^e + c_{{pen}}(\Tens{U}^e - \Tens{Q})^{\mathrm{sym}}}$, the effective elastic law reads $\overline{\Tens{\Sigma}} = \mathbb{C}_{\mathrm{eff}} : \overline{\Tens{U}^e} - c_{{pen}}\,\overline{\Tens{Q}^{\mathrm{sym}}}$ with $\mathbb{C}_{\mathrm{eff}} = \mathbb{C} + c_{{pen}}\,\mathbb{I}^{\mathrm{sym}}$ and $\mathbb{S}_{\mathrm{eff}} = \mathbb{C}_{\mathrm{eff}}^{-1}$. At each time step, the total distortion is updated by the spatially uniform correction
\begin{equation}
    \Tens{U}_{\mathrm{corr}}(\mathbf{x},t) = \Tens{U}(\mathbf{x},t)
    + dt\left[
    \mathbb{S}_{\mathrm{eff}}:\overline{\Tens{\Sigma}}
    - \overline{\Tens{U}}(t)
    + \overline{\Tens{U}^{p}}(t)
    + c_{{pen}}\,\mathbb{S}_{\mathrm{eff}}:\overline{\Tens{Q}^{\mathrm{sym}}}(t)
    \right]
\end{equation}
which adjusts only the macroscopic component of $\Tens{U}$ and leaves the fluctuating field untouched. A pure shear stress $\overline{\Sigma}_{xy} = \overline{\Sigma}_{yx} = \tau$ with $\tau = p\mu$ and $\overline{\Sigma}_{xx} = \overline{\Sigma}_{yy} = 0$ is applied, where $p$ is a prescribed fraction of the shear modulus $\mu$.

\begin{figure}[h!]
  \begin{subfigure}[t]{0.3\textwidth}
    \centering
    \caption{$B=0.095$, $c_{{pen}}=0.1$}
    \includegraphics[width=0.8\linewidth]{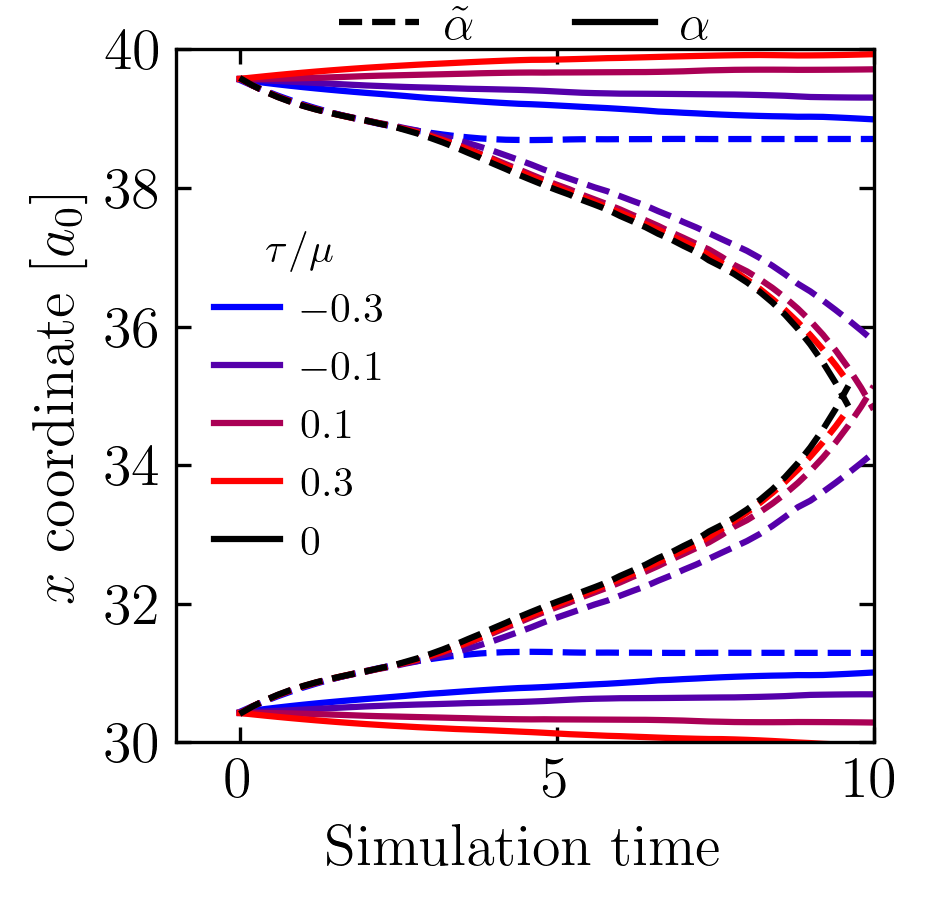}
    \label{fig:FDM_Stress_B0p095_cw0p1}
  \end{subfigure}
  \hspace{0.05cm}
  \begin{subfigure}[t]{0.3\textwidth}
    \centering
    \caption{$B=0.095$, $c_{{pen}}=0.4$}
    \includegraphics[width=0.8\linewidth]{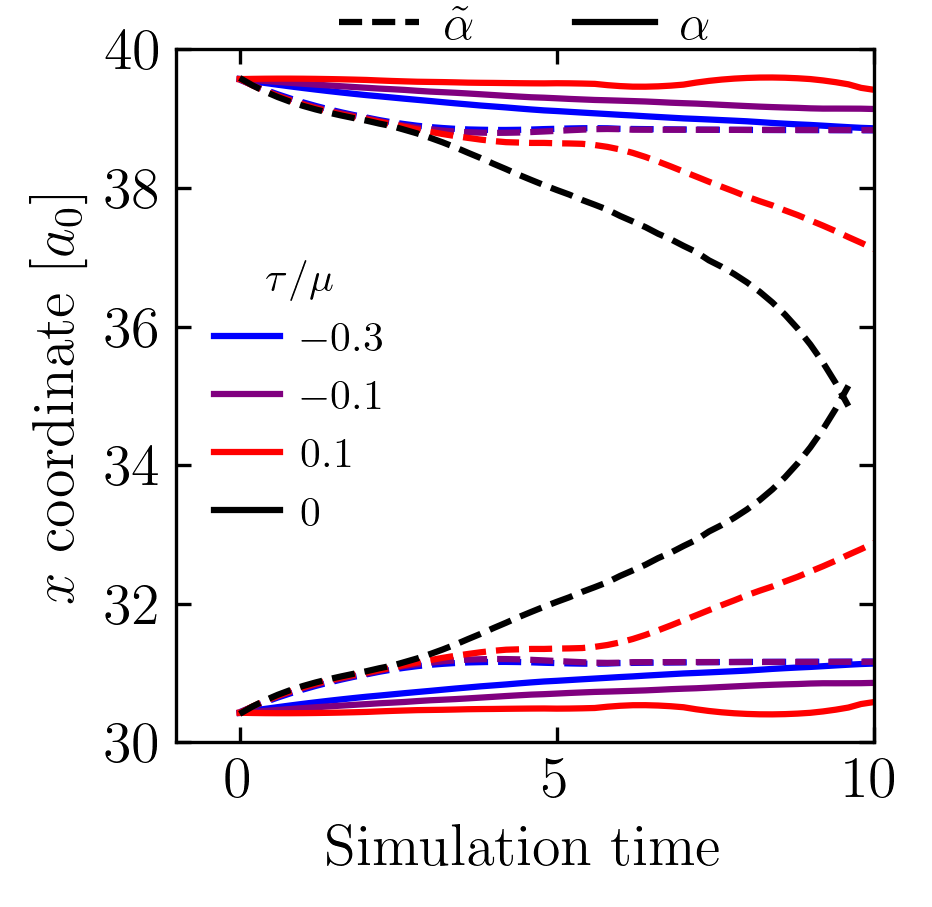}
    \label{fig:FDM_Stress_B0p095_cw0p4}
  \end{subfigure}
  \hspace{0.05cm}
  \begin{subfigure}[t]{0.3\textwidth}
    \centering
    \caption{$B=0.095$, $c_{{pen}}=2$}
    \includegraphics[width=0.8\linewidth]{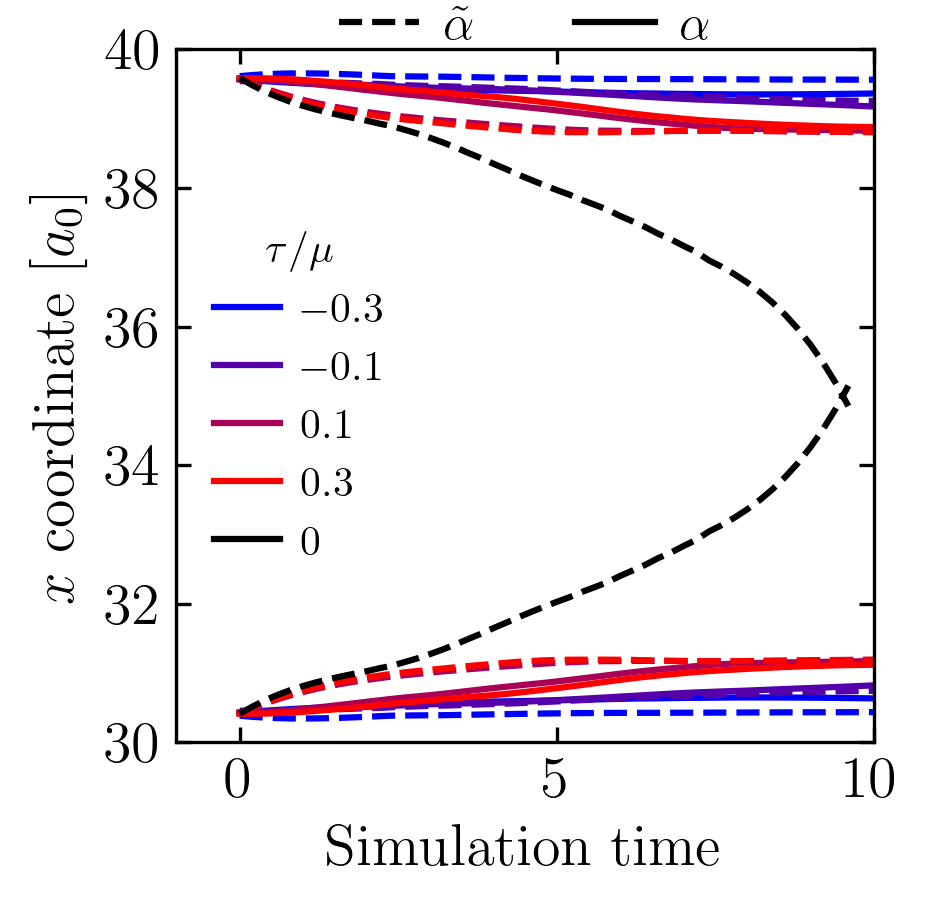}
    \label{fig:FDM_Stress_B0p095_cw2}
  \end{subfigure}

  \vspace{0.6em}

  \begin{subfigure}[t]{0.3\textwidth}
    \centering
    \caption{$B=0.001$, $c_{{pen}}=0.1$}
    \includegraphics[width=0.8\linewidth]{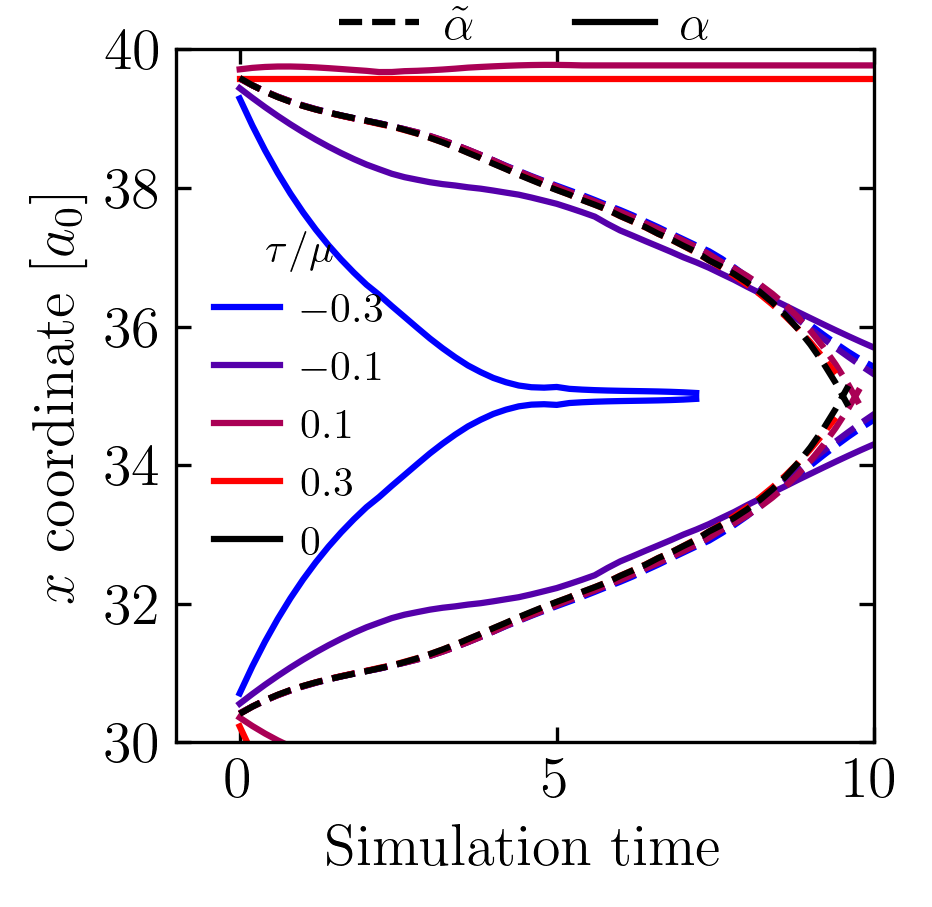}
    \label{fig:FDM_Stress_B0p001_cw0p1}
  \end{subfigure}
  \hspace{0.05cm}
  \begin{subfigure}[t]{0.3\textwidth}
    \centering
    \caption{$B=0.001$, $c_{{pen}}=0.4$}
    \includegraphics[width=0.8\linewidth]{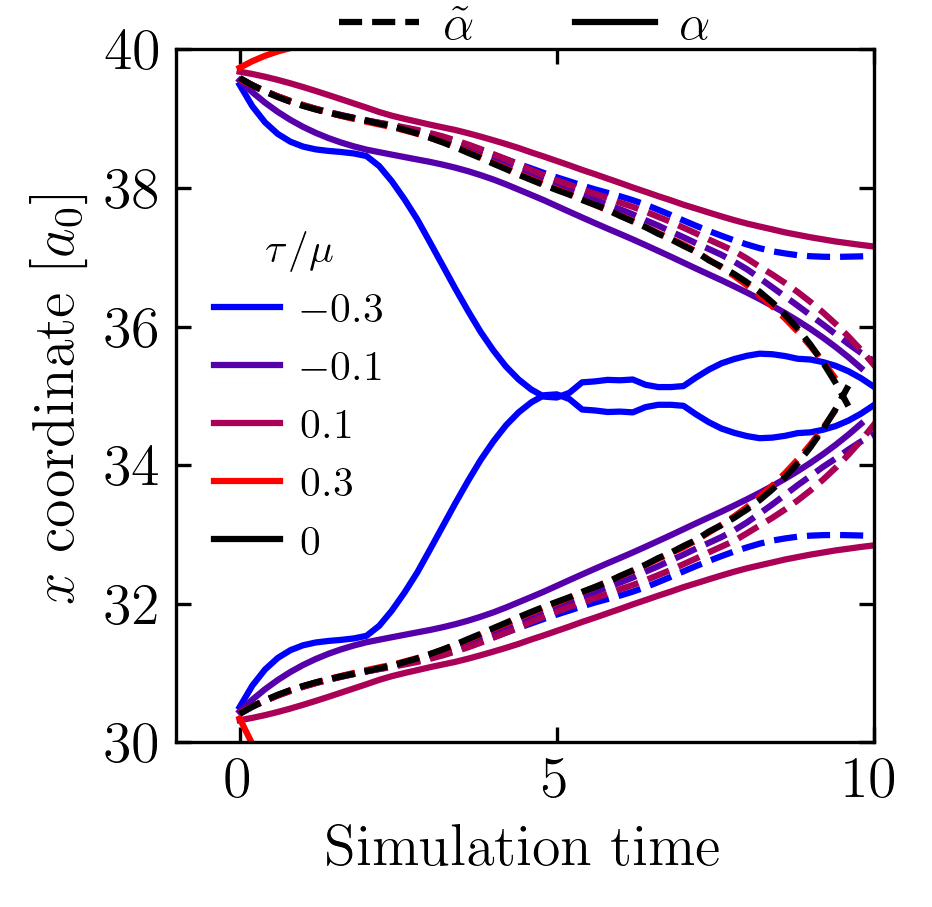}
    \label{fig:FDM_Stress_B0p001_cw0p4}
  \end{subfigure}
  \hspace{0.05cm}
  \begin{subfigure}[t]{0.3\textwidth}
    \centering
    \caption{$B=0.001$, $c_{{pen}}=2$}
    \includegraphics[width=0.8\linewidth]{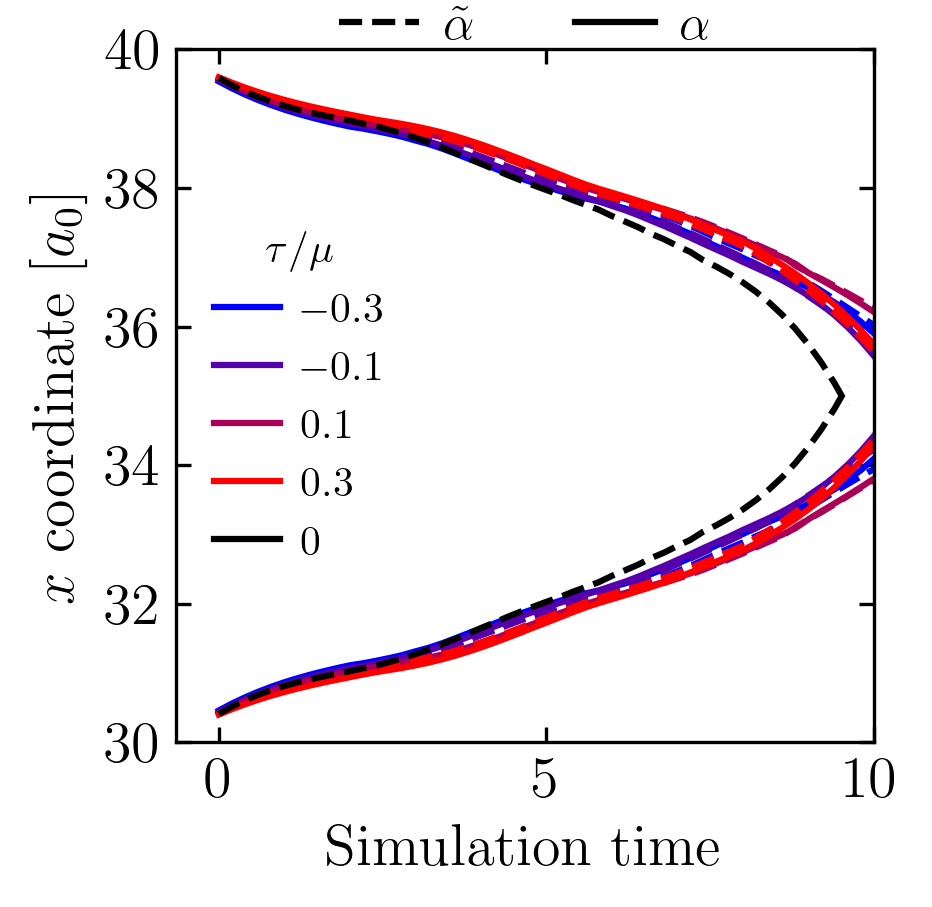}
    \label{fig:FDM_Stress_B0p001_cw2}
  \end{subfigure}
\caption{Dislocation core trajectories, represented by $\Tens{\alpha}$ and $\widetilde{\Tens{\alpha}}$, in the FDM-driven approach under macroscopic shear stress $\tau/\mu$: \subref{fig:FDM_Stress_B0p095_cw0p1} to \subref{fig:FDM_Stress_B0p095_cw2} correspond to $B=0.095$, and \subref{fig:FDM_Stress_B0p001_cw0p1} to \subref{fig:FDM_Stress_B0p001_cw2} to $B=0.001$. From left to right, $c_{pen}=0.1$, $0.4$, and $2$.}\label{fig:FDM_annihilation_Stress}
\end{figure}

The behavior summarized in \cref{fig:FDM_annihilation_Stress} mirrors the unloaded case. 
At $B = B_{{ref}}$, the FDM cores neither attract nor annihilate but spread in place, and $\Tens{\alpha} \neq \widetilde{\Tens{\alpha}}$ in every simulation. 
Small $c_{{pen}}$ allows the $\widetilde{\Tens{\alpha}}$ alone to annihilate (\cref{fig:FDM_Stress_B0p095_cw0p1}), while larger values pull $\widetilde{\Tens{\alpha}}$ towards $\Tens{\alpha}$ and lock the configuration (\cref{fig:FDM_Stress_B0p095_cw0p4,fig:FDM_Stress_B0p095_cw2}). 
Reducing the drag to $B_{{ref}}/95$ 
(\cref{fig:FDM_Stress_B0p001_cw0p1}--\cref{fig:FDM_Stress_B0p001_cw2}) increases the range of $c_{pen}$ for which annihilation of $\boldsymbol{\alpha}$ can occur, yet the spreading observed in \cref{fig:ann_fdm_alpha_stacked} persists.
However, the dislocation velocity during annihilation is slower than the diffusive velocity at which annihilation occurs only with PFC.

Across the scanned range of $B$ and $c_{{pen}}$, dislocation motion is not always triggered; when it occurs, the two density fields evolve inconsistently, as revealed by the gap between the solid and dashed lines. 

\subsubsection{PFC-driven motion}

\begin{figure}[h!]
  \centering

  \begin{subfigure}[t]{0.32\textwidth}
    \centering
    \caption{Dislocation density $\Tens{\alpha}_{xz}$, $t=0$}\label{fig:Glidealpha_0}
    \includegraphics[width=\linewidth]{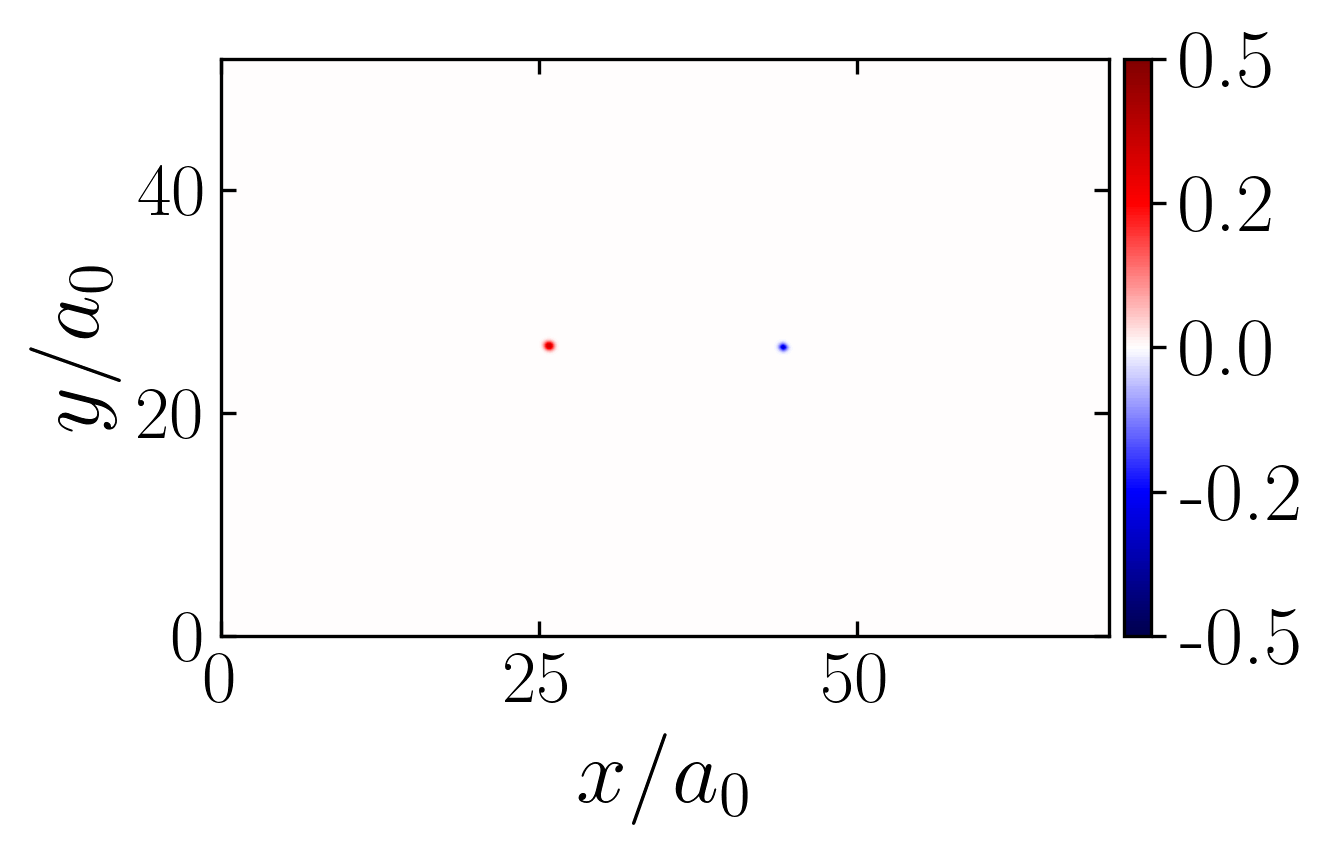}
  \end{subfigure}
  \hspace{0.05cm}
  \begin{subfigure}[t]{0.32\textwidth}
    \centering
    \caption{Dislocation density $\Tens{\alpha}_{xz}$, $t=250$}\label{fig:Glidealpha_250}
    \includegraphics[width=\linewidth]{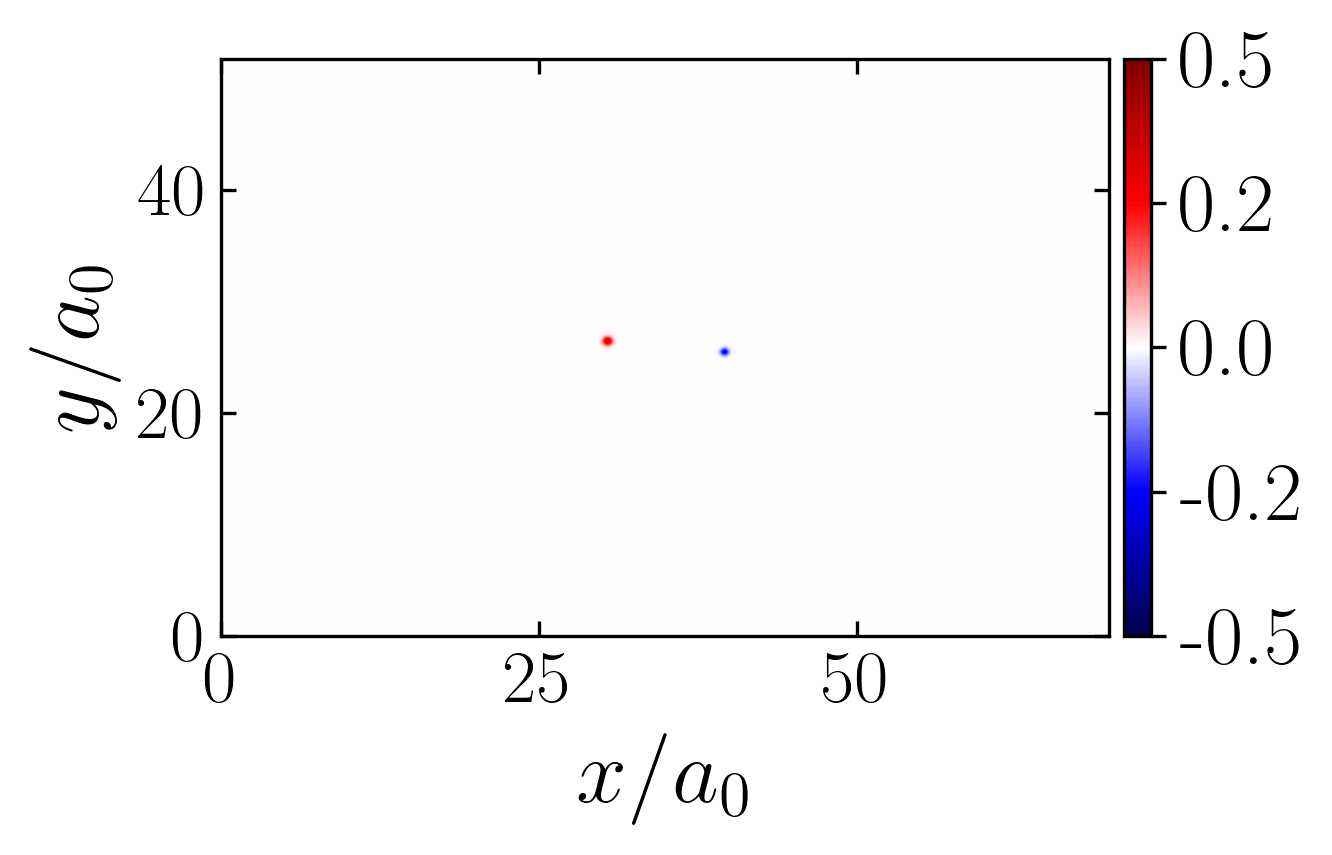}
\end{subfigure}
\hspace{0.05cm}
  \begin{subfigure}[t]{0.32\textwidth}
    \centering
    \caption{Dislocation density $\Tens{\alpha}_{xz}$, $t=t_f$}\label{fig:Glidealpha_-1}
    \includegraphics[width=\linewidth]{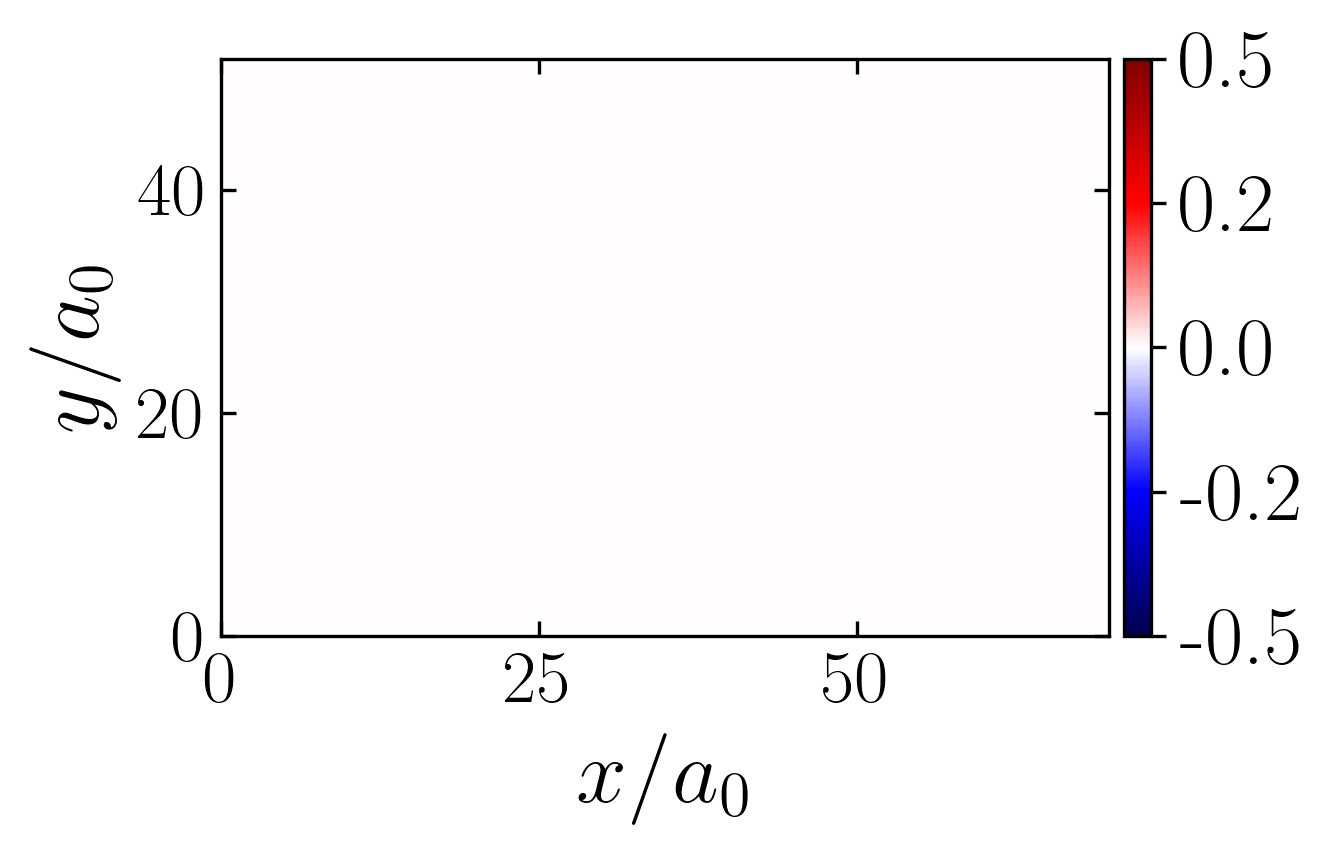}
  \end{subfigure}

  \vspace{0.6em}

  \begin{subfigure}[t]{0.32\textwidth}
    \centering
    \caption{Elastic distortion $\Tens{U}^e_{xx}$, $t=0$}\label{fig:GlideUe_0}
    \includegraphics[width=\linewidth]{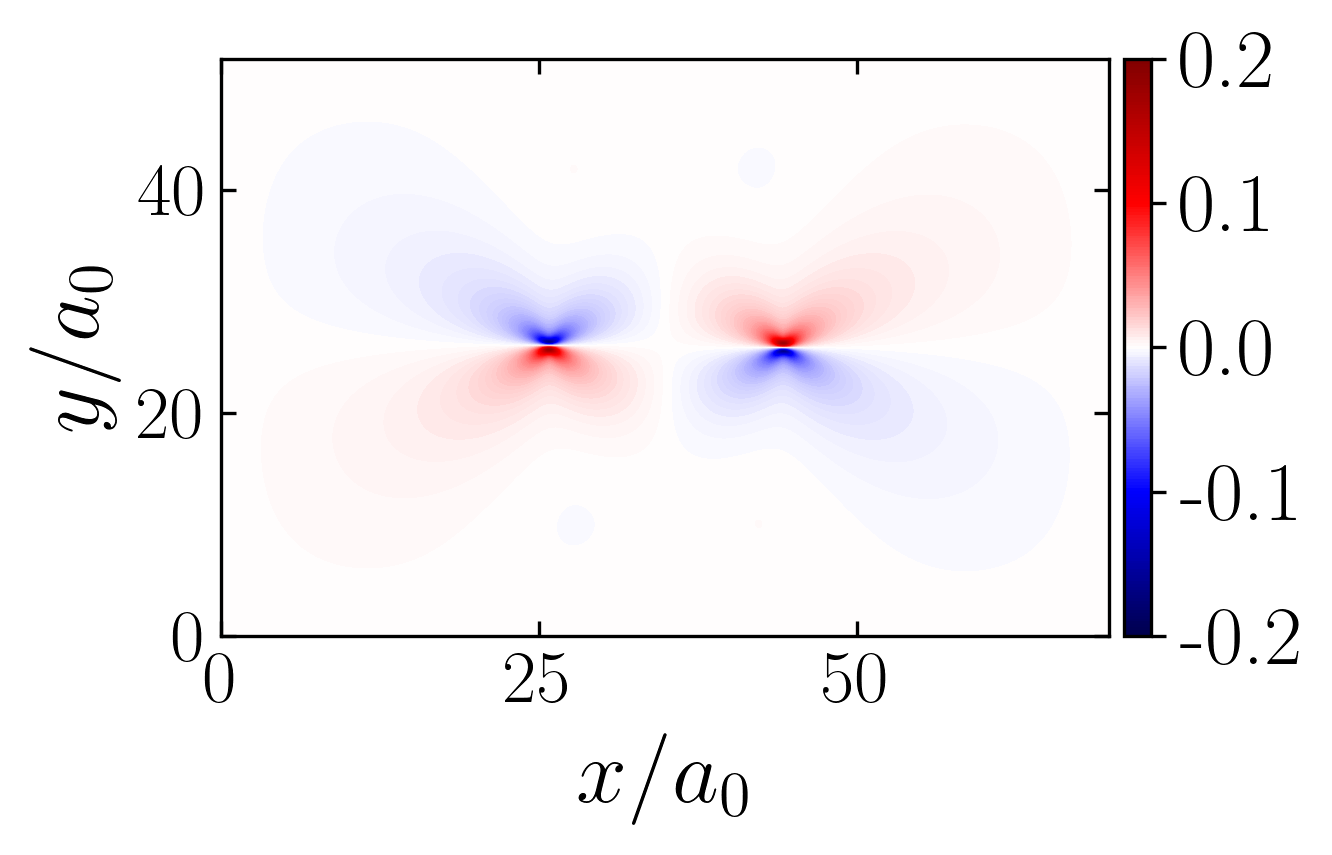}
  \end{subfigure}
  \hspace{0.05cm}
  \begin{subfigure}[t]{0.32\textwidth}
    \centering
    \caption{Elastic distortion $\Tens{U}^e_{xx}$, $t=250$}\label{fig:GlideUe_250}
    \includegraphics[width=\linewidth]{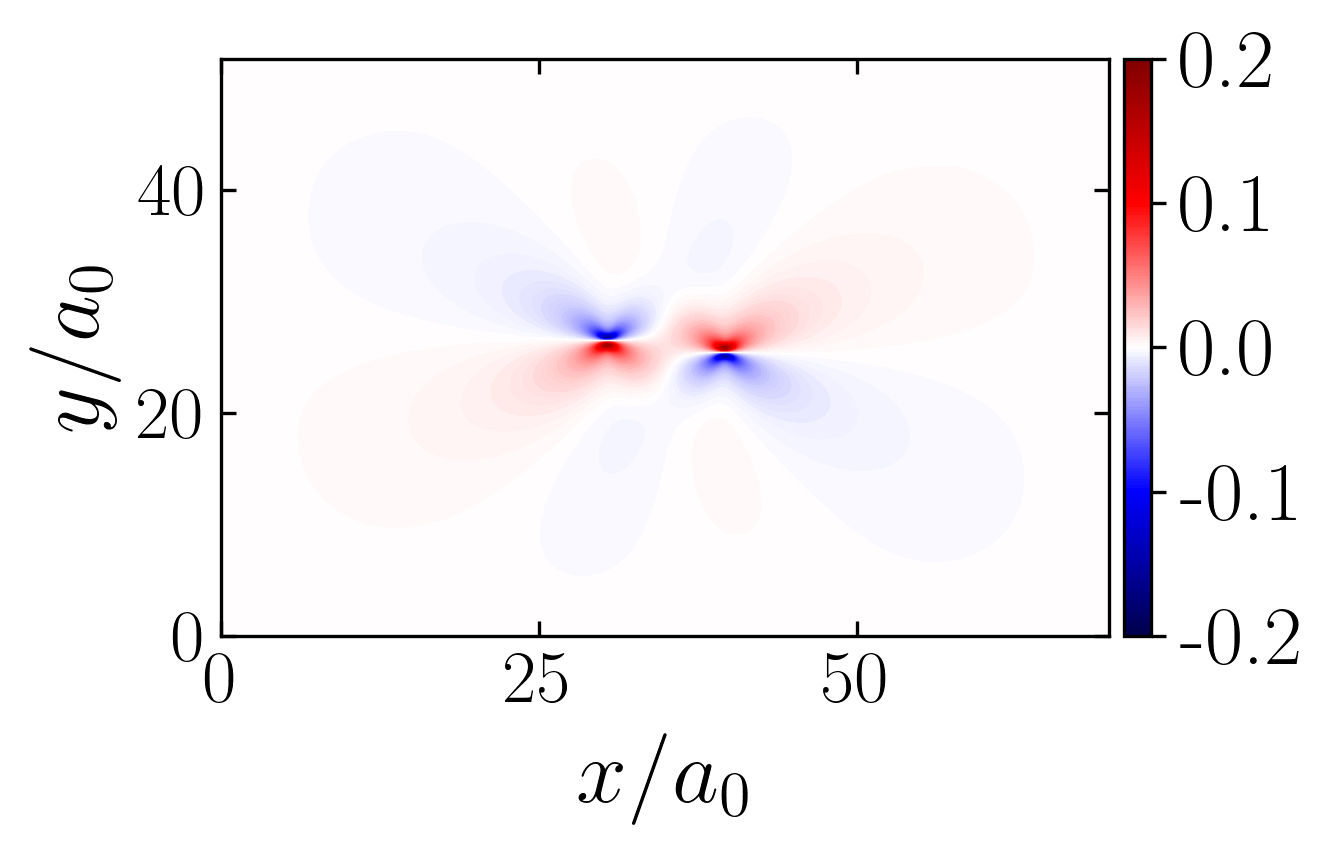}
\end{subfigure}
\hspace{0.05cm}
  \begin{subfigure}[t]{0.32\textwidth}
    \centering
    \caption{Elastic distortion $\Tens{U}^e_{xx}$, $t=t_f$}\label{fig:GlideUe_-1}
    \includegraphics[width=\linewidth]{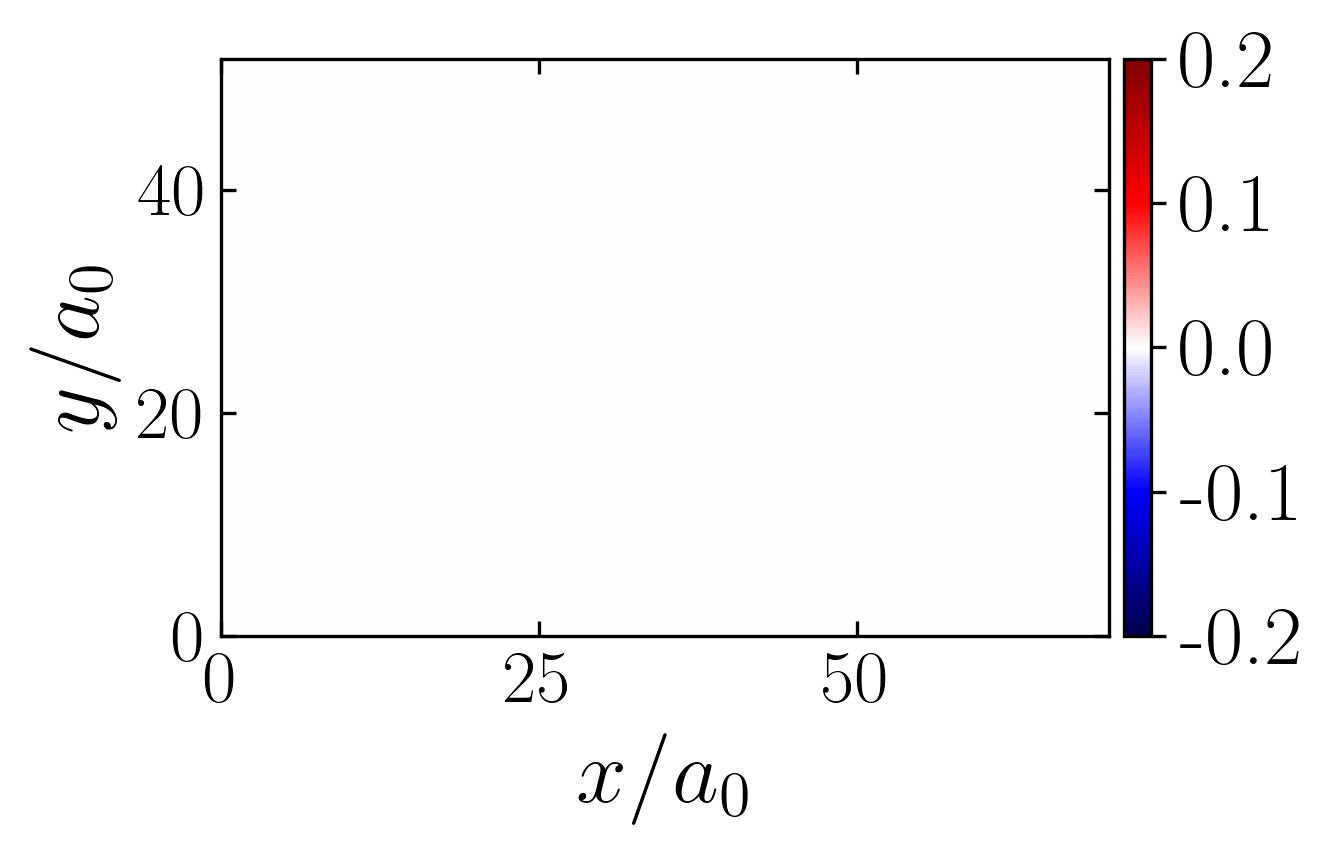}
  \end{subfigure}

    \vspace{0.6em}

    \begin{subfigure}[t]{0.32\textwidth}
    \centering
    \caption{Plastic distortion $\Tens{U^p}_{xy}$, $t=0$}\label{fig:GlideUp_0}
    \includegraphics[width=\linewidth]{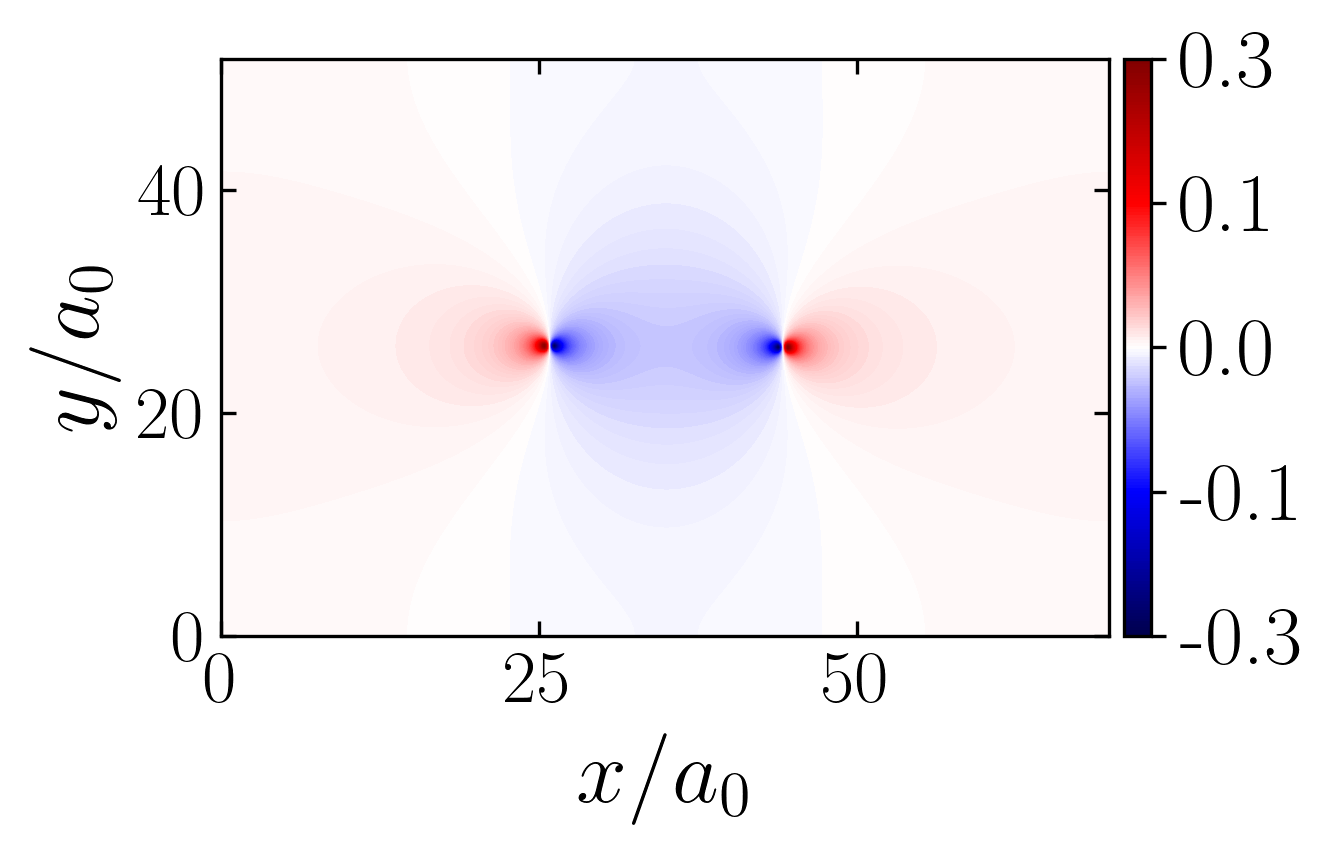}
  \end{subfigure}
  \hspace{0.05cm}
  \begin{subfigure}[t]{0.32\textwidth}
    \centering
    \caption{Plastic distortion $\Tens{U^p}_{xy}$, $t=250$}\label{fig:GlideUp_250}
    \includegraphics[width=\linewidth]{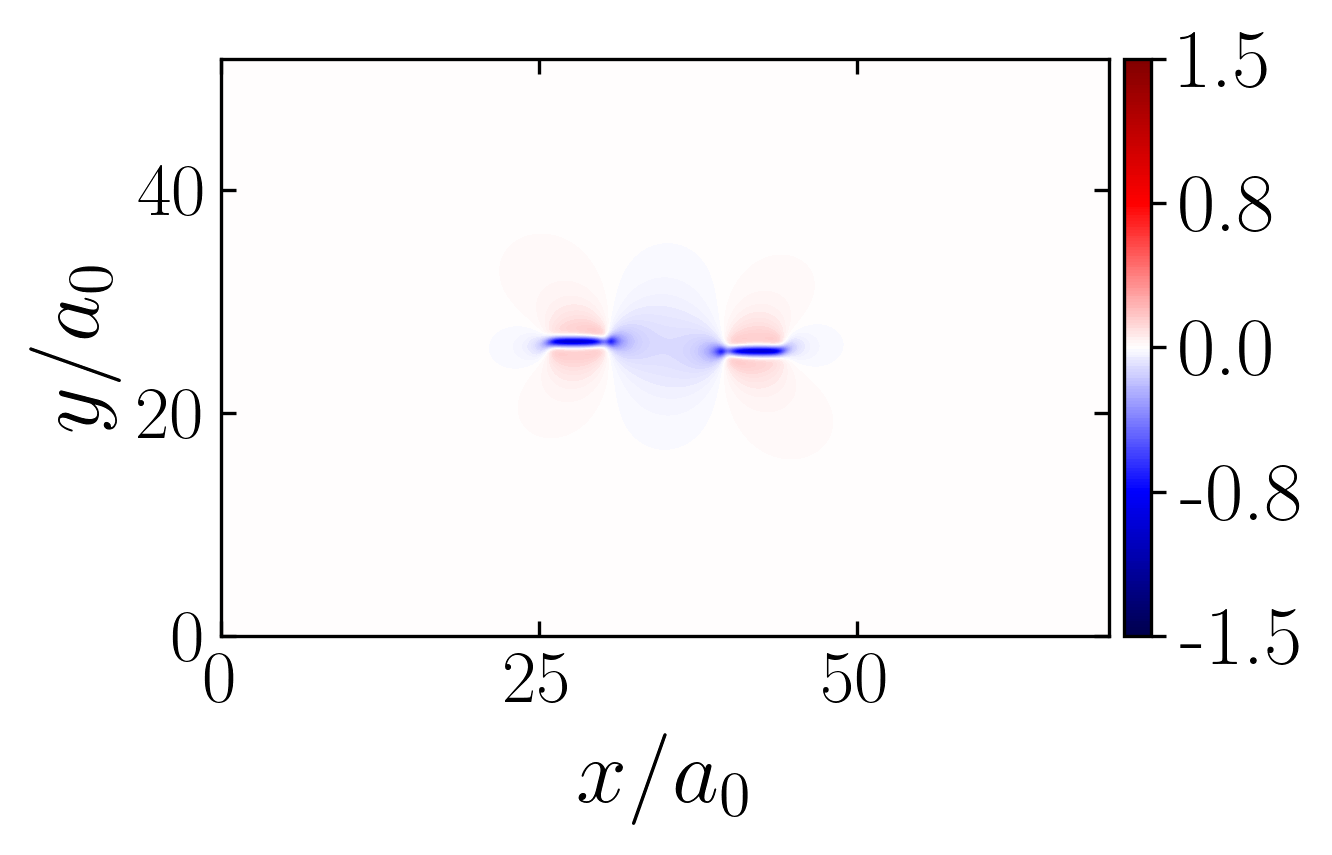}
\end{subfigure}
\hspace{0.05cm}
  \begin{subfigure}[t]{0.32\textwidth}
    \centering
    \caption{Plastic distortion $\Tens{U^p}_{xy}$, $t=t_f$}\label{fig:GlideUp_-1}
    \includegraphics[width=\linewidth]{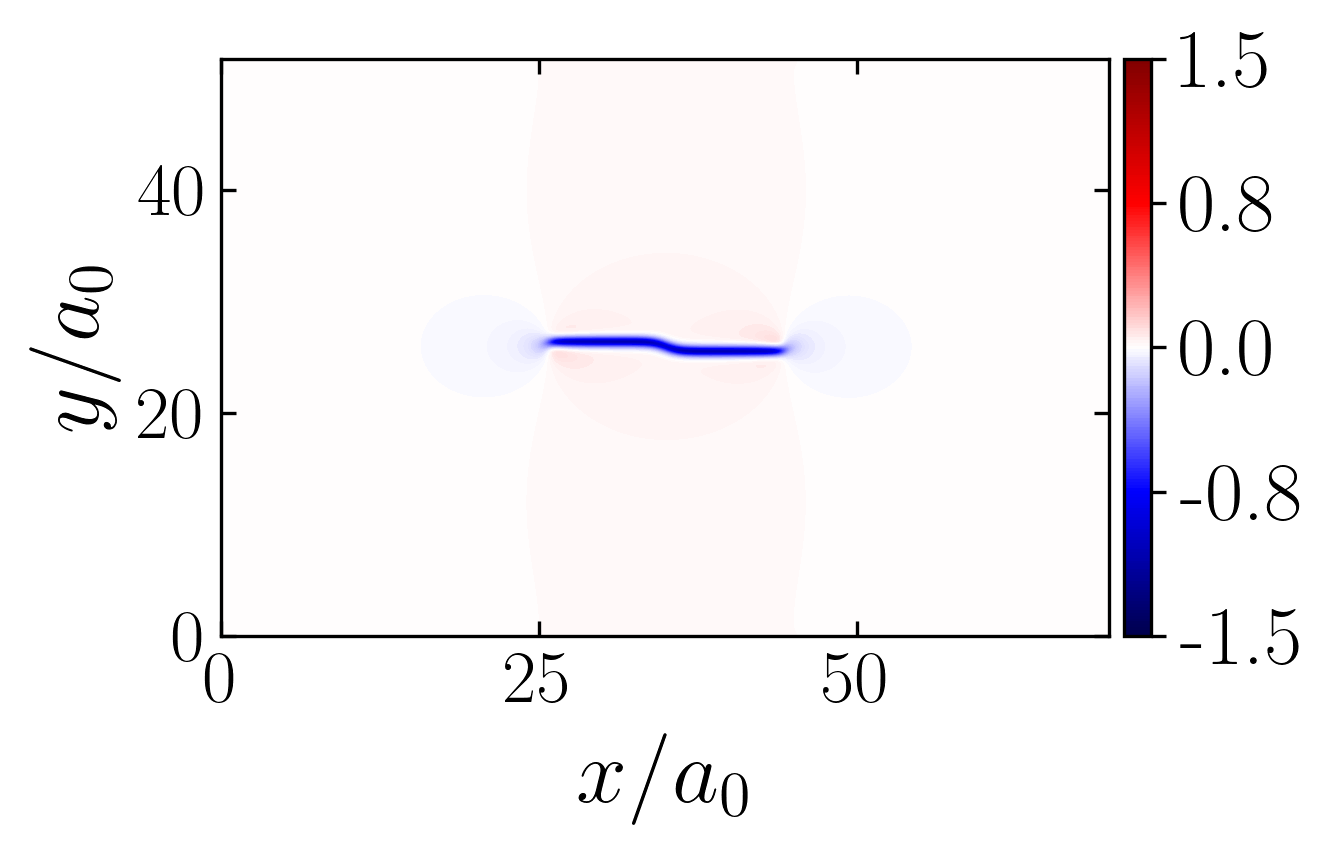}
  \end{subfigure}
\caption{Evolution of a dislocation dipole during PFC-driven glide and annihilation: \subref{fig:Glidealpha_0} to \subref{fig:Glidealpha_-1} dislocation density $\Tens{\alpha}_{xz}$; \subref{fig:GlideUe_0} to \subref{fig:GlideUe_-1} elastic distortion $\Tens{U}^e_{xx}$; and \subref{fig:GlideUp_0} to \subref{fig:GlideUp_-1} plastic distortion $\Tens{U}^p_{xy}$. From left to right, the fields are shown at $t=0$, $t=250$, and $t=t_f$. At $t=t_f$, the dislocations have annihilated and the elastic distortion has relaxed, while $\Tens{U}^p_{xy}$ preserves a permanent slip trace along the glide plane.} \label{fig:PFC_pure_annihilationPFC}

\end{figure}

Since the previously investigated FDM-driven dynamics did not result in observable motion, we shift our focus to PFC-driven evolution. 
Retaining the same glide configuration, we investigate the role of the coupling term by explicitly prescribing $\nabla \times \Tens{U^p} = \Tens{\widetilde{\alpha}}$ and $\dot{\Tens{U^p}} = \Tens{\mathcal{J}^\psi}$. 
This choice is motivated not only by the discrepancy $\Tens{\alpha} \neq \widetilde{\Tens{\alpha}}$, but also by the physical origin of $\Tens{\mathcal{J}^\psi}$. 
For this simulation, the system is setup similar to the one described at the start of section \ref{sec:anni}, but with a separation of $d \approx 20 a_0$.
\cref{fig:Glidealpha_0,fig:GlideUe_0,fig:GlideUp_0} show the initial state of this dislocation configuration.
During evolution, the two dislocations glide towards each other (\cref{fig:Glidealpha_250,fig:GlideUe_250,fig:GlideUp_250}), leaving behind a plastic shear trace as shown in \cref{fig:GlideUp_-1}. 
Following annihilation, the system reaches a dislocation-free (\cref{fig:Glidealpha_-1}) and stress-free state (\cref{fig:GlideUe_-1}). 
The plastic distortion retains a slip trace in the $xy$-component along the trajectories of the cores (\cref{fig:GlideUp_-1}).

This PFC-driven formulation forces the FDM distortions to follow the plastic motion governed by the PFC dynamics. In this regime, no mobility coefficient is required; we fix $c_{sh}=100$ and vary the coupling strength $c_{pen}$, while all other simulation parameters remain consistent with the previous section. For each value of $c_{pen}$, we apply a macroscopic shear stress $\tau$ in the $xy$-direction similarly to the previous section. 

\begin{figure}[h!]
  \centering

  \begin{subfigure}[t]{0.24\textwidth}
    \centering
    \caption{No stress applied}\label{fig:ann_pfc_noStress}
    \includegraphics[width=\linewidth]{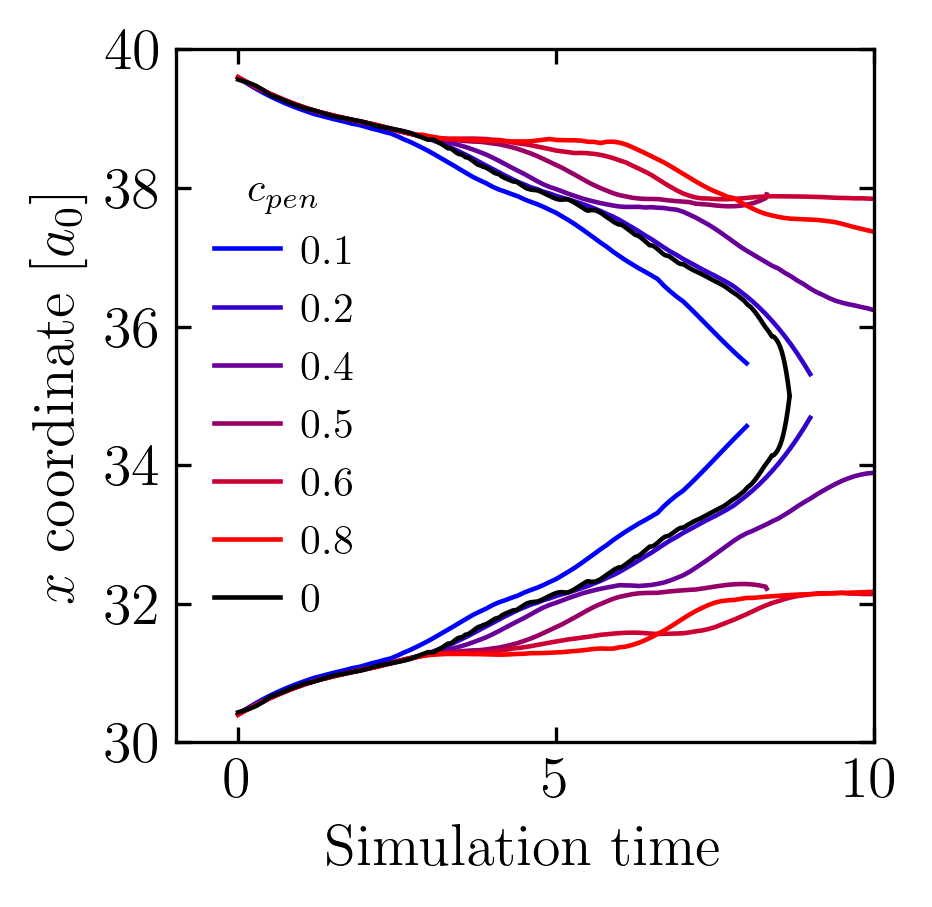}
  \end{subfigure}
    \begin{subfigure}[t]{0.24\textwidth}
    \centering
    \caption{$c_{pen}=0.1$}\label{fig:ann_pFC_01}
    \includegraphics[width=\linewidth]{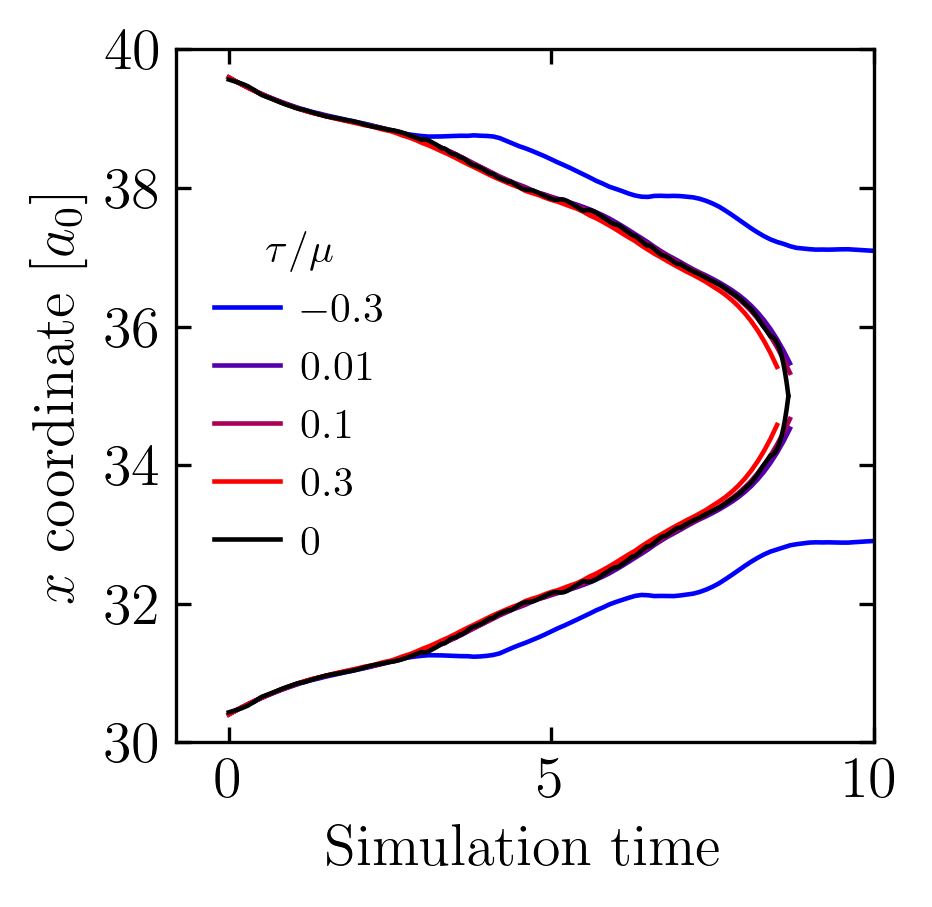}
  \end{subfigure}
      \begin{subfigure}[t]{0.24\textwidth}
    \centering
    \caption{$c_{pen}=0.4$}\label{fig:ann_pFC_04}
    \includegraphics[width=\linewidth]{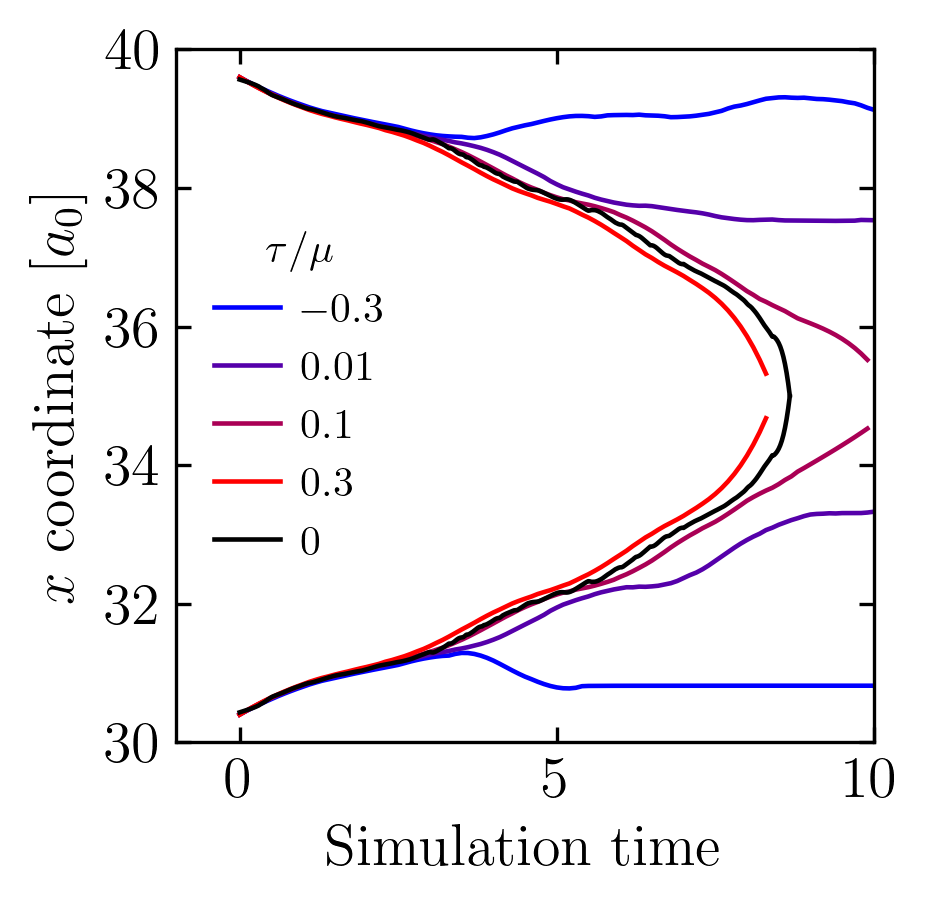}
  \end{subfigure}
      \begin{subfigure}[t]{0.24\textwidth}
    \centering
    \caption{$c_{pen}=0.8$}\label{fig:ann_pFC_08}
    \includegraphics[width=\linewidth]{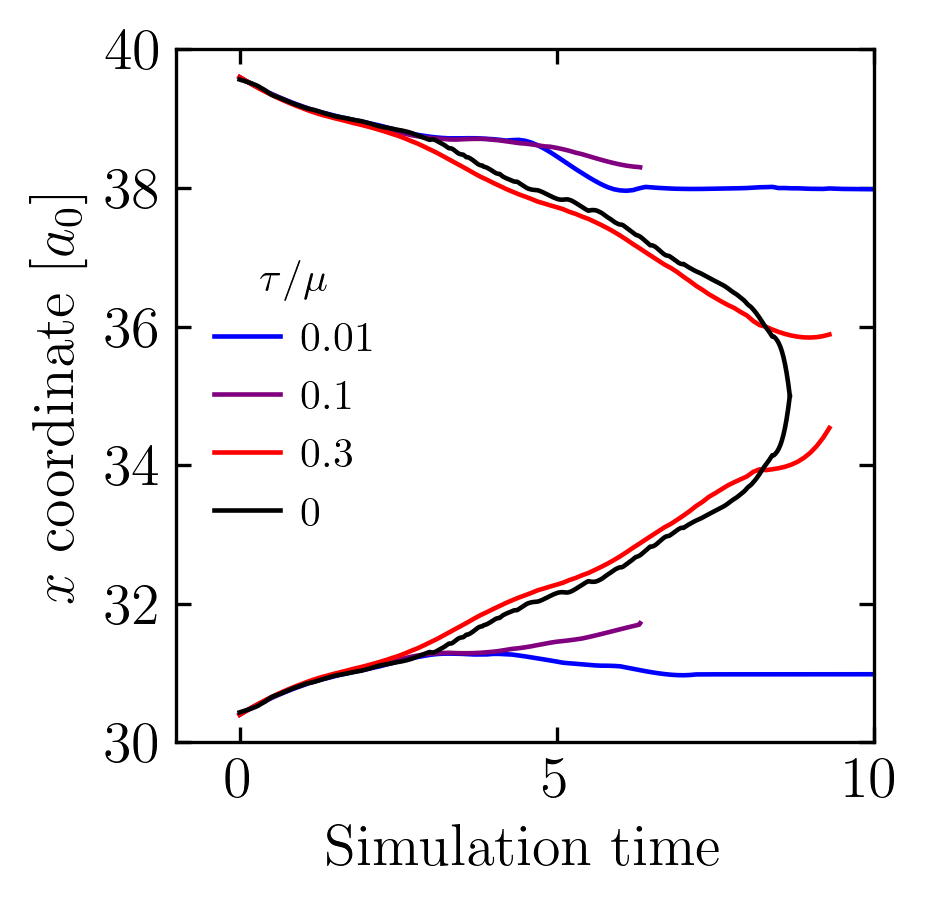}
  \end{subfigure}
  \caption{\subref{fig:ann_pfc_noStress}: Evolution of the coordinate of the cores for different values of $c_{pen}$. \subref{fig:ann_pFC_01} to \subref{fig:ann_pFC_08} same evolution for fixed values of $c_{pen}$ and different applied macroscopic stresses $\tau$. Reference PFC uncoupled motion shown in solid line.}
  \label{fig:annihilationPFC}
\end{figure}

The results are presented in \cref{fig:annihilationPFC}. In \cref{fig:ann_pfc_noStress}, in the absence of any applied stres, increasing $c_{pen}$ causes slower annihilation time and a staggered motion (stick and slip) which corresponds to the relaxation of $\mathcal{F}_{pen}$ that hinders the main carrier $\mathcal{F}_{sh}$. 
Now, applying a macroscopic stress, a very low coupling factor ($c_{pen}=0.1$) results in no significant change, as seen in \cref{fig:ann_pFC_01}, except for a stress field that opposes the motion. 
Larger values of $c_{pen}$ and low applied shears induces dislocation locking after few atomic steps, and annihilation only occurs for large values of $\tau \approx 0.3 \mu$ (\cref{fig:ann_pFC_04,fig:ann_pFC_08}). 
In the cases where motion does occur, dislocation velocities are slower than those observed during purely diffusive PFC evolution, which means that elastic relaxation is never achieved.

\section{Discussion} \label{sec:discussion}
\subsection{Behavior under the coupled framework}
The numerical results of the previous section reveal that increasing the penalty parameter $c_{{pen}}$ produces no appreciable speed up of dislocation dynamics. This behavior is due to how mechanical information propagates through the coupled system, and it can be understood from two complementary perspectives.

First, the coupling does not transmit boundary loading by elastic relaxation but by phase diffusion. The penalty term induces an phase diffusion coefficient $D^{{pen}} = c_{{pen}}/(3A_0^2)$ (\cref{eq:Diffusion_Coeff_PEN}), so that the associated bulk timescale is $\tau^{{pen}} = 3|A_n|^2 L^2 / c_{{pen}}$. Comparing this with the intrinsic Swift-Hohenberg phase diffusion coefficient 
$D_{\mathrm{pfc}}^{\mathrm{phase}} = 4 c_{\mathrm{sh}}$ yields
\begin{equation} \label{eq:TimeComp}
    \frac{\tau^{{pen}}}{\tau^{\mathrm{pfc}}}
    = \frac{12\,|A_n|^2\,c_{\mathrm{sh}}}{c_{{pen}}}
\end{equation}
which evaluates to $4.32\,c_{\mathrm{sh}}/c_{{pen}}$ in the bulk for the simulation parameters used here. Fast relaxation of the penalty forces therefore demands $c_{{pen}} \gg 4.32\,c_{\mathrm{sh}}$; otherwise the Swift-Hohenberg forcing pulls $\psi$ back towards its minimum before the penalty can act meaningfully. The picture is reversed near a core, where $|A_n|\to 0$ and $D^{{pen}}\to\infty$ as $r\to 0$: the core acts as a sink, and the correction reaching its vicinity is rate-limited by the bulk diffusion the forcing must traverse. The resulting timescale far exceeds the annihilation time $t\approx 8$ reported in the previous section, so that the core region is  decoupled from boundary loading on any physically relevant time scale.

A natural choice would be to make $c_{{pen}}$ very large, but this strategy is precluded by a second mechanism: The penalty also acts as a restoring spring on the core itself. Consider a state where the cores in $\Tens{\alpha}$ and $\widetilde{\Tens{\alpha}}$ are located at the same position $\vec{x}_0$, and that the fields are aligned $\Tens{Q}(\vec{x}_0)=\Tens{U^e}(\vec{x}_0)$. If the FDM core advances by an elementary $\delta x$ while the PFC remains frozen, then in this new configuration and in first order: $
  \Tens{U}^e(\bm{x}_0 + \delta x\,\vec{x}) - \Tens{Q}(\bm{x}_0) \approx  \,\delta x\,\partial_x\Tens{U}^e(\bm{x}_0)$, where the 0-th order vanishes due to alignment at $\vec{x}_0$. The penalty energy of this state reads
\begin{equation*}
    \mathcal{F}_{{pen}}(\delta x)
    \approx \frac{c_{{pen}}}{2}\,\delta x^{2}
    \int \|\partial_x \Tens{U}^{e}(\vec{r})\|_{F}^{2}\,\mathrm{d}^{2}r
    \equiv \frac{1}{2}\,K_{\mathrm{eff}}\,\delta x^{2}
\end{equation*}
which is the energy of a harmonic spring with effective stiffness $K_{\mathrm{eff}} = c_{{pen}} 
\int \|\partial_x \Tens{U}^{e}\|_{F}^{2}\,\mathrm{d}^{2}r$. The associated restoring force is
$F_\mathrm{spring}(\delta x) = -\partial_{\delta x}\mathcal{F}_\mathrm{pen}
= -K_\mathrm{eff}\,\delta x$ always directed back toward the aligned state for any $\delta x > 0$. 
The spring stiffness can be evaluated evaluated the analytical displacement field of an edge dislocation with a core cutoff $r_c$:
\begin{equation}\label{eq:Keff}
    K_{\mathrm{eff}}
    = \frac{c_{{pen}}\,b^{2}\left(14-24\nu+16\nu^{2}\right)}
           {32\pi r_c^{2}(1-\nu)^{2}}
\end{equation}
Several properties of this expression are worth noting. First, $K_\mathrm{eff}$ grows linearly with $c_\mathrm{pen}$ and quadratically with $r_c^{-1}$, so that a more localized core produces a stiffer spring. Second, the polynomial $14 - 24\nu + 16\nu^2$ is positive over the physical range of $\nu$. The spring is always restoring, regardless of the elastic constants.

Dislocation glide is possible only when the Peach–Koehler driving force overcomes the spring restoring force. For an edge dislocation dipole at separation $d_0$:
\begin{equation*}
    f_{\mathrm{PK}}
    = \frac{\mu b^{2}}{2\pi(1-\nu)\,d_0}
\end{equation*}
For the dislocation to move one atomic spacing $a_0$, we must have $f_{\mathrm{PK}} >K_{\mathrm{eff}}\,a_0$. Pinning occurs when these two are exactly equal: 
\begin{equation*}
    \frac{c_{{pen}}^{\mathrm{max}}\,b^{2}\,(14-24\nu+16\nu^{2})a_0}
         {32\pi r_c^{2}(1-\nu)^{2}}
    = \frac{\mu b^{2}}{2\pi(1-\nu)\,d_0}
    \;\Longrightarrow\;
    c_{{pen}}^{\mathrm{max}}
    = \frac{16\,\mu(1-\nu)\,r_c^{2}}{(14-24\nu+16\nu^{2})\,d_0\,a_0}
\end{equation*}
which, for the parameters used here, gives an upper bound of $c_{{pen}}^{\mathrm{max}} \approx 0.64$, and the lower bound estimated from \cref{eq:TimeComp} to be $ c_\mathrm{pen}^\mathrm{min} \approx 432$.

Taken together, these two estimates define an empty operating window for the herein studied case: $  c_\mathrm{pen}^\mathrm{min}
  \;\lesssim\; c_\mathrm{pen} \;\lesssim\;
  c_\mathrm{pen}^\mathrm{max},$. More generally, below $c_{{pen}}^{\mathrm{max}}$, the penalty diffuses too slowly to compete with the Swift-Hohenberg relaxation; above it, the spring stiffness locks the PFC dislocation in place, so that motion can only be recovered by lowering the drag coefficient, at the cost of accelerated core spreading. In this regime the FDM core may advance, but it does so without dragging the PFC dislocation along. The PFC-driven approach circumvents      this entirely by focusing the description on a single dislocation density ($\Tens{\alpha}=\widetilde{\Tens{\alpha}}$), it captures both the local core evolution and the macroscopic response to external loading within a unified transport equation. The plastic distortion rate $\Tens{\mathcal{J}}^{\psi}$ proves richer than the classical form $\Tens{\alpha}\times\vec{v}^{d}$, which is to external-field (macroscopic) responses but cannot resolve core relaxation. However, the PFC-driven approach cannot guarantee the positivity of the plastic dissipation or the total dissipation in \cref{eq:PFCDissipation} as shown in the following section.
\subsection{Dissipation comparison}
The dissipation computed from \cref{eq:PFCDissipation} contains two contributions: a plastic contribution, $\int_\Omega \Tens{\sigma}: \dot{\Tens{U}^p} dV$, and a PFC contribution $\dot{\psi}^2$. 
The latter is positive by construction, whereas the former is only guaranteed in an FDM-driven approach but not in the PFC-driven approach, where $\dot{\boldsymbol{U}}^p = \mathcal{J}$ is assumed. 
This assumption does not imply any functional dependence of $\mathcal{J}$ on $\boldsymbol{\sigma}$, $\int_\Omega \boldsymbol{\sigma}: \mathcal{J} dV \ge 0$ is not guaranteed. 
In this section, the dislocation annihilation problem from section \ref{sec:anni} is revisited to study the plastic dissipation in the FDM-driven and PFC-driven cases.

\begin{figure}[h!]
  \centering
      \begin{subfigure}[t]{0.3\textwidth}
    \centering
    \caption{$c_{pen}=0$}\label{fig:Dissip_0}
    \includegraphics[width=\linewidth]{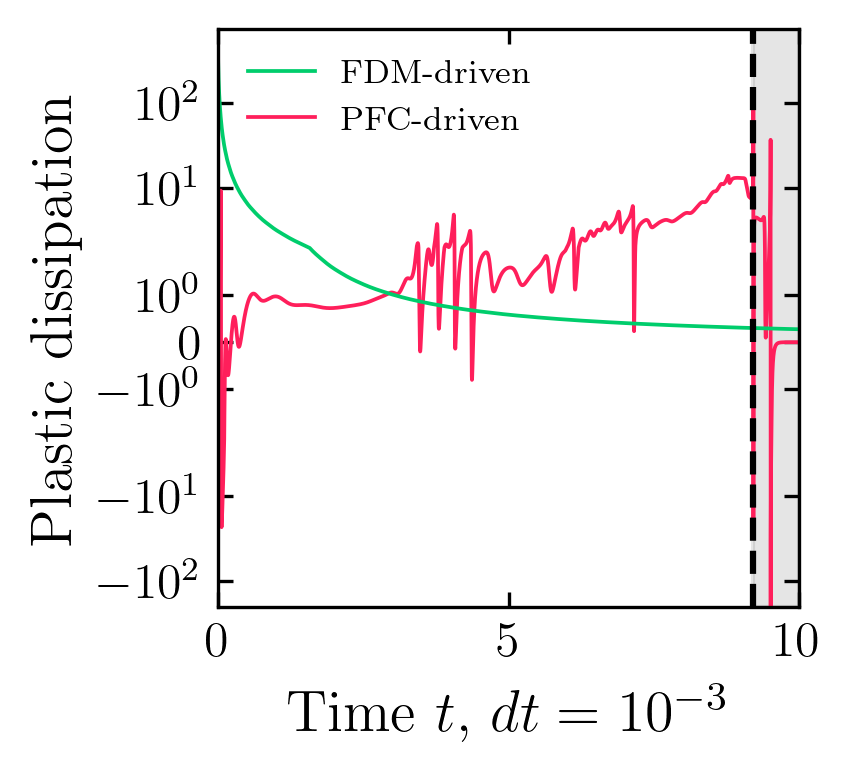}
  \end{subfigure}
    \hspace{1cm}
      \begin{subfigure}[t]{0.3\textwidth}
    \centering
    \caption{$c_{pen}=0.4$}\label{fig:Dissip_04}
    \includegraphics[width=\linewidth]{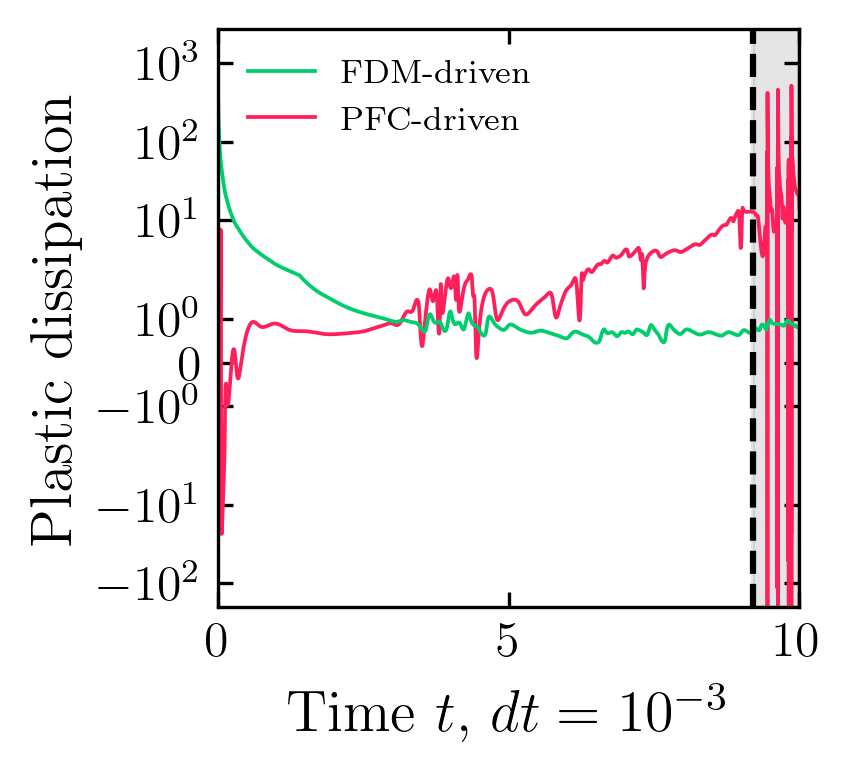}
  \end{subfigure}
  \caption{Plastic and PFC dissipation in the uncoupled case \subref{fig:Dissip_0} and coupled case \subref{fig:Dissip_04} for both PFC and FDM driven cases. Shaded part and dashed black line marks the annihilation start (cores are less than $a_0$ distance). }\label{fig:Dissip}
\end{figure}

In \cref{fig:Dissip}, the plastic dissipation in the FDM-driven case always remains positive but decreases over time, either as the dislocation cores spread in the uncoupled case \cref{fig:Dissip_0}, or as the cores spread and subsequently annihilate in the coupled case \cref{fig:Dissip_04}. 
By contrast, the plastic dissipation in the PFC-driven case is initially negative and changes sign frequently. 
This indicates that the PFC-induced plastic distortion rate $\Tens{\mathcal{J}^\psi}$ is not generally aligned with $\Tens{\sigma}$, so that the resulting motion is not purely stress-driven. 
The fluctuations observed throughout the simulation arise from this misalignment. Even after correction, the evolution of $\psi$ cannot, in general, follow the direction selected by $\Tens{\sigma}$.
The stronger fluctuations observed after the annihilation event, shown by the shaded region in \cref{fig:Dissip}, are attributed to the rapid decay of $\Tens{\sigma}$ toward zero, while the PFC field can only relax diffusively. 
This produces an unphysical oscillatory behavior post-annihilation state.

While the plastic dissipation is negative, the total dissipation after adding the contribution of the PFC dissipation always remains positive for investigated cases using the PFC-driven approach.
However, it is not possible to generalize to stating that the total dissipation for all cases will be non-negative despite the plastic dissipation being negative.
Consequently, there is also a fundamental flaw in the PFC-driven approach.

\section{Conclusion and outlook}

The energetic penalty coupling between the FDM elastic distortion ($\Tens{U}^e$) and PFC configurational distortion ($\Tens{Q}$) proposed in \citep{acharyaFielddislocation2020}, $1/2 \| \Tens{U}^e - \Tens{Q} \|_{L^2}$, is appealing in its simplicity. 
It formulates a thermodynamically consistent attempt to introduce crystallography into an otherwise convex FDM energetic landscape. 
This is accomplished by penalizing the difference between $\Tens{U}^e$ and $\Tens{Q}$ in the $L^2$ sense. 
However, the variational structure reveals a limitation: in the bulk, the forcing depends on $\nabla\cdot(\Tens{Q}-\Tens{U}^e)$, so that divergence free mismatch is invisible to the phase field dynamics. As a result, the coupling can at best drive $\nabla\cdot \Tens{Q} \approx \nabla\cdot \Tens{U}^e$. This amounts to matching primarily the compatible parts of the two distortions, while allowing a mismatch of the incompatible parts that carry all the dislocation information.

Through numerical studies, we have shown that the penalty coupling as discussed does not deliver a reliable bridge between FDM and PFC. 
In the absence of dislocation evolution, the coupled framework can establish an equilibrium between $\Tens{Q}$ and $\Tens{U}^e$, which is reached by pulling the elastic distortion towards the configurational one. 
Allowing the dislocations to evolve according to the dislocation transport equation did not prevent core spreading.
In the dislocation annihilation case, two approaches were followed for modelling plastic distortion rate: (i) the FDM-driven approach that involves solving dislocation transport equation, and (ii) the PFC-driven approach involving setting the FDM plastic distortion rate equal to the dislocation flux from PFC. 
The FDM-driven approach is flawed due to the presence of two distinct dislocation densities $\nabla \times \Tens{Q}$ and $\nabla \times \Tens{U}^e$ that cannot be reconciled due to the curl-free nature of the forcing $\delta \mathcal{F}/\delta \psi$.
The PFC-driven approach uses a single dislocation density field, however, it is more severely flawed because it does not guarantee positive plastic dissipation; negative plastic dissipation was demonstrated for the case of dislocation annihilation.
Furthermore, changing the magnitudes of the penalty coupling factor $c_{pen}$ and the applied stress results either in the numerical scheme becoming unstable, the dislocations remaining pinned, or dislocation movement becoming slower than in the uncoupled diffusive PFC evolution.
In all these cases, elastic relaxation is never achieved.

We conclude that the $L^{2}$ penalty coupling as written does not endow the convex FDM energy with the effects deriving from crystallography, nor consistent mechanics effects into the PFC.

\section*{Acknowledgments}

AG and MVU are grateful to the European Research Council (ERC) for their support through the European Union’s Horizon 2020 research and innovation program for project GAMMA (Grant agreement No. 946959). 
The research of JV is supported by the National Science Foundation, contract no. DMR 2223707.

\clearpage
\begin{appendices}
\section{Notations and Conventions}
\begin{multicols}{2}
\begin{description}[style=nextline,leftmargin=!,labelwidth=1.5cm]
    \item[$\psi$] PFC's order paramter
    \item[$\overline{\psi}$] PFC's domain average
    \item[$r$] PFC's quenching parameter
    \item[$\vec{q_n}$] PFC's $n$-th lattice vector
    \item[$A_n$] Complex amplitude of the $n$-th mode
    \item[$\Tens{\alpha}$] FDM's Dislocation density tensorv
    \item[$\widetilde{\Tens{\alpha}}$] PFC's Dislocation density tensor
    \item[$\Tens{U}^e, \Tens{U}^p$] Elastic distortion and plastic distortion tensors
    \item[$\Tens{U}$] Total distortion tensor
    \item[$\vec{u}$] Displacement vector
    \item[$\Tens{Q}$] PFC's configurational distortion tensor
    \item[$\Tens{\mathcal{J}}$] FDM's plastic distortion rate tensor
    \item[$\Tens{\mathcal{J}^\psi}$] PFC's plastic distortion rate tensor
    \item[$\Tens{\sigma}$] Elastic stress tensor
    \item[$\vec{v}^d$] Dislocation velocity vector
    \item[$1/B$] Dislocation mobility
    \item[$\mathbb{C}$] Fourth-order elastic stiffness tensor
    \item[$\Tens{k}$] Wave vector in Fourier space
\end{description}
\end{multicols}

\medskip
The following typographic conventions are employed throughout this work:
\begin{itemize}
    \item \textbf{Scalars:} Denoted by italic lowercase or Greek letters (e.g., $r, \theta, \psi$).
    \item \textbf{Vectors:} Denoted by boldface italic lowercase letters (e.g., $\boldsymbol{q}, \boldsymbol{u}, \boldsymbol{v}$). Unit vectors are distinguished by a circumflex (e.g., $\hat{\boldsymbol{e}}$).
    \item \textbf{Second-order Tensors:} Denoted by boldface italic uppercase or Greek letters (e.g., $\boldsymbol{U}, \boldsymbol{\alpha}, \boldsymbol{B}$). 
    \begin{itemize}
        \item The identity tensor is denoted by $\mathbf{I}$ (components $\delta_{ij}$).
    \end{itemize}
    \item \textbf{Third-order Tensors:} The Levi-Civita permutation tensor is denoted by $\boldsymbol{\mathcal{X}}$ (components $\varepsilon_{ijk}$).
    \item \textbf{Fourth-order Tensors:} Denoted by blackboard bold (double-stroke) letters (e.g., $\mathbb{C}$).
    \item \textbf{Null Tensor:} Represented by $\mathbf{0}$ regardless of the tensor order.
\end{itemize}

Consider a 2D Cartesian reference frame defined by the orthonormal basis $\{\hat{\boldsymbol{e}}_i\}$ for $i \in \{1, \dots, d\}$ (where $d=2$). Utilizing Einstein summation notation over repeated indices, the following operations are defined:\\

\begin{center}
\begin{small}
\begin{tabular}{@{}ll|ll@{}}
\hline
\multicolumn{2}{l|}{\textbf{Algebraic Products}} & \multicolumn{2}{l}{\textbf{Differential Operators}} \\ \hline
Tensor Prod. & $\boldsymbol{u} \otimes \boldsymbol{v} = u_i v_j \, \hat{\boldsymbol{e}}_i \otimes \hat{\boldsymbol{e}}_j$ & Gradient & $\nabla \psi = \psi_{,i} \hat{\boldsymbol{e}}_i$ \\
Inner Prod. & $\boldsymbol{u} \cdot \boldsymbol{v} = u_i v_i$ & & $\nabla \boldsymbol{u} = u_{i,j} \hat{\boldsymbol{e}}_i \otimes \hat{\boldsymbol{e}}_j$ \\
& $\boldsymbol{\alpha}: \boldsymbol{B} = \alpha_{ij} B_{ij}$ & & $\nabla \boldsymbol{\alpha} = \alpha_{ij,k} \hat{\boldsymbol{e}}_i \otimes \hat{\boldsymbol{e}}_j \otimes \hat{\boldsymbol{e}}_k$ \\ \cline{3-4} 
Dot Prod. & $\boldsymbol{\alpha} \cdot \boldsymbol{B} = \alpha_{ij} B_{jk} \hat{\boldsymbol{e}}_i \otimes \hat{\boldsymbol{e}}_k$ & Divergence & $\nabla \cdot \boldsymbol{u} = u_{i,i}$ \\
& $\boldsymbol{\alpha} \cdot \boldsymbol{u} = \alpha_{ij} u_j \hat{\boldsymbol{e}}_i$ & & $\nabla \cdot \boldsymbol{\alpha} = \alpha_{ij,j} \hat{\boldsymbol{e}}_i$ \\ \cline{3-4} 
Double-Dot & $\boldsymbol{X}: \boldsymbol{\alpha} = \varepsilon_{ijk} \alpha_{jk} \hat{\boldsymbol{e}}_i$ & Curl & $\nabla \times \boldsymbol{u} = \varepsilon_{ijk} u_{k,j} \hat{\boldsymbol{e}}_i$ \\
& $\mathbb{C}: \boldsymbol{\alpha} = C_{ijkl} \alpha_{kl} \hat{\boldsymbol{e}}_i \otimes \hat{\boldsymbol{e}}_j$ & & $\nabla \times \boldsymbol{\alpha} = \varepsilon_{jkl} \alpha_{il,k} \hat{\boldsymbol{e}}_i \otimes \hat{\boldsymbol{e}}_j$ \\ \cline{1-2} 
Cross Prod. & $\boldsymbol{u} \times \boldsymbol{v} = \varepsilon_{ijk} u_j v_k \hat{\boldsymbol{e}}_i$ & Laplacian & $\nabla^2 \psi = \psi_{,ii}$ \\
& $\boldsymbol{\alpha} \times \boldsymbol{u} = \varepsilon_{jkl} \alpha_{ik} u_l \hat{\boldsymbol{e}}_i \otimes \hat{\boldsymbol{e}}_j$ & & $\nabla^2 \boldsymbol{\alpha} = \alpha_{ij,kk} \hat{\boldsymbol{e}}_i \otimes \hat{\boldsymbol{e}}_j$ \\ \hline
\end{tabular}
\end{small}
\end{center}

\newpage
 \section[Derivation of dFudAn]{Derivation of $\delta\mathcal{F}_{pen}/\delta A_n$} \label{app:Var_der_derivation}

First state the analytical expression for the configurational distortion $\Tens{Q}$, as defined in \citep{acharyaFielddislocation2020}:
\begin{equation}
    \Tens{Q} = - \frac{d}{N} \sum_n \vec{q}^n \otimes \text{Im} \left( \frac{\nabla A_n}{A_n} \right) = - \frac{d}{2N i} \sum_n \vec{q}^n \otimes \left[ \frac{\nabla A_n}{A_n} - \frac{\nabla A_n^*}{A_n^*} \right]
\end{equation}
where $d$ is the dimensionality and $N$ is the number of principal lattice vectors.

We now proceed to derive the explicit expression for $\delta\mathcal{F}_{pen}/\delta A_n$.
The variation of $\mathcal{F}_{pen}$ with respect to $A_n$ is given by:
\begin{equation}
    \delta_{A_n} \mathcal{F}_{pen}
    = \int_\Omega \frac{\partial f_{pen}}{\partial \Tens Q}: \delta \Tens Q
    = \int_\Omega \frac{\partial f_{pen}}{\partial \Tens Q}:
    \left[
        \frac{\partial \Tens{Q}}{\partial A_n} \, \delta A_n
        + \frac{\partial \Tens{Q}}{\partial \nabla A_n} \cdot
        \delta (\nabla A_n)
    \right] \mathrm{d}x
\end{equation}

The second term, which contains $\nabla A_n$, is treated using the divergence theorem:
\begin{align}
\int_\Omega
    \frac{\partial f_{pen}}{\partial \Tens Q}:
    \left(
        \frac{\partial \Tens{Q}}{\partial \nabla A_n} \cdot
        \delta (\nabla A_n)
    \right)
    \mathrm{d}x
    &= - \int_\Omega
    \nabla \cdot
    \left(
        \frac{\partial f_{pen}}{\partial \Tens{Q}}:
        \frac{\partial \Tens{Q}}{\partial \nabla A_n}
    \right)
    \delta A_n \, \mathrm{d}x \notag\\
    &\quad +
    \int_{\partial\Omega}
    \left[
        \left(
            \frac{\partial f_{pen}}{\partial \Tens{Q}}:
            \frac{\partial \Tens{Q}}{\partial \nabla A_n}
        \right)
        \cdot \vec{n}
    \right]
    \delta A_n \, \mathrm{d}s
\end{align}

Substituting this into the previous expression of $\delta_{A_n} \mathcal{F}_{pen}$ gives:
\begin{equation}
    \delta_{A_n} \mathcal{F}_{pen}
    = \int_\Omega
    \left[
        \frac{\partial f_{pen}}{\partial \Tens Q}:
        \frac{\partial \Tens{Q}}{\partial A_n}
        - \nabla \cdot
        \left(
            \frac{\partial f_{pen}}{\partial \Tens{Q}}:
            \frac{\partial \Tens{Q}}{\partial \nabla A_n}
        \right)
    \right]
    \delta A_n \, \mathrm{d}x
    + \int_{\partial\Omega}
    \left[
        \left(
            \frac{\partial f_{pen}}{\partial \Tens{Q}}:
            \frac{\partial \Tens{Q}}{\partial \nabla A_n}
        \right)
        \cdot \vec{n}
    \right]
    \delta A_n \, \mathrm{d}s
\end{equation}

By the variational principle, we can identify the bulk contribution and the boundary term:
\begin{equation}
        \forall n, \quad
        \frac{\delta \mathcal{F}_{pen}}{\delta A_n}
        =
        \frac{\partial f_{pen}}{\partial \Tens Q}:
        \frac{\partial \Tens{Q}}{\partial A_n}
        - \nabla \cdot
        \left(
            \frac{\partial f_{pen}}{\partial \Tens{Q}}:
            \frac{\partial \Tens{Q}}{\partial \nabla A_n}
        \right)
        \quad \text{on } \Omega,
        \qquad
        \left(
            \frac{\partial f_{pen}}{\partial \Tens{Q}}:
            \frac{\partial \Tens{Q}}{\partial \nabla A_n}
        \right)
        \cdot \vec{n} = 0
        \quad \text{on } \partial\Omega
\end{equation}

The bulk term can be further expanded as follows:
\begin{align}
    \frac{\delta \mathcal{F}_{pen}}{\delta A_n}
    &=
    \frac{\partial f_{pen}}{\partial \Tens Q}:
    \frac{\partial \Tens{Q}}{\partial A_n}
    - \nabla \cdot
    \left(
        \frac{\partial f_{pen}}{\partial \Tens{Q}}:
        \frac{\partial \Tens{Q}}{\partial \nabla A_n}
    \right) \\
    &=
    \frac{\partial f_{pen}}{\partial \Tens Q}:
    \frac{\partial \Tens{Q}}{\partial A_n}
    - \nabla
    \left(
        \frac{\partial f_{pen}}{\partial \Tens{Q}}
    \right):
    \frac{\partial \Tens{Q}}{\partial \nabla A_n}
    - \frac{\partial f_{pen}}{\partial \Tens{Q}}:
    \nabla \cdot
    \left(
        \frac{\partial \Tens{Q}}{\partial \nabla A_n}
    \right) \\
    &=
    \frac{\partial f_{pen}}{\partial \Tens Q}:
    \underbrace{
        \left[
            \frac{\partial \Tens{Q}}{\partial A_n}
            - \nabla \cdot
            \left(
                \frac{\partial \Tens{Q}}{\partial \nabla A_n}
            \right)
        \right]
    }_{=0}
    -
    \nabla
    \left(
        \frac{\partial f_{pen}}{\partial \Tens{Q}}
    \right):
    \frac{\partial \Tens{Q}}{\partial \nabla A_n}
\end{align}
The term in brackets vanishes due to the structure of the configurational distortion tensor $\Tens Q$ in the small deformation regime:
\begin{equation}\label{eq:dQ_ddkA}
    \Tens{Q}
    = - \frac{d}{2N \mathrm{i}}
    \sum_n \vec{q}^{\,n} \otimes
    \left[
        \frac{\nabla A_n}{A_n} - \frac{\nabla A_n^*}{A_n^*}
    \right]
    \quad \Rightarrow \quad
    \frac{\partial \Tens{Q}_{ij}}{\partial (\partial_k A_n)}
    = - \frac{d}{2N \mathrm{i}}
    q_i^n \frac{1}{A_n} \delta_{jk}
\end{equation}

Hence,
\begin{equation}
    \partial_k
    \left(
        \frac{\partial \Tens{Q}_{ij}}{\partial (\partial_k A_n)}
    \right)
    =
    \frac{d}{2N \mathrm{i}} q_i^n
    \frac{\partial_k A_n}{A_n^2} \delta_{jk}
    =
    \frac{d}{2N \mathrm{i}} q_i^n
    \frac{\partial_j A_n}{A_n^2}
    =
    \frac{\partial \Tens{Q}_{ij}}{\partial A_n}
    \quad \Rightarrow \quad
    \frac{\partial \Tens{Q}}{\partial A_n}
    = \nabla \cdot
    \left(
        \frac{\partial \Tens{Q}}{\partial \nabla A_n}
    \right)
\end{equation}
Thus, the variational derivative depends only on spatial derivatives of
$\partial f / \partial \Tens{Q}$:
\begin{equation}
        \frac{\delta \mathcal{F}_{pen}}{\delta A_n}
        = - \nabla
        \left(
            \frac{\partial f_{pen}}{\partial \Tens{Q}}
        \right):
        \frac{\partial \Tens{Q}}{\partial \nabla A_n}
\end{equation}

Using the structure of $\Tens Q$, the variational derivative simplifies to:
\begin{equation}
    \frac{\delta \mathcal{F}_{pen}}{\delta A_n}
    = - \nabla
    \left(
        \frac{\partial f_{pen}}{\partial \Tens{Q}}
    \right):
    \frac{\partial \Tens{Q}_{ij}}{\partial \nabla A_n}
    = - \sum_{ijk}
    \partial_k
    \left(
        \frac{\partial f_{pen}}{\partial \Tens{Q}_{ij}}
    \right)
    \frac{\partial \Tens{Q}_{ij}}{\partial (\partial_k A_n)}
    =
    \frac{d}{2N \mathrm{i}}
    \sum_{ijk}
    \partial_k
    \left(
        \frac{\partial f_{pen}}{\partial \Tens{Q}_{ij}}
    \right)
    q_i^n \frac{1}{A_n} \delta_{jk}
\end{equation}

Finally, we obtain:
\begin{equation}
    \boxed{
        \frac{\delta \mathcal{F}_{pen}}{\delta A_n}
        = \frac{d}{2N \mathrm{i}}
        \frac{1}{A_n}
        \vec{q}^{\,n} \cdot
        \nabla \cdot
        \frac{\partial f_{pen}}{\partial \Tens{Q}}
        = \frac{d}{2N \mathrm{i}}
        \frac{1}{|A_n|^2}
        \vec{q}^{\,n} \cdot
        \nabla \cdot \left(
        \frac{\partial f_{pen}}{\partial \Tens{Q}} \right) \, A_n^* \qquad \text{in }  \, \Omega}
\end{equation}
We treat the boundary term in a similar fashion. We write:

\begin{equation}
            \left(
            \frac{\partial f_{pen}}{\partial \Tens{Q}}:
            \frac{\partial \Tens{Q}}{\partial \nabla A_n}
        \right)_k = \sum_{ij} \frac{\partial f_{pen}}{\partial \Tens{Q}_{ij}} \frac{\partial \Tens{Q}_{ij}}{\partial (\partial_k A_n)} = - \frac{d}{2N \mathrm{i}}
    \sum_{ij} \frac{\partial f_{pen}}{\partial \Tens{Q}_{ij}}  q_i^n \frac{1}{A_n} \delta_{jk} = - \frac{d}{2N \mathrm{i} A_n}
    \sum_{i} \frac{\partial f_{pen}}{\partial \Tens{Q}_{ik}}  q_i^n
\end{equation}
Thus

\begin{equation}
            \left(
            \frac{\partial f_{pen}}{\partial \Tens{Q}}:
            \frac{\partial \Tens{Q}}{\partial \nabla A_n}
        \right) =  - \frac{d}{2N \mathrm{i} A_n} \left(\frac{\partial f_{pen}}{\partial \Tens{Q}}\right)^t  \cdot \vec{q^n}
\end{equation}
Finally we obtain on the boundary:

\begin{equation}
    \boxed{
        \left(
            \frac{\partial f_{pen}}{\partial \Tens{Q}}:
            \frac{\partial \Tens{Q}}{\partial \nabla A_n}
        \right)
        \cdot \vec{n} = 0
        \qquad \Longleftrightarrow \qquad
        \vec{n}\cdot \left(\frac{\partial f_{pen}}{\partial \Tens{Q}}\right)^t  \cdot \vec{q^n} = 0
        \quad \text{on } \partial\Omega}
\end{equation}

    \newpage
\section{Numerical methods}\label{app:Methods}
The previous simulations were all performed in periodic boxes using spectral methods with the \texttt{scipy.fft} package in \texttt{python}, with the exception of the dislocation transport equation, which was solved using a finite volume method. All details are given below.

\subsection{Phase field crystal}
\subsubsection{Evolution equation}
The evolution equation of the phase field crystal for a hexagonal lattice reads:
\begin{equation}
    \cfrac{\partial \psi}{\partial t} =  -\left[ \frac{\delta \mathcal{F}_{sh}}{\delta \psi} + c_{pen}\frac{\delta \mathcal{F}_{pen}}{\delta \psi}\right] = -\left[  (1-\varepsilon)\psi + \Delta^2\psi+2 \Delta \psi +\psi^3   + c_{pen}\frac{\delta \mathcal{F}_{pen}}{\delta \psi}\right]
\end{equation}
In Fourier space, we have:
\begin{equation}\label{eq:expanded}
    \widehat{\cfrac{\partial  \psi}{\partial t}} = -\left[ \left( (1-\epsilon) +|\Tens{k}|^4 -2|\Tens{k}|^2 \right) \widehat \psi   + \widehat{\psi^3} + c_{pen} \widehat{\frac{\delta \mathcal{F}_{pen}}{\delta \psi}}\right] = -\widehat{\mathcal{L}}(\Tens{k})\widehat{\psi}-\widehat{\mathcal{N}}(\psi),
\end{equation}
where we define:
\begin{equation}
\widehat{\mathcal{L}}(\Tens{k})=(1-\varepsilon)+|\Tens{k}|^4-2|\Tens{k}|^2,\qquad
\widehat{\mathcal{N}}(\psi)=\widehat{\psi^3}+c_{pen}\,\widehat{\frac{\delta\mathcal{F}_{pen}}{\delta\psi}}
\end{equation}
The evolution equation is integrated using exponential time differencing (ETD1) \citep{coxExponentialTimeDifferencing2002}: 
\begin{equation}\label{eq:PFC_evol_disct}
    \begin{aligned}
&\widehat{\psi}^{\,n+1}(\Tens{k})=e^{-\widehat{\mathcal{L}}(\Tens{k})\Delta t}\widehat{\psi}^{\,n}(\Tens{k})
-\frac{1-e^{-\widehat{\mathcal{L}}(\Tens{k})\Delta t}}{\widehat{\mathcal{L}}(\Tens{k})}\,
\widehat{\mathcal{N}}(\psi^n)(\Tens{k}) = I_0 \widehat{\psi}^{\,n} + I_1 \widehat{\mathcal{N}}(\psi^n)\\
&\psi^{\,n+1} = \mathcal{F}^{-1}(\widehat{\psi}^{\,n+1})
    \end{aligned}
\end{equation}
The integration constants $I_0 = e^{-\widehat{\mathcal{L}}(\Tens{k})\Delta t}$ and $I_1 = -(1-I_0)/\widehat{\mathcal{L}}(\Tens{k})$ are computed once at the beginning of the simulation. The nonlinear term is evaluated pseudospectrally, forming ${\mathcal{N}}(\psi^n)$ in real space, followed by an FFT to obtain $\widehat{\mathcal{N}}(\psi^n)$.\\
It's worth nothing that using non conserved dynamics can result in localized stripping and loss of the hexagonal phase . A numerical solution to stabilize the hexagonal lattice was to correct for the total average instead thus:
\begin{equation}\label{eq:AVG_corr}
    \psi^{n+1} = \overline{\psi} - \frac{1}{|\Omega|} \int_\Omega \psi^{n+1}  \dd{x}
\end{equation}
where $\overline{\psi}$ is the initial average that we keep constant.

\subsubsection{Demodulation and field computation}

The configurational distortion computed from $\Tens Q$ in \cref{eq:Q_def} is highly singular exhibits variation on the sub-lattice scale which are not physical. Thus, we use a gaussian filter of width $a_0/4$ to smooth the configurational distortion field:
\begin{equation}\label{eq:smoothq}
    \widetilde{\Tens{Q}}(\vecc{r}) = \mathcal{G}_{a_0/4} * \Tens{Q}(\vecc{r}) 
\end{equation}
The dislocation density tensor is given by:
\begin{equation}
    \widetilde{\Tens{\alpha}} = \nabla \times \widetilde{\Tens{\Tens{Q}}}
\end{equation}
These smoothing operations \cref{eq:demodulation,eq:smoothq} are performed in Fourier space.\\

\begin{figure}[H]
  \centering
    \includegraphics[width=0.4\linewidth]{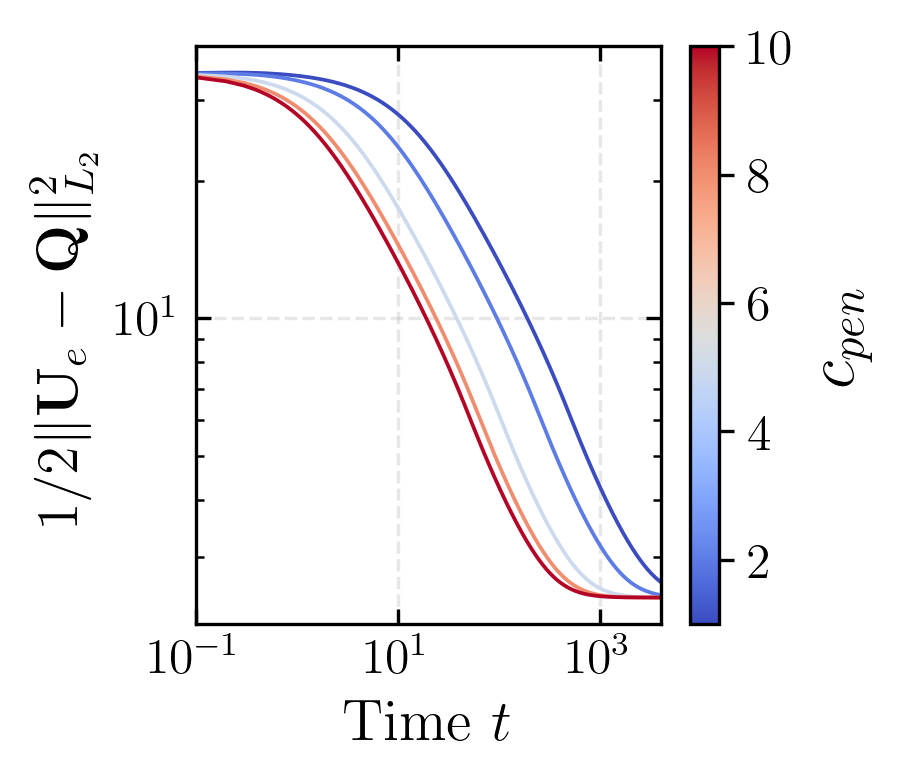}
    \caption{Decay of $\mathcal{F}_{pen}$ in penalty only case for for an immobile dipole in climb. For numerical validation.}\label{eq:PenOnly}
  \end{figure}

The nonlinear coupling term $\delta\mathcal{F}_{pen}/\delta\psi$ is computed from \cref{eq:var^psi_opt,eq:Func_derv_Fu} using Fourier transform:
\begin{equation}
    \frac{\delta \mathcal{F}_{pen}}{\delta \psi (\vec{x})} = 2 \Re \sum_n \left[ \mathcal{F}^{-1}\left(\widehat{\frac{\delta\mathcal{F}_{pen}}{\delta A_n}} \widehat{\mathcal{G}_{a_0}} \right) \, e^{-i \vec{q}_n \cdot \vec{x}} \right]
\end{equation}
In quest of efficiency, an automatic differentiation library was used in the implementation based on \texttt{jax.autograd} \citep{jax2018github}. The implementation was tested in the immobile case of \cref{sec:immobile} with the same same spatial and time discretizations. However, the elastic distortion $\Tens{U}^e$ is fixed and the evolution solved is only $\partial_t \psi = -c_{pen} \delta \mathcal{F}_{pen}/\delta \psi (\vec{x})$ with no Swift-Hohenberg energy contribution. The system stably decreases $\mathcal{F}_{pen}$ for the tested $c_{pen}$ from 1 to 10 as shown in \cref{eq:PenOnly}. This validates our implementation of $\delta \mathcal{F}_{pen}/\delta \psi (\vec{x})$ .

\subsection{Dislocation Transport: Edge Dislocation in $2$D}
In the framework of Field Dislocation Mechanics (FDM), the evolution of the dislocation density tensor $\Tens{\alpha}$ is governed by the transport equation:
\begin{equation}\label{eq:transport_annex}
\dot{\Tens{\alpha}} + \nabla \times \big(\Tens{\alpha} \times \vec{v}^{d}\big) = \Tens{0}.
\end{equation}
We consider a 2D setting in the $(x,y)$ plane, assuming invariance along the $z$-axis and an in-plane dislocation velocity $\vec{v}^{d} = (v_x, v_y, 0)$. For an edge dislocation with a line direction along $z$, the non-zero components of $\Tens{\alpha}$ are $\alpha_{xz}$ and $\alpha_{23}$. Under these conditions, the transport equation reduces to two independent scalar advection equations in conservation form:
\begin{equation}
\dot \alpha_{k3} + \nabla \cdot (\alpha_{k3} \vec{v}^{d}) = 0, \quad \text{for } k \in \{1, 2\}.
\end{equation}
\subsubsection{Numerical Implementation}
These equations are solved using a vertex-centered finite volume method on a uniform grid ($\Delta x, \Delta y$) with MUSCL reconstruction \citep{FiniteVolumeMethods2000}, an MC slope limiter \citep{vanleerUltimateConservativeDifference1979}, time integration is done via third-order Strong Stability Preserving Runge-Kutta (SSPRK3) scheme \citep{SSPRKbook}.

Let $\alpha$ represent $\alpha_{xz}$ or $\alpha_{23}$. The semi-discrete scheme at node $(i,j)$ is:
\begin{equation}\label{eq:semi-discrete}
\frac{d\alpha_{i,j}}{dt} = -\frac{1}{\Delta x \Delta y } \left( \Phi^{L}_{i,j} - \Phi^{R}_{i,j} + \Phi^{T}_{i,j} - \Phi^{B}_{i,j} \right),
\end{equation}
where $\Phi$ are the numerical fluxes across the edges.
\begin{align}
\Phi^{L}_{i,j} &= \left(v_{x,i+\frac12,j}\,\Delta y\right) \alpha^{up}_{i+\frac12,j}, &
\Phi^{R}_{i,j} &= \left(v_{x,i-\frac12,j}\,\Delta y\right) \alpha^{up}_{i-\frac12,j}, \\
\Phi^{T}_{i,j} &= \left(v_{y,i,j+\frac12}\,\Delta x\right) \alpha^{up}_{i,j+\frac12}, &
\Phi^{B}_{i,j} &= \left(v_{y,i,j-\frac12}\,\Delta x\right) \alpha^{up}_{i,j-\frac12}.
\end{align}
Face velocities are calculated via arithmetic averaging:
\begin{equation}
v_{x,i+\frac12,j} = \frac{v_{x,i,j} + v_{x,i+1,j}}{2}, \quad v_{y,i,j+\frac12} = \frac{v_{y,i,j} + v_{y,i,j+1}}{2}.
\end{equation}
he upwind states $\alpha^{up}$ are determined using MUSCL reconstruction with limited slopes. For the horizontal faces:
\begin{equation}
\alpha^{up}_{i+\frac12,j} = \begin{cases} 
\alpha_{i,j} + \frac{1}{2}s^x_{i,j} & \text{if } v_{x,i+\frac12,j} \ge 0 \\
\alpha_{i+1,j} - \frac{1}{2}s^x_{i+1,j} & \text{if } v_{x,i+\frac12,j} < 0
\end{cases},\qquad
\alpha^{up}_{i-\frac12,j} = \begin{cases} 
\alpha_{i-1,j} + \frac{1}{2}s^x_{i-1,j} & \text{if } v_{x,i-\frac12,j} \ge 0 \\
\alpha_{i,j} - \frac{1}{2}s^x_{i,j} & \text{if } v_{x,i-\frac12,j} < 0
\end{cases}
\end{equation}
And for the vertical faces:
\begin{equation}
\alpha^{up}_{i,j+\frac12} = \begin{cases} 
\alpha_{i,j} + \frac{1}{2}s^y_{i,j} & \text{if } v_{y,i,j+\frac12} \ge 0 \\
\alpha_{i,j+1} - \frac{1}{2}s^y_{i,j+1} & \text{if } v_{y,i,j+\frac12} < 0
\end{cases}, \qquad
\alpha^{up}_{i,j-\frac12} = \begin{cases} 
\alpha_{i,j-1} + \frac{1}{2}s^y_{i,j-1} & \text{if } v_{y,i,j-\frac12} \ge 0 \\
\alpha_{i,j} - \frac{1}{2}s^y_{i,j} & \text{if } v_{y,i,j-\frac12} < 0
\end{cases}
\end{equation}
The slopes $s^x$ and $s^y$ are computed using the Monotonized Central (MC) limiter $\Lambda$:
\begin{equation}
s^x_{i,j} = \Lambda(\alpha_{i,j}-\alpha_{i-1,j}, \alpha_{i+1,j}-\alpha_{i,j}), \qquad
s^y_{i,j} = \Lambda(\alpha_{i,j}-\alpha_{i,j-1}, \alpha_{i,j+1}-\alpha_{i,j})
\end{equation}

Denoting by $\mathcal{L}(\alpha)$ the finite-volume right-hand in \cref{eq:semi-discrete}, we advance in time with SSPRK3:
\begin{equation}\label{eq:FDM_transport_full}
    \begin{aligned}
        &\alpha^{(1)} &= \alpha^{n} + \Delta t\,\mathcal{L}\!\left(\alpha^{n}\right),\\
        &\alpha^{(2)} &= \tfrac34 \alpha^{n} + \tfrac14\Big(\alpha^{(1)} + \Delta t\,\mathcal{L}\!\left(\alpha^{(1)}\right)\Big),\\
        &\alpha^{n+1} &= \tfrac13 \alpha^{n} + \tfrac23\Big(\alpha^{(2)} + \Delta t\,\mathcal{L}\!\left(\alpha^{(2)}\right)\Big).
    \end{aligned}
\end{equation}

Stability is ensured by the CFL condition:
\begin{equation}
\Delta t \le C_{\text{cfl}} \left( \frac{\max|v_x|}{\Delta x} + \frac{\max|v_y|}{\Delta y} \right)^{-1}
\end{equation}
where $C_{\text{cfl}} = 0.5$ is used.

\subsection{Plastic distortion}
At each time step $t_{n+1}$, we use a Stokes-Helmholtz decomposition:
\begin{equation}
\Tens{U}^p_{n+1}=\Tens{U}^{p,\perp}_{n+1}+\Tens{U}^{p,\parallel}_{n+1},\qquad
\Tens{U}^{p,\parallel}_{n+1}=\nabla \vec{z}^p_{n+1},
\end{equation}
where $\nabla\times\Tens{U}^{p,\parallel}_{n+1}=\Tens{0}$ and $\nabla\cdot\Tens{U}^{p,\perp}_{n+1}=0$.

\subsubsection{Incompatible part}
The incompatible part depend only on the current dislocation state. It is defined by the div-curl system:
\begin{equation}
\nabla\times\Tens{U}^{p,\perp}_{n+1}=-\Tens{\alpha}_{n+1},\qquad
\nabla \cdot \Tens{U}^{p,\perp}_{n+1}=0,
\end{equation}
It can be recast into a Poisson-type equation:
\begin{equation}
   \nabla^2 \Tens{U^{p,\perp}} = - \nabla \times \Tens \alpha
\end{equation}
It is solved spectrally:
\begin{equation}\label{eq:divcurlFT}
   \Tens{U^{p,\perp}} = \mathcal{F}^{-1}\left(
      \begin{cases}
         i\cfrac{(\Tens{\hat{\alpha}}\cdot\Tens{X})\cdot \vecc{k}}{|\vecc{k}|}&, \: |\vecc{k}|\neq 0\\
         0&, \: |\vecc{k}|= 0\\
      \end{cases}
   \right)
\end{equation}
where $\Tens{X}$ is the third order levi-civita tensor, and $\Tens{\hat{\alpha}}=\mathcal{F}(\Tens{\alpha})$ the fourier transorm.
\subsubsection{Compatible part}
Given the plastic distortion rate $\Tens{\mathcal{J}}_{n+1}$, either computed from FDM or from PFC, the compatible part rate satisfies $\Delta(\dot{\Tens{\vec{z}}}^{\Tens{p}})=\nabla\cdot\Tens{\mathcal{J}}_{n+1}$. In Fourier space:
\begin{equation}
    \dot{\Tens{\vec{z}}}^{\Tens{p}}  = \mathcal{F}^{-1}\left[-\ii \frac{\hat{\Tens{\mathcal{J}}}\cdot \vec{k}}{|\vec{k}|^2} \right]
\end{equation}
and we update using a forward Euler scheme:
\begin{equation}\label{eq:comp}
    \Tens{U^{p \para}}_{n+1} =\Tens{U^{p \para}}_{n} +dt \, \dot  \nabla \Tens{z^p} 
\end{equation}
Finally, the plastic distortion is obtained by summing \cref{eq:divcurl} and \cref{eq:comp}
\begin{equation}\label{eq:UPfull}
    \Tens{U^p}_{n+1} = \Tens{U^{p \para}}_{n+1} +\Tens{U^{p \perp}}_{n+1}
\end{equation} 

\subsection{Mechanical Equilibrium}
At a given timestep and given $\Tens{Q}$ and $\Tens{U^p}$, mechanical equilibrium $\nabla \cdot \Tens{\sigma} = \Tens{0}$ is solved for the displacement $\vec{u}^{n+1}$. The stress is:
\begin{equation}
\Tens{\sigma} = \mathbb{C}: (\Tens{U} - \Tens{U}^p) + c_{pen} (\Tens{U} - \Tens{U}^p - \Tens{Q})_{\text{sym}}.
\end{equation}
In Fourier space, defining $\widehat{\Tens{A}}_{ij} = \mathbb{C}_{ijkl} \widehat{\Tens{U}}^p + c_{pen} (\widehat{\Tens{U}}^p + \widehat{\Tens{Q}})_s$ and noting $\Tens{U} = \nabla \vec{u} \rightarrow \widehat{U}_{il} = i k_l \widehat{u}_i$, the equilibrium equation becomes:
\begin{equation}
    \left[ k_j k_l \mathbb{C}_{ijkl} + 1/2 |\Tens{k}|^2 c_{pen} \delta_{ik} \right] \widehat{u}_k = -i k_j \widehat{A}_{ij}.
\end{equation}
Defining the matrix $N_{ik} = k_j k_l \mathbb{C}_{ijkl}$, the displacement is found by inversion:
\begin{equation}\label{eq:Mec_eq}
      \widehat{u}_k = -i \left( N_{ik} + 1/2 |\Tens{k}|^2 c_{pen} \delta_{ik} \right)^{-1} k_j \widehat{A}_{ij}\quad \text{and} \quad {U}_{kl} = \mathcal{F}^{-1}(i k_l \widehat{u}_k)
\end{equation}
The total distortion $\Tens{U}$ is then computed as $\widehat{U}_{kl} = i k_l \widehat{u}_k$ and transformed back to real space.
\subsection{Algorithm}
The generic simulation algorithm is shown bellow. In the case of PFC-driven motion we compute the flux and the dislocation density from the current PFC state: the red equation replace the FDM updates. Finally, and optionally, we apply a macroscopic distortion or stress.\\

\SetKwComment{Comment}{/* }{ */}
\RestyleAlgo{ruled}
\begin{algorithm}[H]
   \caption{Coupled PFC-FDM, small deformation}\label{alg:CoupledPFCFDM}
   \KwData{$\Omega,dx,dy,dt,\mathbb{C},\psi(t=0),\overline{psi},r,c_{sh}, c_{w}$}
   \KwResult{$\Tens{\alpha}, \Tens Q,\Tens{U^p},\Tens U$, $\psi $}
   $t \gets 0$ \;
   $\psi^{0} \gets \psi(0)$\;
    $A_n^{0} \gets$ RHS of \cref{eq:CmpA}\;
   $\Tens{Q}^{0} \gets$ RHS of \cref{eq:smoothq}\;
   ${{\Tens{\alpha}}}^{0} \gets \nabla \times \Tens{\Tens{Q}}^{0} $ \;
   $(\Tens{U^p}^{\para})^{0}\gets 0$ \Comment*[r]{Assume no plastic history in the beginning}
   $(\Tens{U^p}^{\perp})^{0}\gets $ solution of \cref{eq:divcurl}\;
   $(\Tens{U^p})^{0}\gets (\Tens{U^p}^{\perp})^{0} $\;
   $\Tens{U}^{0} \gets$ solution of \cref{eq:Mec_eq}\;
   $\Tens{U}^{e,0} \gets \Tens{U}^{0}-\Tens{U}^{p,0}$\;
   \While{$t \leq T$}{
      $\psi^{n+1} \gets$ solution of \cref{eq:PFC_evol_disct}\;
      $\psi^{n+1} \gets$ RHS \cref{eq:AVG_corr}\Comment*[r]{Average correction}
      $A_n^{n+1} \gets$ RHS of \cref{eq:CmpA}\;
      $\Tens{Q}^{n+1} \gets$ RHS of \cref{eq:smoothq}\;
      ${\Tens{\mathcal{J}}} \gets \Tens{\alpha}^{t}\times (\vec{v^d})^{t}$ RHS of \cref{eq:current} \textbf{or} \textcolor{red}{${\Tens{\mathcal{J}}} \gets {\Tens{\mathcal{J}^\psi} =}$ RHS of \cref{eq:current}}\;
      $\Tens{\alpha}^{n+1} \gets $ RHS of \cref{eq:FDM_transport_full} \textbf{or} \textcolor{red}{$\Tens{\alpha}^{n+1} \gets \nabla \times \Tens{\Tens{Q}}^{n+1}$}\;
      $\Tens{U^p}^{\perp\,, n+1}\gets $ solution of \cref{eq:divcurl}\;
      $\Tens{U^p}^{\para\,, n+1}\gets $ solution of \cref{eq:comp}\;
      $\Tens{U^p}^{n+1}\gets $ RHS of \cref{eq:UPfull}\;
      $\Tens{U}^{n+1} \gets$ RHS of \cref{eq:Mec_eq}\;
      \textcolor{blue}{(Optional): Apply macroscopic stress} \;
      $\Tens{U}^{e,n+1} \gets \Tens{U}^{n+1}-\Tens{U}^{p,n+1}$\;
      $t \gets t+dt$\;
      
   }
\end{algorithm}
\newpage
\section{Additional figures}
\subsection{Peach-Koehler force fit PFC reference for FDM drag coefficient}\label{app:PKfit2PFC}
\begin{figure}[H]
  \centering
    \includegraphics[width=0.4\linewidth]{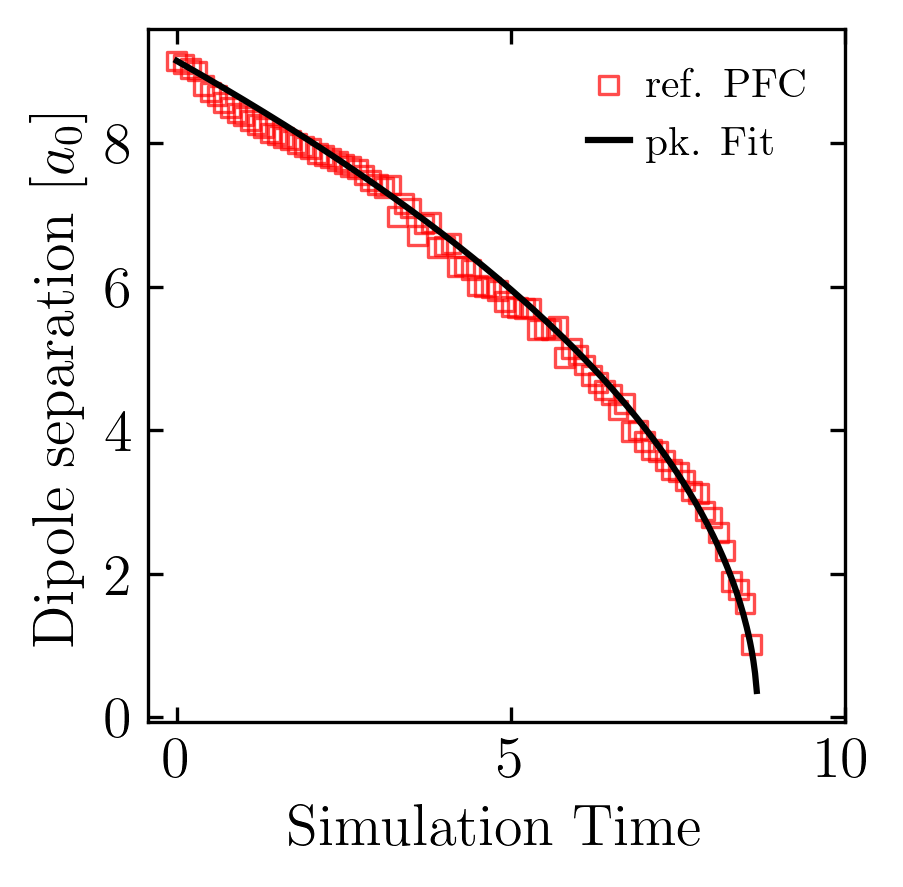}
    \caption{Time evolution of $\widetilde{\boldsymbol{\alpha}}$ and the Peach-Koehler (pk) fit for $B$ in \cref{eq:PK_distance}}
  \end{figure}
\end{appendices}

\bibliography{Library}
\bibliographystyle{abbrv}
\end{document}